\pdfoutput=1

\documentclass[11pt,twoside,a4paper,cmspaper,final,collab]{cms-tdr}
\def\svnVersion{449177P}\def\cmsCernNoTag{CERN-EP-2017-181}\def\cmsCernDate{\today}\def\cmsMessage{Published in Physics Letters B as \href{http://dx.doi.org/10.1016/j.physletb.2018.02.004}{\doi{10.1016/j.physletb.2018.02.004.}}}

\begin{document}\cmsNoteHeader{HIG-16-043}

\hyphenation{had-ron-i-za-tion}
\hyphenation{cal-or-i-me-ter}
\hyphenation{de-vices}
\RCS$Revision: 433631 $
\RCS$HeadURL: svn+ssh://svn.cern.ch/reps/tdr2/papers/HIG-16-043/trunk/HIG-16-043.tex $
\RCS$Id: HIG-16-043.tex 433631 2017-11-09 17:41:03Z alverson $
\newlength\cmsFigWidth
\ifthenelse{\boolean{cms@external}}{\setlength\cmsFigWidth{0.48\textwidth}}{\setlength\cmsFigWidth{0.65\textwidth}}
\ifthenelse{\boolean{cms@external}}{\providecommand{\cmsLeft}{top}}{\providecommand{\cmsLeft}{left}}
\ifthenelse{\boolean{cms@external}}{\providecommand{\cmsRight}{bottom}}{\providecommand{\cmsRight}{right}}
\providecommand{\NA}{\ensuremath{\text{ --- }}\xspace}
\newlength\cmsTabSkip\setlength\cmsTabSkip{1.6ex}
\newcommand{\MT}{\ensuremath{m_\mathrm{T}}\xspace}
\newcommand{\mvis}{\ensuremath{m_\text{vis}}\xspace}
\newcommand{\mtt}{\ensuremath{m_{\Pgt\Pgt}}\xspace}
\newcommand{\hww}{\ensuremath{\PH\to\PW\PW}\xspace}
\newcommand{\pth}{\ensuremath{\pt^{\tau\tau}}\xspace}
\newcommand{\mjj}{\ensuremath{m_\mathrm{jj}}\xspace}
\newcommand{\aMCATNLO} {\textsc{MG5}\_a\MCATNLO\xspace}
\newcommand{\emu}{\ensuremath{\Pe\Pgm}\xspace}
\providecommand{\mH}{\ensuremath{m_{\PH}}\xspace}

\cmsNoteHeader{HIG-16-043}
\title{Observation of the Higgs boson decay to a pair of $\tau$ leptons with the CMS detector}

\date{\today}

\abstract{
A measurement of the $\PH\to\Pgt\Pgt$ signal strength is performed using events recorded in proton-proton collisions by the CMS experiment at the LHC in 2016 at a center-of-mass energy of 13\TeV.
The data set corresponds to an integrated luminosity of 35.9\fbinv. The $\PH\to \Pgt \Pgt$ signal is established with a significance of 4.9 standard deviations, to be compared to an expected significance of 4.7 standard deviations. The best fit of the product of the observed $\PH\to \Pgt \Pgt$ signal production cross section and branching fraction is $1.09^{+0.27}_{-0.26}$ times the standard model expectation. The combination with the corresponding measurement performed with data collected by the CMS experiment at center-of-mass energies of 7 and 8\TeV leads to an observed significance of 5.9 standard deviations, equal to the expected significance. This is the first observation of Higgs boson decays to $\Pgt$ leptons by a single experiment.
}

\hypersetup{%
pdfauthor={CMS Collaboration},%
pdftitle={Observation of the Higgs boson decay to a pair of tau leptons with the CMS detector},%
pdfsubject={CMS},%
pdfkeywords={CMS, physics, tau, Higgs, observation, LHC}}

\maketitle

\section{Introduction}

In the standard model (SM) of particle physics~\cite{Glashow:1961tr,SM1,SM3},
electroweak symmetry breaking is achieved via the Brout--Englert--Higgs
mechanism~\cite{Englert:1964et,Higgs:1964ia,Higgs:1964pj,Guralnik:1964eu,Higgs:1966ev,Kibble:1967sv},
leading, in its minimal version, to the prediction of the existence of one physical neutral scalar particle,
commonly known as the Higgs boson ($\PH$).
A particle compatible with such a boson was observed by the ATLAS and CMS experiments at the CERN LHC
in the $\cPZ\cPZ$, $\Pgg \Pgg$, and $\PWp\PWm$ decay channels~\cite{Aad:2012tfa, Chatrchyan:2012xdj, Chatrchyan:2013lba},
during the proton-proton ($\Pp\Pp$) data taking period in 2011 and 2012
at center-of-mass energies of $\sqrt{s} = 7$ and 8\TeV, respectively.
Subsequent results from both experiments, described in
Refs.~\cite{Aad:2015gba, Khachatryan:2014jba, Chatrchyan:2012jja, Aad:2013xqa, Khachatryan:2014kca,Sirunyan:2017exp},
established that the measured properties of the new particle,
including its spin, CP properties,
and coupling strengths to SM particles, are consistent with those expected for the Higgs boson predicted by the SM.
The mass of the Higgs boson has been determined to be
$125.09\pm0.21\stat\pm0.11\syst\GeV$, from a combination of
ATLAS and CMS measurements~\cite{Aad:2015zhl}.

To establish the mass generation mechanism for fermions,
 it is necessary to probe the direct coupling of
the Higgs boson to such particles.
The most promising decay channel is $\Pgt^+\Pgt^-$,
because of the large event rate expected in the SM compared to the $\Pgm^+\Pgm^-$ decay channel ($\mathcal{B}(\PH\to\Pgt^+\Pgt^-)=6.3$\% for a mass of 125.09\GeV), and of the smaller contribution from background events
with respect to the $\bbbar$ decay channel.

Searches for a Higgs boson decaying to a $\Pgt$ lepton pair were performed at the LEP~\cite{Barate:2000ts,Abdallah:2003ip,Achard:2001pj,Abbiendi:2000ac},
Tevatron~\cite{Aaltonen:2012jh, Abazov:2012zj}, and LHC colliders.
Using $\Pp\Pp$ collision data at $\sqrt{s}=7$ and $8\TeV$, the CMS Collaboration showed evidence for this process with an observed\,(expected)
significance of 3.2\,(3.7) standard deviations (s.d.)~\cite{Chatrchyan:2014nva}. The ATLAS
experiment reported evidence for Higgs bosons decaying into pairs
of $\Pgt$ leptons with an observed (expected) significance of 4.5 (3.4)
s.d. for a Higgs boson mass of 125\GeV~\cite{Aad:2015vsa}.
The combination of the results from both experiments yields an observed (expected)
significance of 5.5\,(5.0) s.d.~\cite{Khachatryan:2016vau}.

This Letter reports on a measurement of the $\PH\to\Pgt\Pgt$ signal strength.
The analysis targets both the gluon fusion and the vector boson fusion production mechanisms.
The analyzed data set corresponds to an integrated luminosity of 35.9\fbinv, and was collected in 2016 in $\Pp\Pp$ collisions at a center-of-mass energy of
13\TeV.
In the following, the symbol $\ell$ refers to electrons or muons,
the symbol $\tauh$ refers to $\Pgt$ leptons reconstructed in their hadronic decays, and
$\PH\to\Pgt^+\Pgt^-$  and $\PH\to\PW^+\PW^-$ are simply denoted as $\PH\to\Pgt\Pgt$  and $\PH\to\PW\PW$, respectively.
All possible $\Pgt\Pgt$ final states are studied, except for those with two muons or two electrons because of the low branching fraction and large background contribution. The analysis covers about 94\% of all possible $\Pgt\Pgt$ final states.

\section{The CMS detector}

The central feature of the CMS apparatus is a superconducting solenoid of 6\unit{m} internal diameter, providing a magnetic field of 3.8\unit{T}. Within the solenoid volume, there are a silicon pixel and strip tracker, a lead tungstate crystal electromagnetic calorimeter (ECAL), and a brass and scintillator hadron calorimeter (HCAL), each composed of a barrel and two endcap sections. Forward calorimeters extend the pseudorapidity coverage provided by the barrel and endcap detectors. Muons are detected in gas-ionization chambers embedded in the steel flux-return yoke outside the solenoid.

Events of interest are selected using a two-tiered trigger system~\cite{Khachatryan:2016bia}. The first level (L1), composed of custom hardware processors, uses information from the calorimeters and muon detectors to select events at a rate of around 100\unit{kHz} within a time interval of less than 4\mus. The second level, known as the high-level trigger (HLT), consists of a farm of processors running a version of the full event reconstruction software optimized for fast processing, and reduces the event rate to about 1\unit{kHz} before data storage.

Significant upgrades of the L1 trigger during the first long shutdown of the LHC have benefitted this analysis, especially in the $\tauh\tauh$ channel. These upgrades improved the $\tauh$ identification at L1 by giving more flexibility to object isolation, allowing new techniques to suppress the contribution from additional $\Pp\Pp$ interactions per bunch
crossing, and to reconstruct the L1 $\tauh$ object in a fiducial region that matches more closely that of a true hadronic $\Pgt$ decay. The flexibility is achieved by employing high bandwidth optical links for data communication and large field-programmable gate arrays (FPGAs) for data processing.

A more detailed description of the CMS detector, together with a definition of the coordinate system used and the relevant kinematic variables, can be found in Ref.~\cite{Chatrchyan:2008zzk}.

\section{Simulated samples}

Signal and background processes are modeled with samples of simulated events.
The signal samples with a Higgs boson produced through gluon fusion ($\cPg\cPg\PH$), vector boson fusion (VBF),
or in association with a $\PW$ or $\PZ$ boson ($\PW\PH$ or $\PZ\PH$), are generated at next-to-leading order (NLO) in perturbative quantum chromodynamics (pQCD) with the \POWHEG 2.0~\cite{Nason:2004rx,Frixione:2007vw, Alioli:2010xd, Alioli:2010xa, Alioli:2008tz} generator. The \textsc{minlo hvJ}~\cite{Luisoni:2013kna} extension of \POWHEG 2.0 is used for the $\PW\PH$ and $\PZ\PH$ simulated samples. The set of parton distribution functions (PDFs) is NNPDF30\_nlo\_as\_0118~\cite{Ball:2011uy}. The $\ttbar\PH$ process is negligible.
The various production cross sections and branching fractions for the SM Higgs boson production, and their corresponding uncertainties are taken from Refs.~\cite{deFlorian:2016spz,Denner:2011mq,Ball:2011mu} and references therein.

The \aMCATNLO~\cite{Alwall:2014hca} generator is used for $\PZ+\text{jets}$ and $\PW+\text{jets}$ processes. They are simulated at leading order (LO) with the MLM jet matching and merging~\cite{Alwall:2007fs}.
The \aMCATNLO generator is also used for diboson production simulated at next-to-LO (NLO) with the FxFx jet matching and merging~\cite{Frederix:2012ps}, whereas $\POWHEG$ 2.0 and 1.0 are used for $\ttbar$ and single top quark production, respectively.
The generators are interfaced with \PYTHIA 8.212 ~\cite{Sjostrand:2014zea} to model the parton showering and fragmentation, as well as the decay of the $\Pgt$ leptons.
The \PYTHIA parameters affecting the description of the underlying event are set to the {CUETP8M1} tune~\cite{Khachatryan:2015pea}.

Generated events are processed through a simulation of the CMS detector based on
\GEANTfour~\cite{Agostinelli:2002hh}, and are reconstructed with the same algorithms used for data.
The simulated samples include additional $\Pp\Pp$ interactions per bunch
crossing, referred to as ``pileup''.
The effect of pileup is taken into account by generating concurrent minimum bias collision events generated with \PYTHIA.
The simulated events are weighted such that the distribution of the number of additional pileup interactions, estimated from the measured instantaneous luminosity for each bunch crossing, matches that in data, with an average of approximately 27 interactions per bunch crossing.

\section{Event reconstruction}
\label{sec:reconstruction}

The reconstruction of observed and simulated events relies on the particle-flow (PF) algorithm~\cite{Sirunyan:2017ulk},
which combines the information from the CMS subdetectors to identify
and reconstruct the particles emerging from $\Pp\Pp$ collisions:
charged hadrons, neutral hadrons, photons, muons, and electrons.
Combinations of these PF objects are used to reconstruct
higher-level objects such as jets, $\tauh$ candidates, or
missing transverse momentum.
The reconstructed vertex with the largest value of summed physics-object $\pt^2$ is taken to be the primary $\Pp\Pp$ interaction vertex. The physics objects are the objects constructed by a jet finding algorithm~\cite{Cacciari:2008gp,Cacciari:2011ma} applied to all charged tracks associated with the vertex, including tracks from lepton candidates, and the corresponding associated missing transverse momentum.

Muons are identified with requirements on the quality of
the track reconstruction and on the number of measurements in the
tracker and the muon systems~\cite{Chatrchyan:2012xi}.
Electrons are identified with a multivariate discriminant
combining several quantities describing the track quality,
the shape of the energy deposits in the ECAL,
and the compatibility of the measurements from the tracker and the
ECAL~\cite{Khachatryan:2015hwa}.
To reject non-prompt or misidentified leptons, a relative lepton isolation is defined as:
\begin{equation}
I^{\ell} \equiv \frac{\sum_\text{charged}  \PT + \max\left( 0, \sum_\text{neutral}  \PT
                                         - \frac{1}{2} \sum_\text{charged, PU} \PT  \right )}{\PT^{\ell}}.
\label{eq:reconstruction_isolation}
\end{equation}
In this expression, $\sum_\text{charged}  \PT$ is the scalar sum of the
transverse momenta of the charged particles originating from
the primary vertex and located in a cone of size
$\Delta R = \sqrt{\smash[b]{(\Delta \eta)^2 + (\Delta \phi)^2}} = 0.4$\,(0.3)
centered on the muon (electron) direction. The sum
$\sum_\text{neutral}  \PT$  represents
a similar quantity for neutral particles.
The contribution of photons and neutral hadrons originating from pileup vertices is estimated from the scalar sum of the transverse
momenta of charged hadrons in the cone originating from pileup vertices,
$\sum_\text{charged, PU} \PT$. This sum is multiplied by a factor of
$1/2$, which corresponds approximately to the ratio of neutral to charged
hadron production in the hadronization process
of inelastic $\Pp\Pp$ collisions, as estimated from simulation.
The expression $\PT^{\ell}$ stands for the $\pt$ of the lepton. Isolation requirements used in this analysis, based on $I^{\ell}$, are listed in Table~\ref{tab:inclusive_selection}.

Jets are reconstructed with an anti-\kt clustering algorithm implemented
in the \FASTJET library~\cite{Cacciari:2011ma, Cacciari:fastjet2}.
It is based on the clustering of neutral and charged PF candidates within a distance parameter of 0.4. Charged PF candidates
not associated with the primary vertex of the interaction
are not considered when building jets. An offset correction is applied to jet energies to take into account the contribution from additional $\Pp\Pp$ interactions within the same or nearby bunch crossings. The energy of a jet is calibrated based on simulation and
data through correction factors~\cite{CMS-JME-10-011}.
In this analysis, jets are required to
have $\pt$ greater than 30\GeV and $\abs{\eta}$ less than 4.7, and
are separated from the selected leptons by a $\Delta R$ of at least 0.5.
The combined secondary vertex (CSV) algorithm is used to identify jets that are likely to originate from a b quark (``b jets"). The algorithm exploits the track-based lifetime information together with the secondary vertices associated with the jet to provide a likelihood ratio discriminator for the b jet identification. A set of $\pt$-dependent correction
factors are applied to simulated events to account for differences in the b tagging efficiency
between data and simulation. The working point chosen in this analysis gives an efficiency for real b jets of about 70\%, and for about 1\% of light flavor or quark jets being misidentified.

Hadronically decaying $\Pgt$ leptons
are reconstructed with the hadron-plus-strips (HPS)
algorithm~\cite{Khachatryan:2015dfa, CMS-PAS-TAU-16-002}, which is
seeded with anti-\kt jets.
The HPS algorithm reconstructs $\tauh$ candidates on the basis of the
number of tracks and of the number of ECAL strips in the $\eta$-$\phi$ plane with energy deposits, in the 1-prong,
1-prong + $\PGpz$(s), and 3-prong decay modes. A
multivariate (MVA) discriminator~\cite{Hocker:2007ht}, including isolation
and lifetime information, is used to reduce the rate for  quark- and gluon-initiated jets
to be identified as $\tauh$ candidates. The working point used in this analysis
has an efficiency of about 60\% for genuine $\tauh$,
with about 1\% misidentification rate for quark- and gluon-initiated jets, for a $\pt$ range typical of $\tauh$ originating from a $\PZ$ boson.
Electrons and muons misidentified as $\tauh$ candidates are suppressed using dedicated criteria
based on the consistency between the measurements in the tracker, the calorimeters, and the muon detectors~\cite{Khachatryan:2015dfa, CMS-PAS-TAU-16-002}.
The working points of these discriminators depend on the
decay channel studied.
The $\tauh$ energy scale in simulation is corrected per decay mode, on the basis of a measurement in $\cPZ\to\Pgt\Pgt$ events. The rate and the
energy scale of electrons and muons misidentified as $\tauh$ candidates are also corrected in simulation, on the basis of a tag-and-probe measurement~\cite{CMS:2011aa} in $\PZ\to\ell\ell$ events.

All particles reconstructed in the event are used to determine the missing transverse momentum,
\ptvecmiss. The missing transverse momentum is defined as the negative vectorial sum of the transverse momenta of
all PF candidates~\cite{Khachatryan:2014gga}. It is adjusted for the effect of jet energy corrections.
Corrections to the $\ptvecmiss$ are applied to reduce the mismodeling of the simulated
$\PZ+\text{jets}$, $\PW+\text{jets}$ and Higgs boson samples.
The corrections are applied to the simulated events on the basis of the vectorial difference
of the measured missing transverse momentum and total transverse momentum of neutrinos
originating from the decay of the $\PZ$, $\PW$, or Higgs boson. Their average effect is the reduction of the $\ptmiss$ obtained from simulation by a few \GeV.

The visible mass of the $\Pgt\Pgt$ system, $\mvis$, can be used to separate
the $\PH\to \Pgt \Pgt$ signal events
from the large contribution of irreducible $\PZ \to \Pgt \Pgt$ events.
However, the neutrinos from the $\Pgt$ lepton decays carry a large fraction of
the $\Pgt$ lepton energy and reduce the discriminating power of this variable.
The \textsc{svfit} algorithm combines the \ptvecmiss with the four-vectors of both $\Pgt$ candidates
to calculate a more accurate estimate of the mass of the parent boson, denoted as $\mtt$. The resolution of $\mtt$ is between 15 and 20\% depending on the $\Pgt\Pgt$ final state.
A detailed description of the algorithm can be found
in Ref.~\cite{Bianchini:2014vza}. Both variables are used in the analysis, as detailed in Section.~\ref{sec:categories}, and $\mvis$ is preferred over $\mtt$ when the background from $\PZ \to \ell\ell$ events is large.

\section{Event selection}
\label{sec:selection}

Selected events are classified into the various decay channels according to
the number of selected electrons, muons, and $\tauh$ candidates.
The resulting event samples are made mutually exclusive by discarding events that have additional loosely identified
and isolated muons or electrons.
Leptons must meet the minimum requirement
that the distance of closest approach to the primary vertex satisfies $\abs{d_z}<0.2\unit{cm}$
along the beam direction, and $\abs{d_{xy}}<0.045\unit{cm}$ in the transverse plane.
The two leptons assigned to the Higgs boson decay are required to have opposite-sign electric charges.
In the $\Pgm\tauh$ channel, events are selected with a combination of online criteria that require at least one isolated muon trigger candidate, or at least one isolated muon and one $\tauh$ trigger candidate, depending on the offline muon $\pt$. In the $\Pe\tauh$ channel, the trigger system requires at least one isolated
electron object, whereas in the $\Pe\Pgm$ channel, the triggers  rely on the presence of both an electron and a muon, allowing
lower online $\pt$ thresholds.
In the $\tauh\tauh$ channel, the trigger selects events with two loosely isolated $\tauh$ objects.
The selection criteria are summarized in Table~\ref{tab:inclusive_selection}.

\begin{table*}[htbp]
\centering
\topcaption{Kinematic selection requirements for the four di-$\Pgt$ decay channels.
The trigger requirement is defined by a combination of trigger candidates with \pt over a given threshold (in \GeV), indicated inside parentheses. The pseudorapidity thresholds come from trigger and object reconstruction constraints. The $\pt$ thresholds for the lepton selection are driven by the trigger requirements, except for the leading $\tauh$ candidate in the $\tauh\tauh$ channel, the $\tauh$ candidate in the $\Pgm\tauh$ and $\Pe\tauh$ channels, and the muon in the $\Pe\Pgm$ channel, where they have been optimized to increase the significance of the analysis.
\label{tab:inclusive_selection}
}
\begin{tabular}{lllll}
  Channel           &         Trigger requirement              &    \multicolumn{3}{c}{Lepton selection}                 \\ \cline{3-5}
 & & $\pt$ ($\GeVns{}$) & $\eta$ & Isolation \\
\hline
 $\tauh\tauh$    &         $\tauh (35)\,\&\,\tauh (35)$              &     $\pt^{\tauh}>50\,\&\,40$ & $\abs{\eta^{\tauh}}<2.1$  &    MVA $\tauh$ ID              \\
\hline
  $\mu\tauh$       &         $\Pgm(22)$     &     $\pt^\Pgm>23$  &  $\abs{\eta^\Pgm}<2.1$   &   $I^{\Pgm}<0.15$       \\
                          &       &     $\pt^{\tauh}>30$ &  $\abs{\eta^{\tauh}}<2.3$ &   MVA $\tauh$ ID  \\[\cmsTabSkip]

                   &         $\Pgm(19)\,\&\,\tauh (21)$     &     $20<\pt^\Pgm<23$  &  $\abs{\eta^\Pgm}<2.1$   &   $I^{\mu}<0.15$       \\
                   &           &     $\pt^{\tauh}>30$ &  $\abs{\eta^{\tauh}}<2.3$ &   MVA $\tauh$ ID  \\
\hline
  $\Pe\tauh$        &         $\Pe (25)$     &     $\pt^\Pe>26$  & $\abs{\eta^\Pe}<2.1$  &   $I^{\Pe}<0.1$  \\
                          &       &     $\pt^{\tauh}>30$ &  $\abs{\eta^{\tauh}}<2.3$ &   MVA $\tauh$ ID  \\
\hline
  $\Pe\Pgm$        &         $\Pe(12)\,\&\,\Pgm (23)$    &     $\pt^{\Pe}>13$ & $\abs{\eta^\Pe}<2.5$             & $I^{\Pe}<0.15$   \\
                          &              &     $\pt^{\Pgm}>24$ & $\abs{\eta^\Pgm}<2.4$  & $I^{\Pgm}<0.2$ \\[\cmsTabSkip]
         &         $\Pe(23)\,\&\,\Pgm (8)$    &     $\pt^{\Pe}>24$ & $\abs{\eta^\Pe}<2.5$             & $I^{\Pe}<0.15$   \\
                          &          &     $\pt^{\Pgm}>15$ & $\abs{\eta^\Pgm}<2.4$  & $I^{\Pgm}<0.2$    \\
\hline
\end{tabular}
\end{table*}

In the $\ell \tauh$ channels, the large $\PW+\text{jets}$ background is reduced by requiring the transverse mass,
$\MT$, to satisfy
\begin{equation}
\MT \equiv \sqrt{\smash[b]{2 \pt^\ell \ptmiss [1-\cos(\Delta\phi)]}} < 50\GeV,
\end{equation}
where $\pt^\ell$ is the transverse momentum of the lepton $\ell$,
and $\Delta\phi$ is the azimuthal angle between its direction and the \ptvecmiss.

In the $\emu$ channel, the \ttbar background is reduced by requiring $p_\zeta - 0.85 \, p_\zeta^{\text{vis}} > -35$ or $-10$\GeV depending on the category,
where $p_\zeta$ is the component of the \ptvecmiss along the bisector of the transverse momenta of the two leptons
and $p_\zeta^{\text{vis}}$ is the sum of the components of the lepton transverse momenta along
the same direction~\cite{Khachatryan:2014wca}.
This selection criterion has a high signal efficiency because the \ptvecmiss is typically oriented in the same direction as the visible di-$\Pgt$ system in signal events.
In addition, events with a b-tagged jet are discarded to further suppress the \ttbar background in the $\emu$ channel.

\section{Categorization}
\label{sec:categories}

The event sample is split into three mutually exclusive categories per decay channel.
In each category the two variables that maximize the $\PH\to\Pgt\Pgt$ sensitivity are chosen to build two-dimensional (2D) distributions.

\begin{figure*}[htbp]
\centering
     \includegraphics[width=0.4\textwidth]{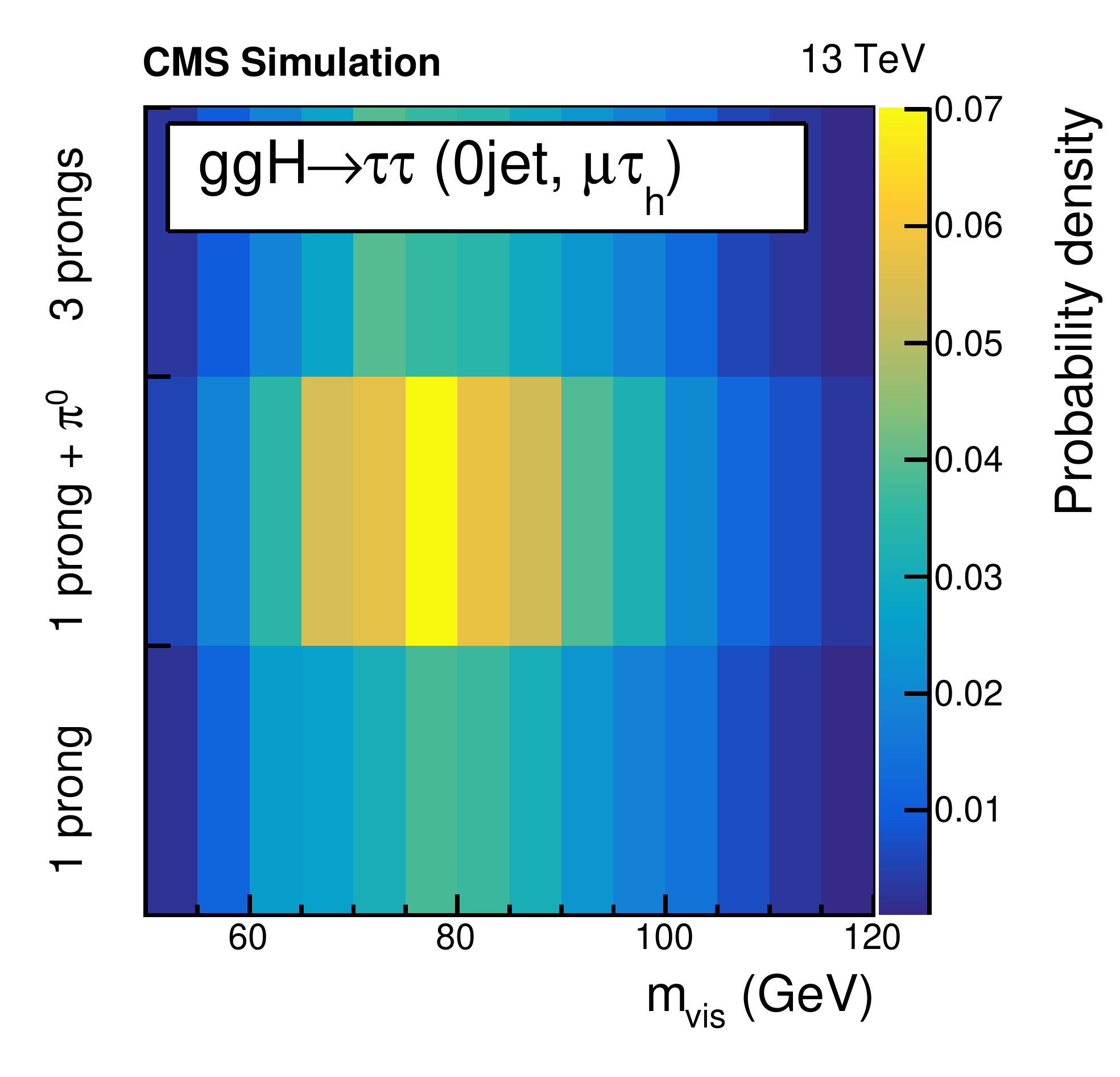}
     \includegraphics[width=0.4\textwidth]{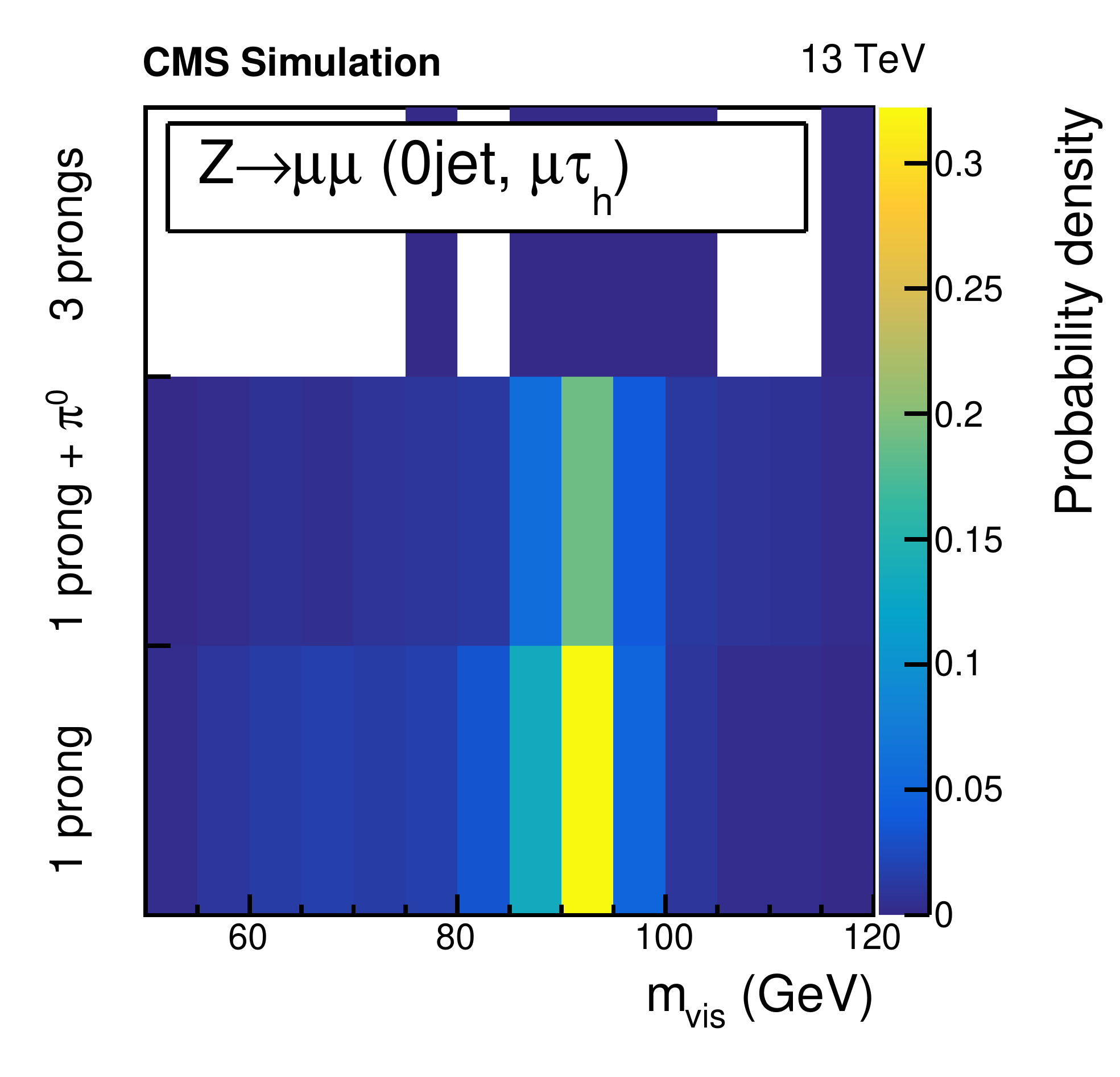}\\
     \includegraphics[width=0.4\textwidth]{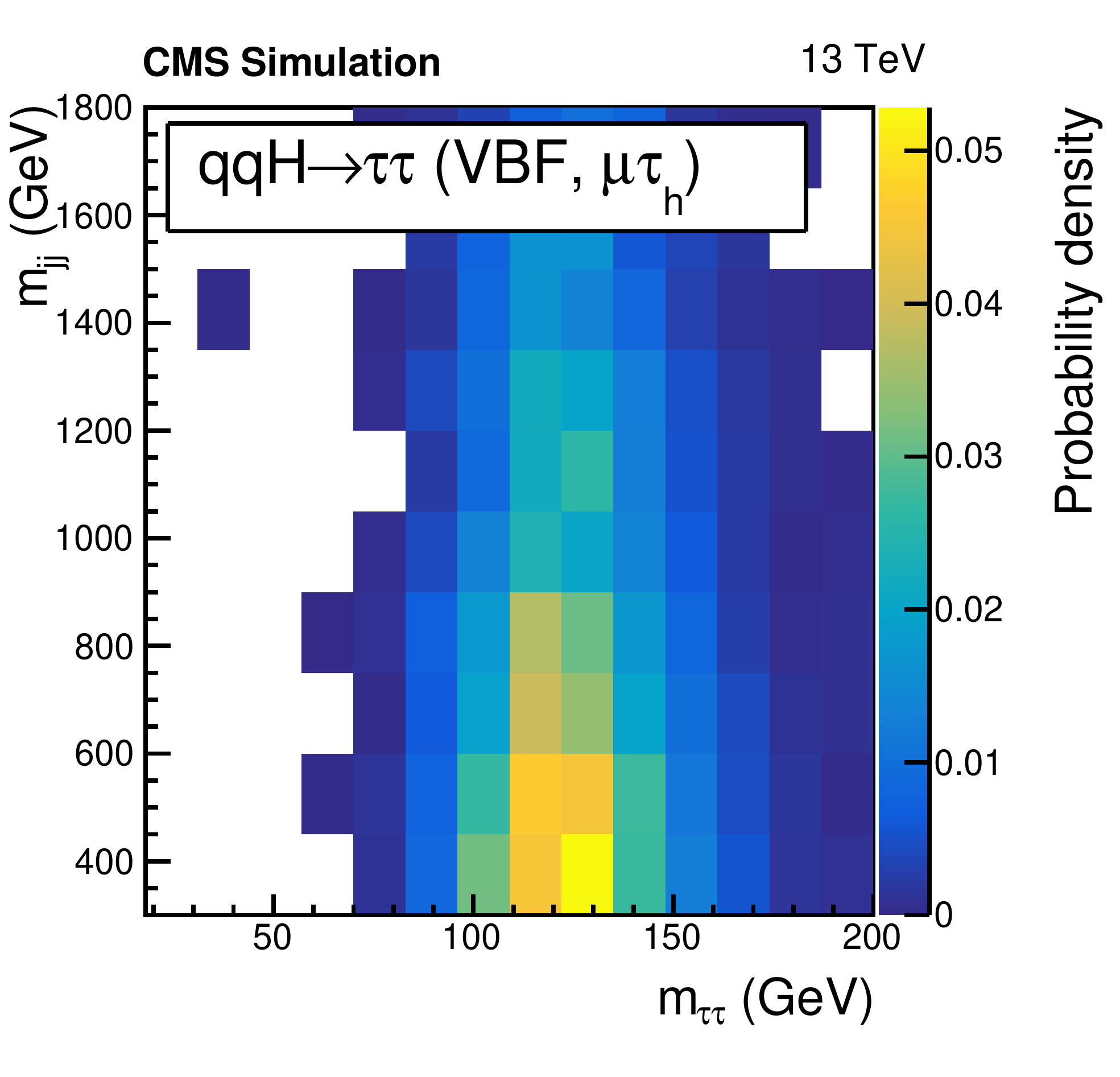}
     \includegraphics[width=0.4\textwidth]{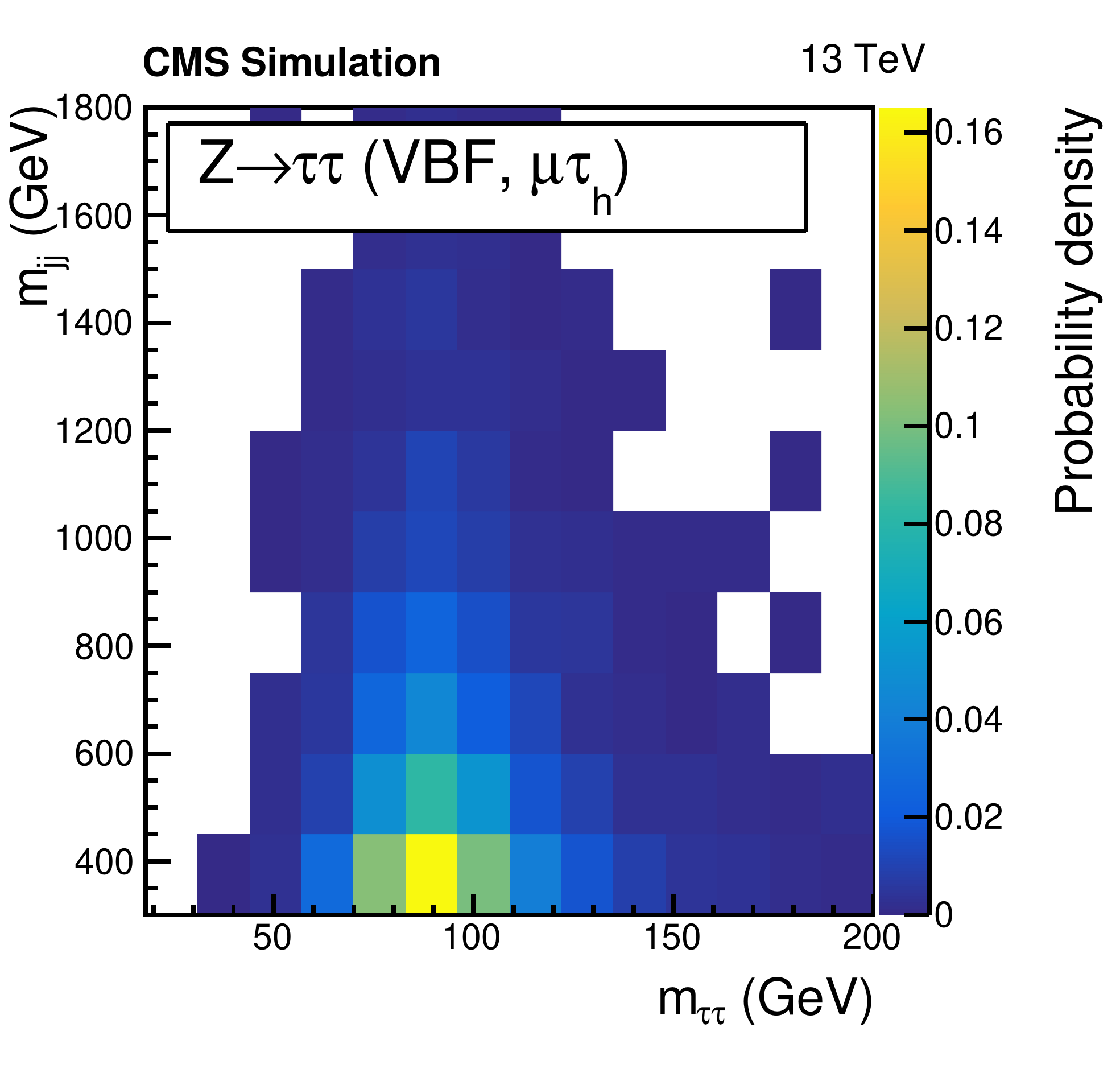}\\
     \includegraphics[width=0.4\textwidth]{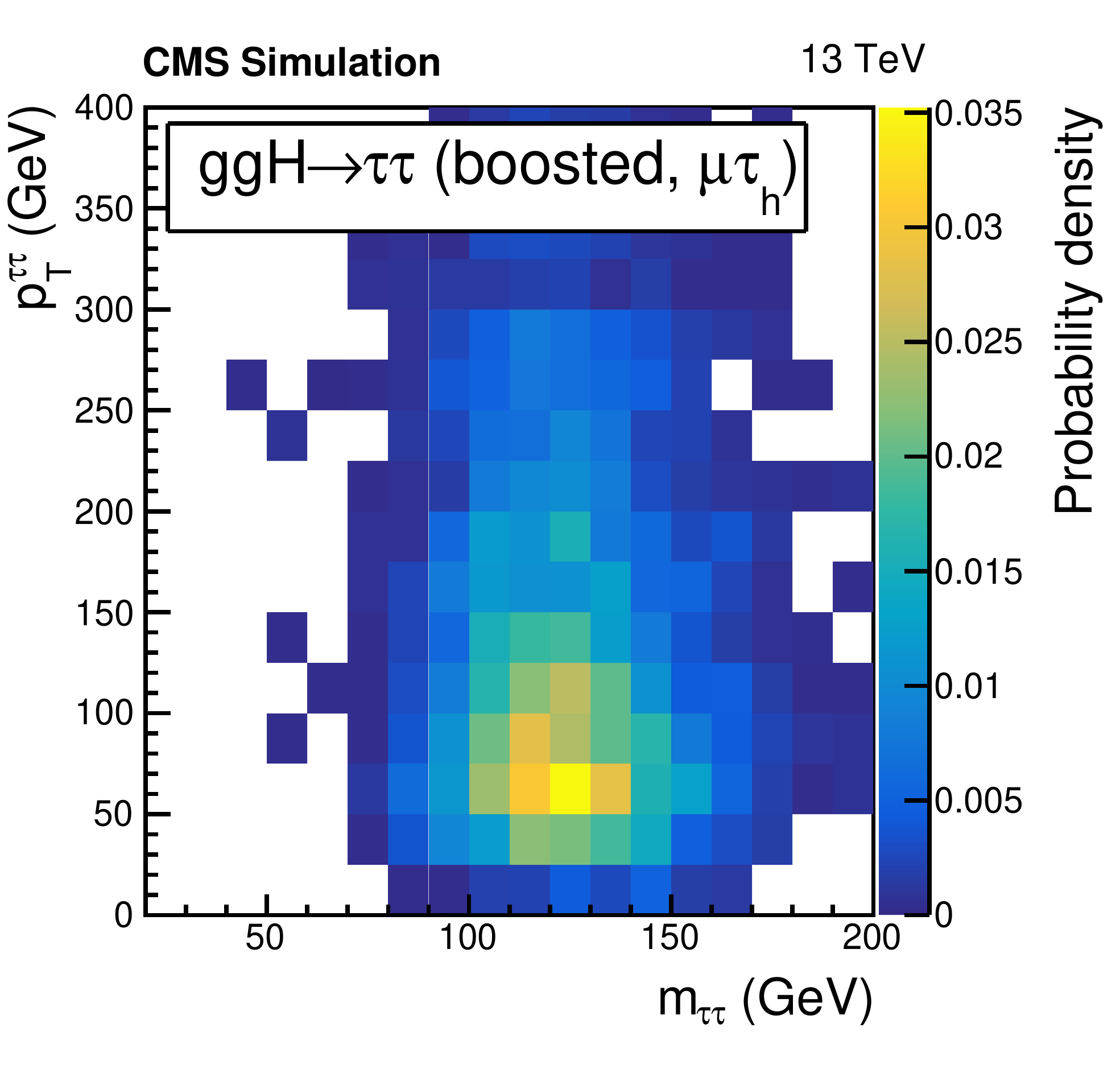}
     \includegraphics[width=0.4\textwidth]{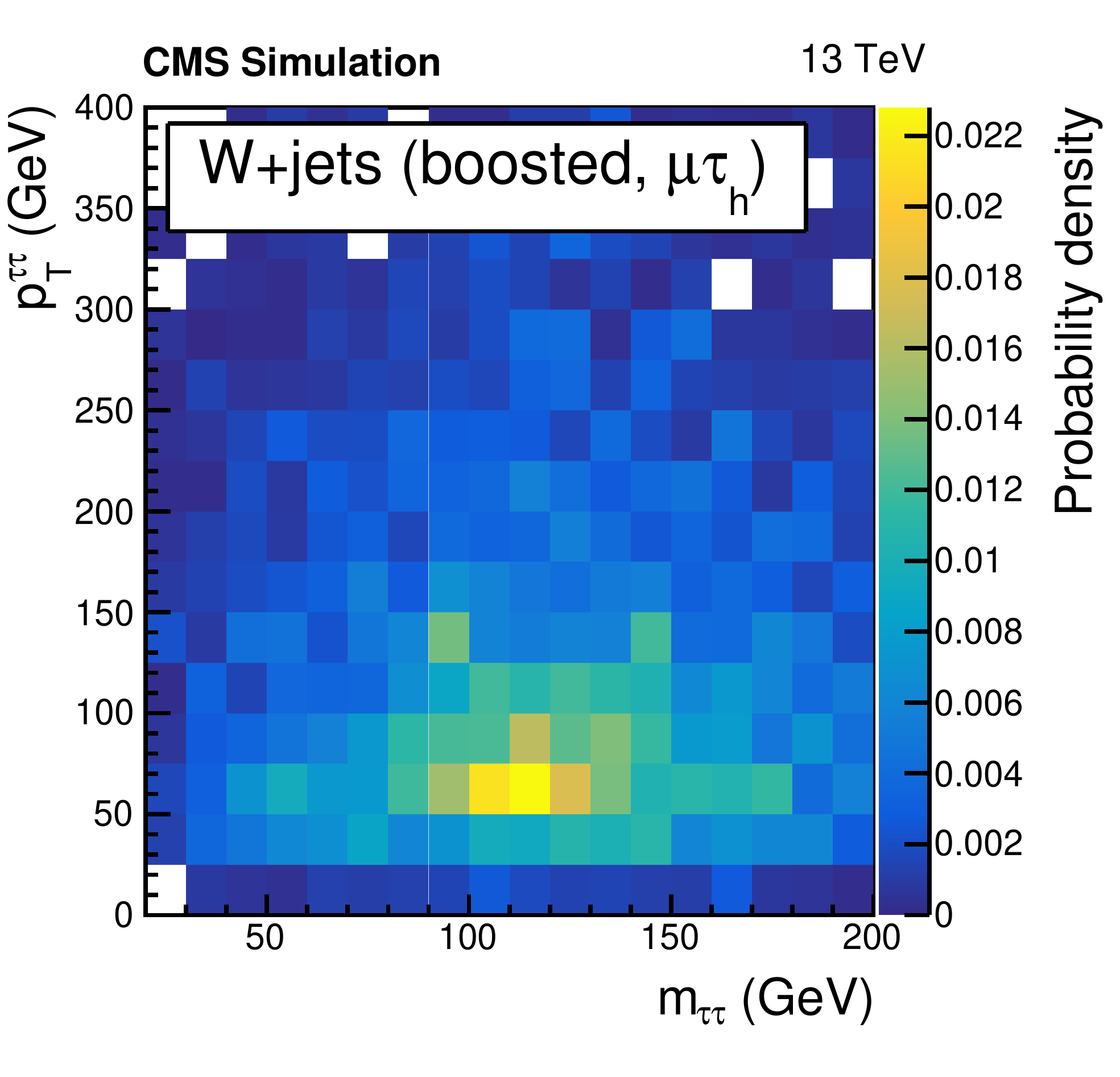}
     \caption{Distributions for the signal (left) and for some dominant background processes (right) of the two observables chosen in the 0-jet (top), VBF (center), and boosted
(bottom) categories in the $\Pgm\tauh$ decay channel. The background processes are chosen for illustrative purpose for their separation from the signal. The $\PZ\to\Pgm\Pgm$ background in the 0-jet category is concentrated in the regions where the visible mass is close to 90\GeV and is negligible when the $\tauh$ candidate is reconstructed in the 3-prong decay mode. The $\PZ\to\Pgt\Pgt$ background in the VBF category mostly lies at low $\mjj$ values whereas the distribution of VBF signal events extends to high $\mjj$ values. In the boosted category, the W+jets background, which behaves similarly to the QCD multijet background, is rather flat with respect to $\mtt$, and is concentrated at low $\pth$ values. These distributions are not used as such to extract the results.}
     \label{fig:2Dcategories}
\end{figure*}

The three categories are defined as:
\begin{itemize}
\item {0-jet}: This category targets Higgs boson events produced via gluon fusion.
The two variables chosen to extract the results are $\mvis$ and
the reconstructed $\tauh$ candidate decay mode (in the $\Pgm\tauh$ and $\Pe\tauh$ decay channels)
or the $\pt$ of the muon (in the $\Pe\Pgm$ channel). The $\PZ\to\ell\ell$ background is large in the 1-prong and 1-prong + $\PGpz$(s) $\tauh$
decay modes in the
$\Pgm\tauh$ and $\Pe\tauh$ channels.
The $\mvis$ variable is used as a final discriminant in the fit instead of $\mtt$ because it separates the signal from the $\PZ\to\ell\ell$ background, which peaks
around the $\PZ$ boson mass. The reconstructed $\tauh$ candidate decay mode is used as the other discriminant in the $\Pgm\tauh$ and $\Pe\tauh$ decay channels because the $\PZ\to\ell\ell$ background is negligible for $\tauh$
reconstructed in the 3-prong decay mode, leading to an increased signal-to-background ratio for this particular decay mode, and several systematic uncertainties related to the $\tauh$ decay mode can be constrained with more precision. The 2D distributions for the signal and $\PZ\to\ell\ell$ background
in the 0-jet category of the $\Pgm\tauh$ decay channel are shown in Fig.~\ref{fig:2Dcategories} (top).
In the $\tauh\tauh$ decay channel, only one observable, $\mtt$, is considered because of the low event yields
due to the relatively high $\pt$ thresholds on the $\tauh$ at trigger level, and because of the sharply falling $\tauh$ $\pt$ distribution. Simulations indicate that about 98\% of signal events in the 0-jet category correspond to the gluon fusion production mechanism.\\
\item {VBF}: This category targets Higgs boson events produced via VBF.
Events are selected with at least two (exactly two) jets with $\pt>30$\GeV in the
$\tauh\tauh$, $\Pgm\tauh$, and $\Pe\tauh$ ($\Pe\Pgm$) channels.
In the $\Pgm\tauh$, $\Pe\tauh$, and $\Pe\Pgm$ channels, the two leading jets are required to have an invariant mass, $\mjj$, larger than 300\GeV. The variable $\pth$, defined as the magnitude of the vectorial sum of the $\ptvec$ of the visible decay products of the $\Pgt$ leptons and $\ptvecmiss$, is required to be greater than 50 (100)\GeV in the $\Pgm\tauh$
 and $\Pe\tauh$ ($\tauh\tauh$) channels to reduce the contribution from $\PW+\text{jets}$ backgrounds. This selection criterion also suppresses the background from SM events composed uniquely of jets produced through the strong interaction, referred to as quantum chromodynamics (QCD) multijet events.
In addition, the $\pt$ threshold on the $\tauh$ candidate is raised to 40\GeV in the $\Pgm\tauh$ channel, and the
two leading jets in the $\tauh\tauh$ channel should be separated in pseudorapidity by $\Delta\eta>2.5$.
The two observables in the VBF category are $\mtt$ and $\mjj$. The 2D distributions for the signal and $\PZ\to\Pgt\Pgt$ background
in the VBF category of the $\Pgm\tauh$ decay channel are shown in Fig.~\ref{fig:2Dcategories} (center). Integrating over the whole $\mjj$ phase space, up to 57\% of the signal events in the VBF category are produced in the VBF production mode, but this proportion increases with $\mjj$.\\
\item {Boosted}: This category contains all the events that do not
enter one of the previous categories, namely events with one jet and events with several jets that fail the specific requirements of the VBF category.
It contains gluon fusion events produced in association with one or more jets (78--80\% of signal events),
VBF events where one of the jets has escaped detection or has low $\mjj$ (11--13\%), as well as
Higgs bosons produced in association with a $\PW$ or a $\PZ$ boson decaying hadronically (4--8\%).
While $\mtt$ is chosen as one of the dimensions of the distributions, $\pth$ is taken as the second dimension to specifically target Higgs boson events produced in gluon fusion,
with a Lorentz-boosted boson recoiling against jets. Most background processes, including $\PW+\text{jets}$ and QCD multijet events, typically have low $\pth$. The 2D
distributions for the signal and $\PW+\text{jets}$ background in the boosted category of the $\Pgm\tauh$ decay channel are shown in Fig.~\ref{fig:2Dcategories} (bottom).
\end{itemize}

The categories and the variables used to build the 2D distributions are summarized in
Table~\ref{tab:categories}. The results of the analysis are extracted with a global maximum likelihood fit based on  the 2D distributions in the various signal regions, and on some control regions, detailed in Section~\ref{sec:background_estimation}, that constrain the normalizations of the main backgrounds.

\begin{table*}
\centering
\topcaption{ Category selection and observables used to build the 2D kinematic distributions. The events neither selected in the 0-jet nor in the VBF category are included in the boosted category, as denoted by ``Others".
\label{tab:categories}
}
\begin{tabular}{llll}
 & 0-jet & VBF & Boosted \\
\hline
 & \multicolumn{3}{c}{Selection} \\ \cline{2-4}
$\tauh\tauh$ & No jet &  $\geq$2 jets, $\pth>100\GeV$, $\Delta\eta_{\mathrm{jj}}>2.5$ & Others\\
$\Pgm\tauh$ & No jet &  $\geq$2 jets, $\mjj>300\GeV$, $\pth>50\GeV$, $\pt^{\tauh}>40\GeV$ & Others\\
$\Pe\tauh$ & No jet &  $\geq$2 jets, $\mjj>300\GeV$, $\pth>50\GeV$ & Others\\
$\Pe\Pgm$ & No jet & 2 jets, $\mjj>300\GeV$ & Others \\
\hline
 & \multicolumn{3}{c}{Observables}\\ \cline{2-4}
$\tauh\tauh$ & $\mtt$                 &    $\mjj$, $\mtt$  &   $\pth$, $\mtt$  \\
$\Pgm\tauh$ & $\tauh$ decay mode, $\mvis$   &    $\mjj$, $\mtt$  &  $\pth$, $\mtt$  \\
$\Pe\tauh$ & $\tauh$ decay mode, $\mvis$   &    $\mjj$, $\mtt$  &  $\pth$, $\mtt$ \\
$\Pe\Pgm$ & $\pt^{\Pgm}$, $\mvis$   &     $\mjj$, $\mtt$  &   $\pth$, $\mtt$  \\
\hline
\end{tabular}
\end{table*}

\section{Background estimation}
\label{sec:background_estimation}

The largest irreducible source of background is the Drell--Yan production
of $\PZ/\Pgg^*\to\Pgt\Pgt, \ell\ell$.
In order to correct the yield and distributions of the $\PZ/\Pgg^*\to\Pgt\Pgt, \ell\ell$ simulations to better reproduce the Drell--Yan process in data, a dedicated control sample of $\PZ/\Pgg^*\to\Pgm\Pgm$
events is collected in data with a single-muon trigger, and compared to simulation. The control sample is composed of events with two well-identified
and well-isolated opposite-charge muons with $\pt$ greater than 25\GeV and an invariant mass between 70 and 110\GeV.
More than 99\% of events in this region come from $\PZ/\Pgg^*\to\Pgm\Pgm$ decays.
Differences in the distributions of $m_{\ell\ell/\Pgt\Pgt}$ and $\pt(\ell\ell/\Pgt\Pgt)$ in data and in simulations are observed in this control region, and 2D weights based on these variables are derived and applied to simulated $\PZ/\Pgg^*\to\Pgt\Pgt, \ell\ell$ events in the signal region of the analysis. In addition, corrections depending on $\mjj$ are derived from the $\PZ/\Pgg^*\to\Pgm\Pgm$ region and applied to the $\PZ/\Pgg^*\to\Pgt\Pgt, \ell\ell$ simulation for events with at least two jets passing the VBF category selection criteria. After this reweighting, good agreement between data in the $\PZ/\Pgg^*\to\Pgm\Pgm$ region and simulation is found for all other variables.
The simulated sample is split, on the basis of the matching between objects at the generator and at the detector levels,
into events with prompt leptons (muons or electrons), hadronic decays of the $\Pgt$ leptons,
and jets or misidentified objects at the detector level that do not have corresponding objects
at generator level within $\Delta R < 0.2$.
The electroweak production of $\PZ$ bosons in association with two jets is also taken into account in the analysis; it
contributes up to 8\% of the $\PZ$ boson production in the VBF category.

\begin{figure*}[!htbp]
\centering
     \includegraphics[width=0.19\textwidth]{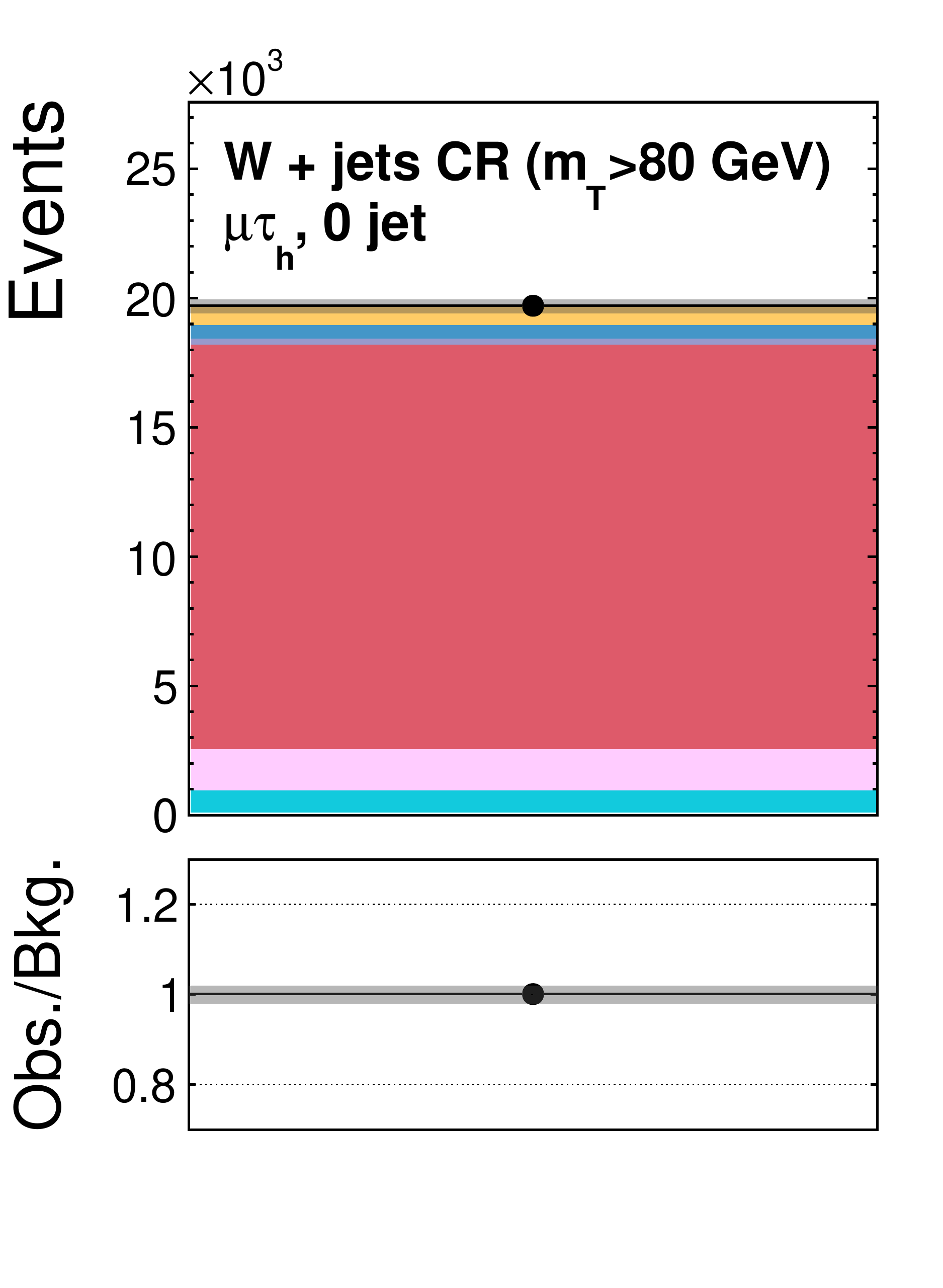}
     \includegraphics[width=0.19\textwidth]{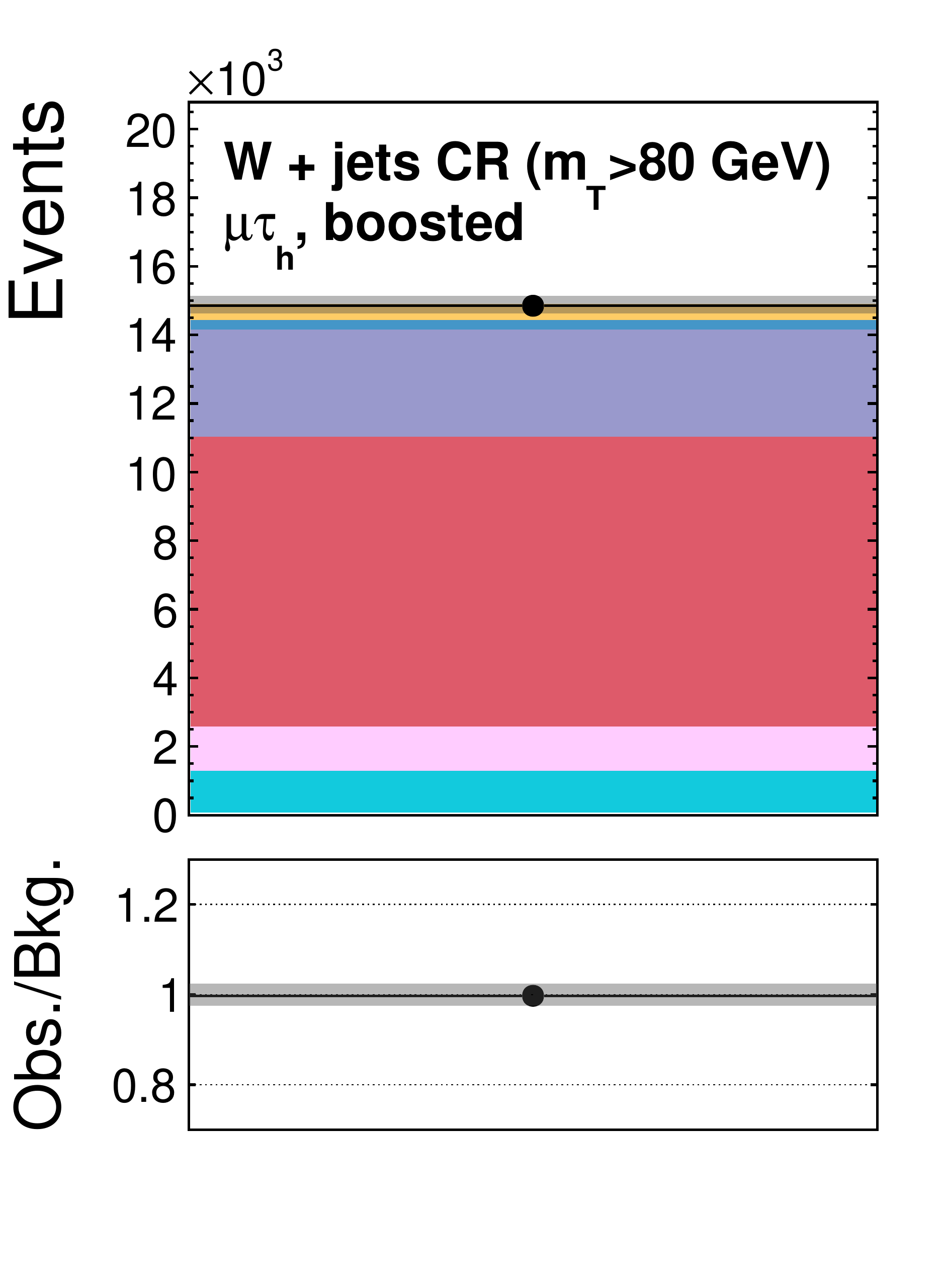}
     \includegraphics[width=0.19\textwidth]{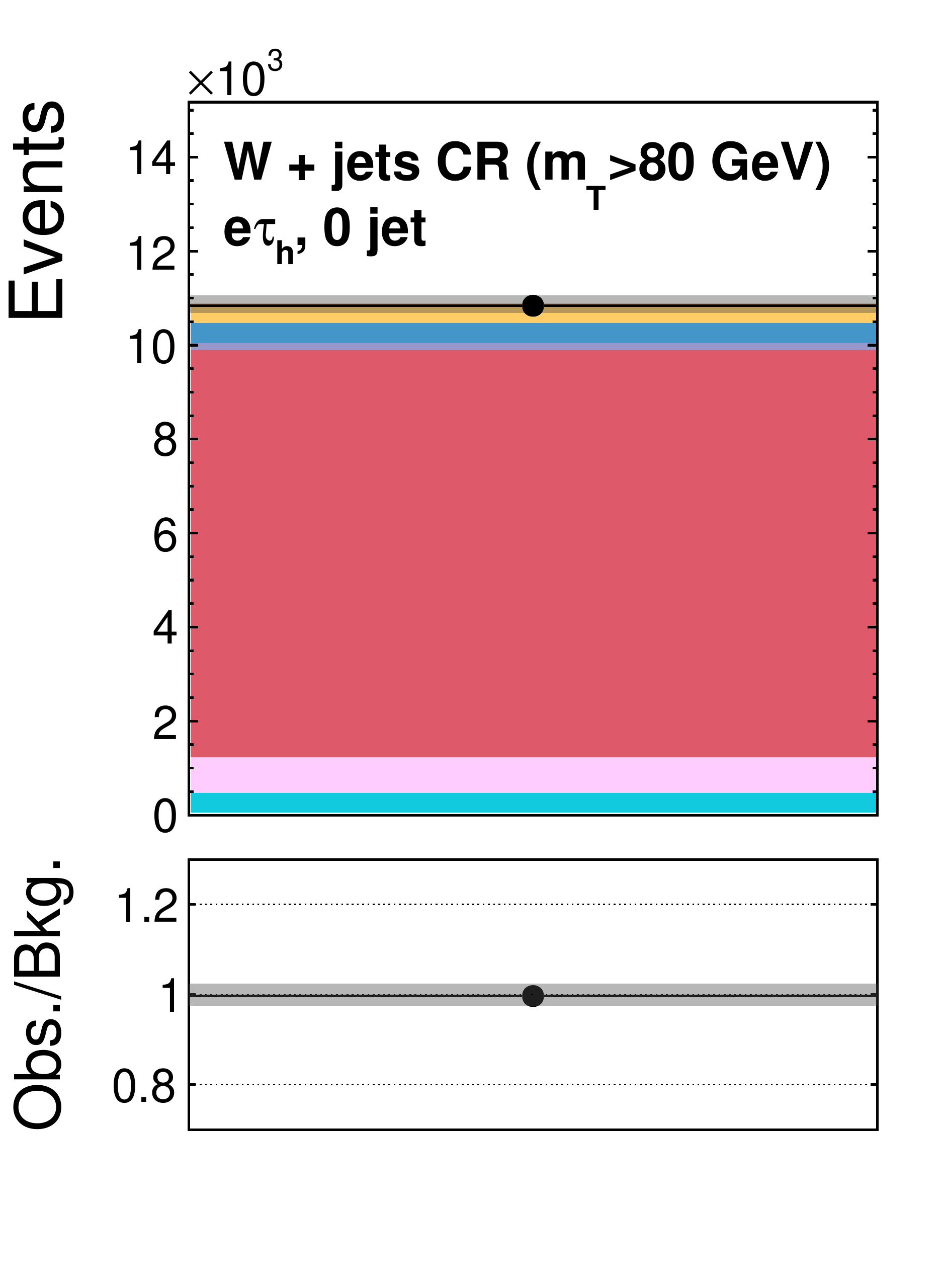}
     \includegraphics[width=0.19\textwidth]{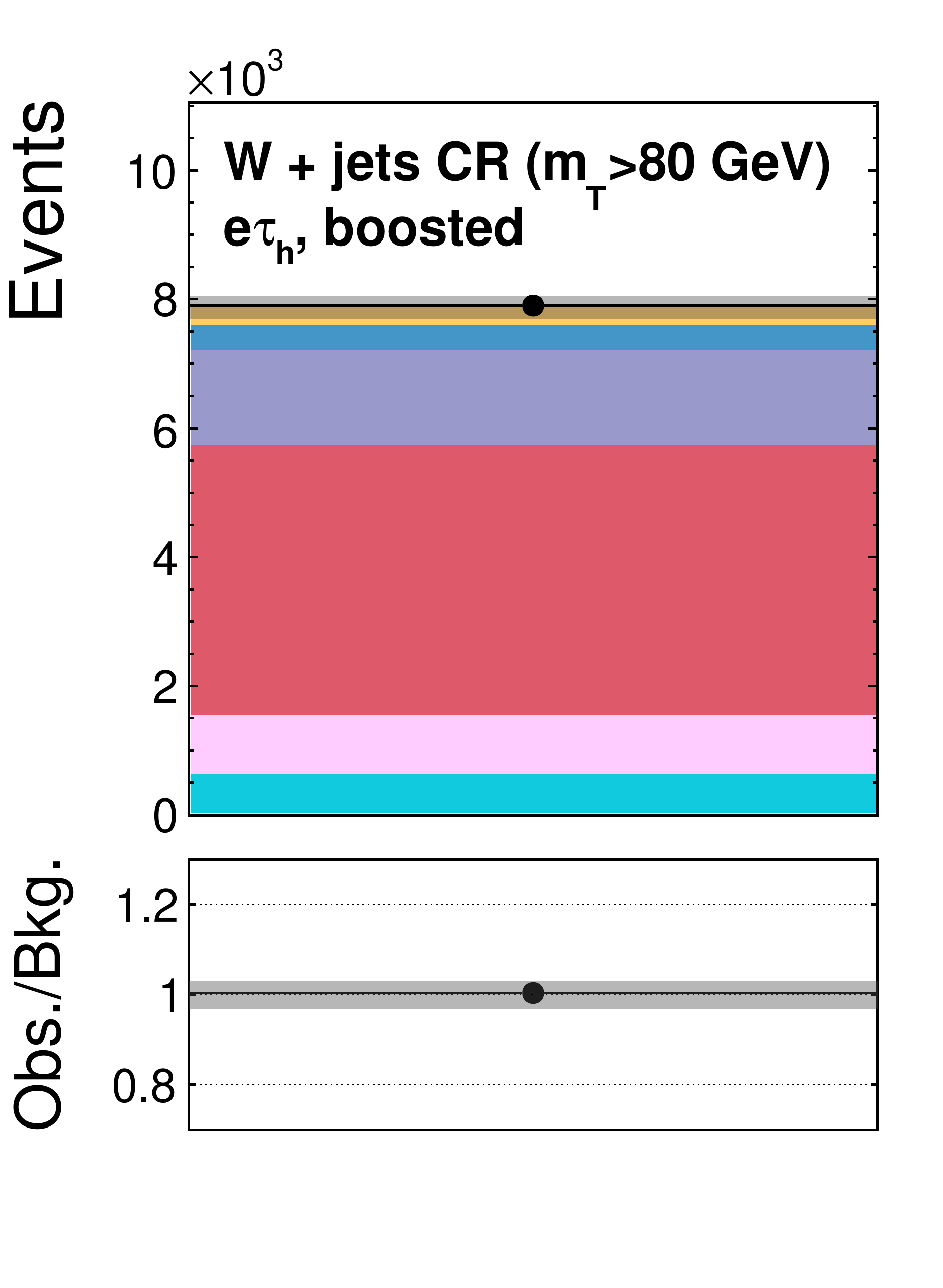}
     \includegraphics[width=0.19\textwidth]{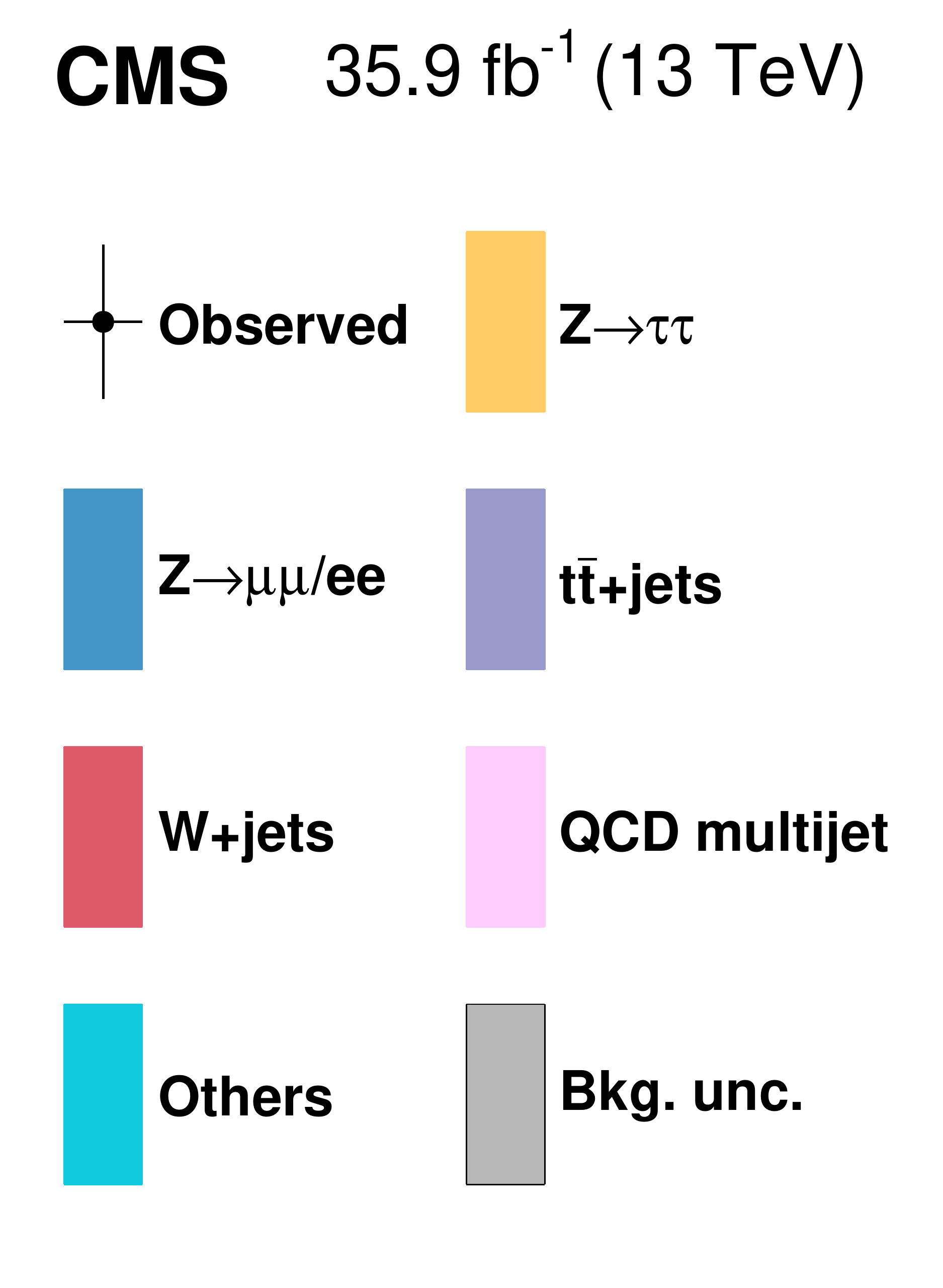}
     \caption{Control regions enriched in the $\PW+\text{jets}$ background used in the maximum likelihood fit, together with the signal regions, to extract the results. The normalization of the predicted background distributions corresponds to the result of the global fit. These regions, defined with $\MT>80$\GeV,  control the
yields of the $\PW+\text{jets}$ background in the $\Pgm\tauh$ and $\Pe\tauh$ channels.  The constraints obtained in the boosted categories are propagated to the VBF categories of the corresponding channels.}
     \label{fig:CR1}
\end{figure*}

The background from $\PW+\text{jets}$ production contributes significantly to the
$\Pgm\tauh$ and $\Pe\tauh$ channels, when the $\PW$ boson decays leptonically and
a jet is misidentified as a $\tauh$ candidate.
The $\PW+\text{jets}$ distributions are modelled using simulation, while their yields are estimated using data, as detailed below. In the boosted and VBF categories, statistical fluctuations in the distributions from simulations are reduced by relaxing the isolation of the $\tauh$ and $\ell$ candidates, which has been checked not to bias the distributions.
The simulated sample is normalized in such a way as to obtain agreement between the yields in data and the predicted backgrounds in a control region enriched in the $\PW+\text{jets}$ background,
which is obtained by applying all selection criteria,
with the exception that $\MT$ is required to be greater than 80\GeV instead of less than 50\GeV.
The $\PW+\text{jets}$ event purity in this
region varies from about 50\% in the boosted category to 85\% in the 0-jet category.
The high-$\MT$ sidebands described above, for each category, are considered
as a control regions in this fit.
 The constraints obtained in the boosted category are extrapolated to the VBF category of the corresponding decay channel because
the topology of the boosted and VBF events is similar, and few data events would pass the high-$\MT$ sideband selection in the VBF category. Figure~\ref{fig:CR1} shows the control regions with $\MT>80$\GeV in the 0-jet and boosted categories of the $\Pgm\tauh$ and $\Pe\tauh$ channels. These control regions are composed of only one bin because they are used solely to constrain the normalization of the $\PW+\text{jets}$ process.
In the $\Pe\Pgm$ and $\tauh\tauh$ decay channels, the $\PW+\text{jets}$
background is small compared to other backgrounds, and its contribution is
estimated from simulations.

\begin{figure*}[!htbp]
\centering
     \includegraphics[width=0.3\textwidth]{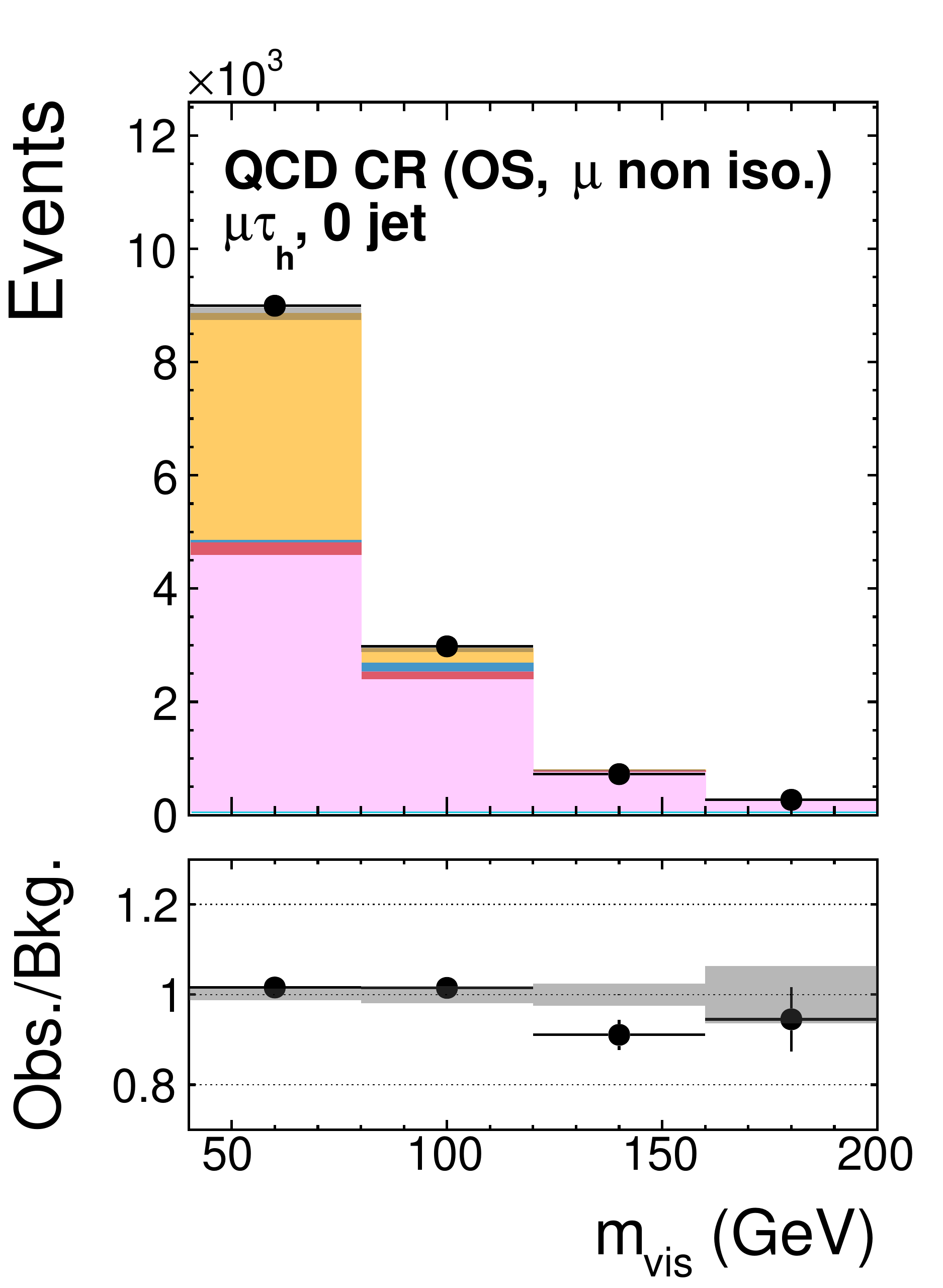}
     \includegraphics[width=0.3\textwidth]{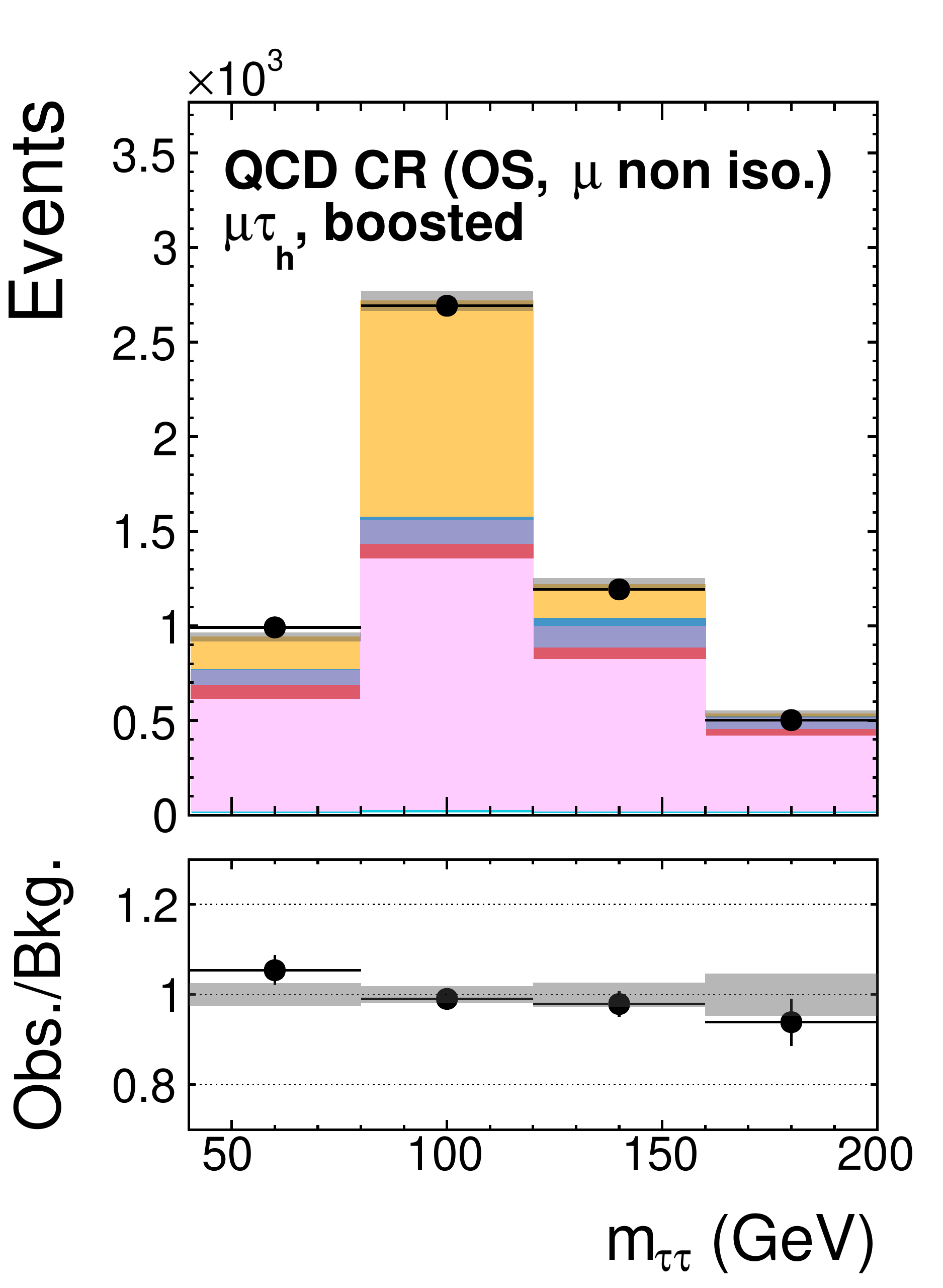}
     \includegraphics[width=0.3\textwidth]{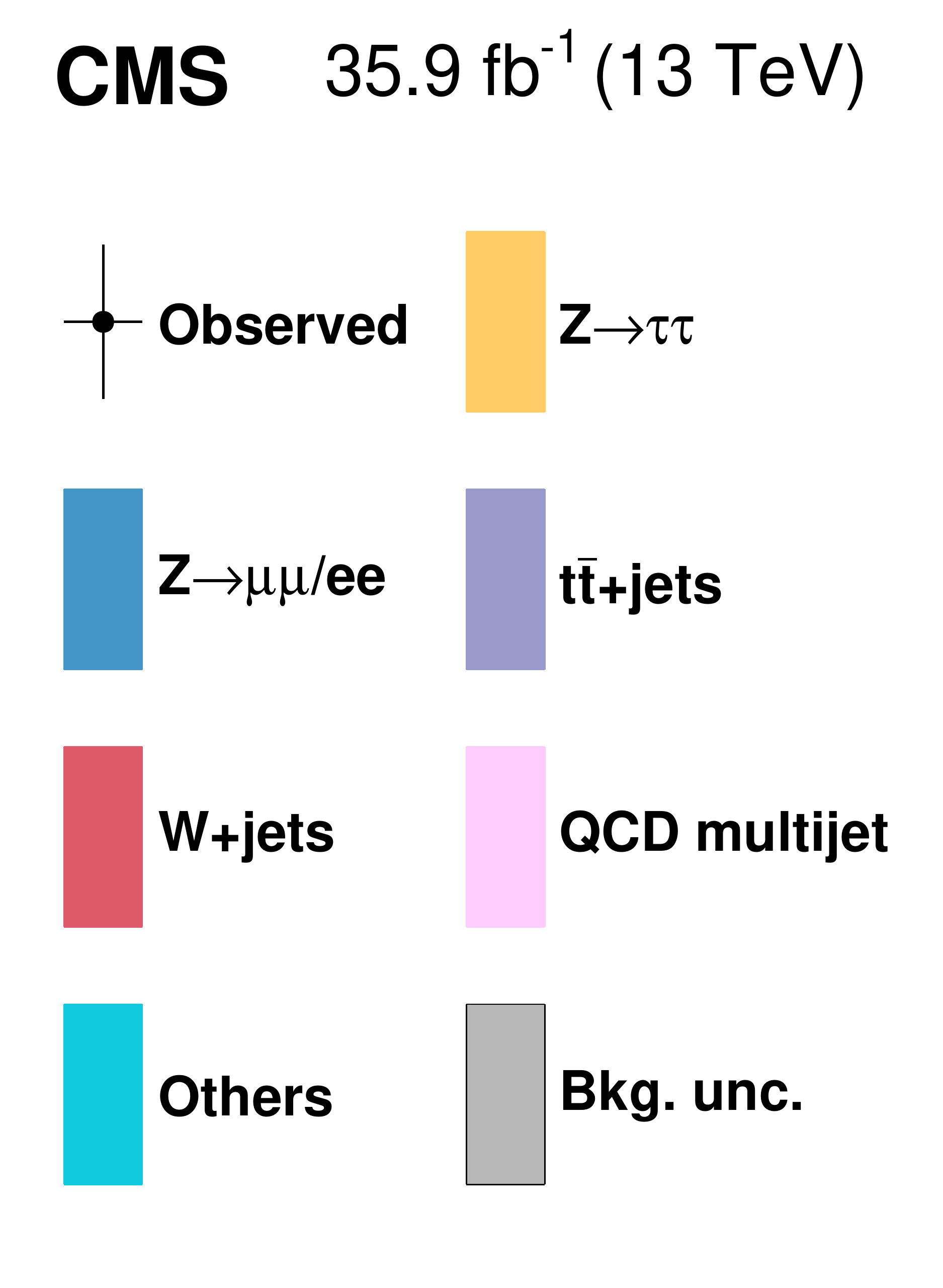}
     \includegraphics[width=0.3\textwidth]{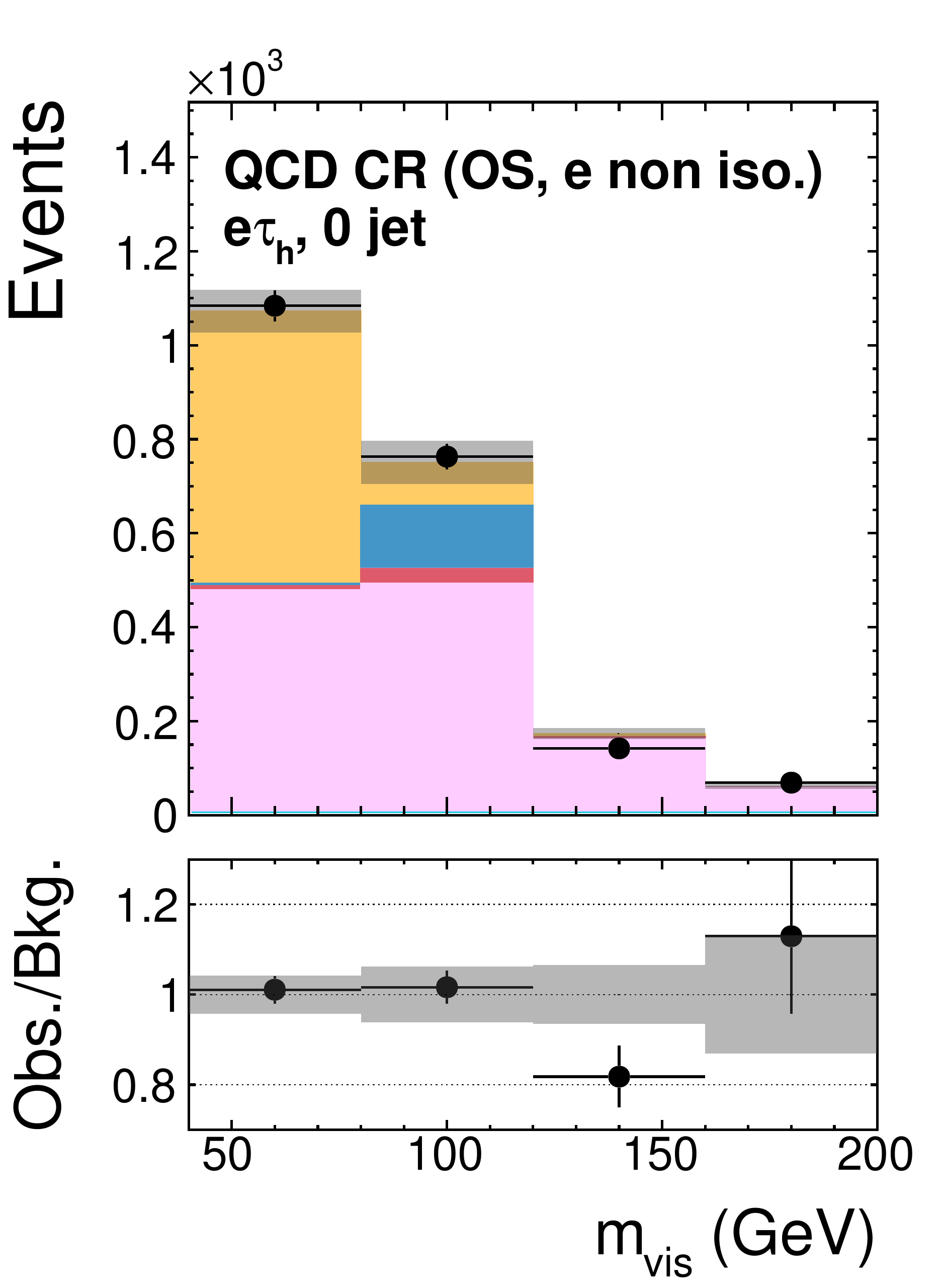}
     \includegraphics[width=0.3\textwidth]{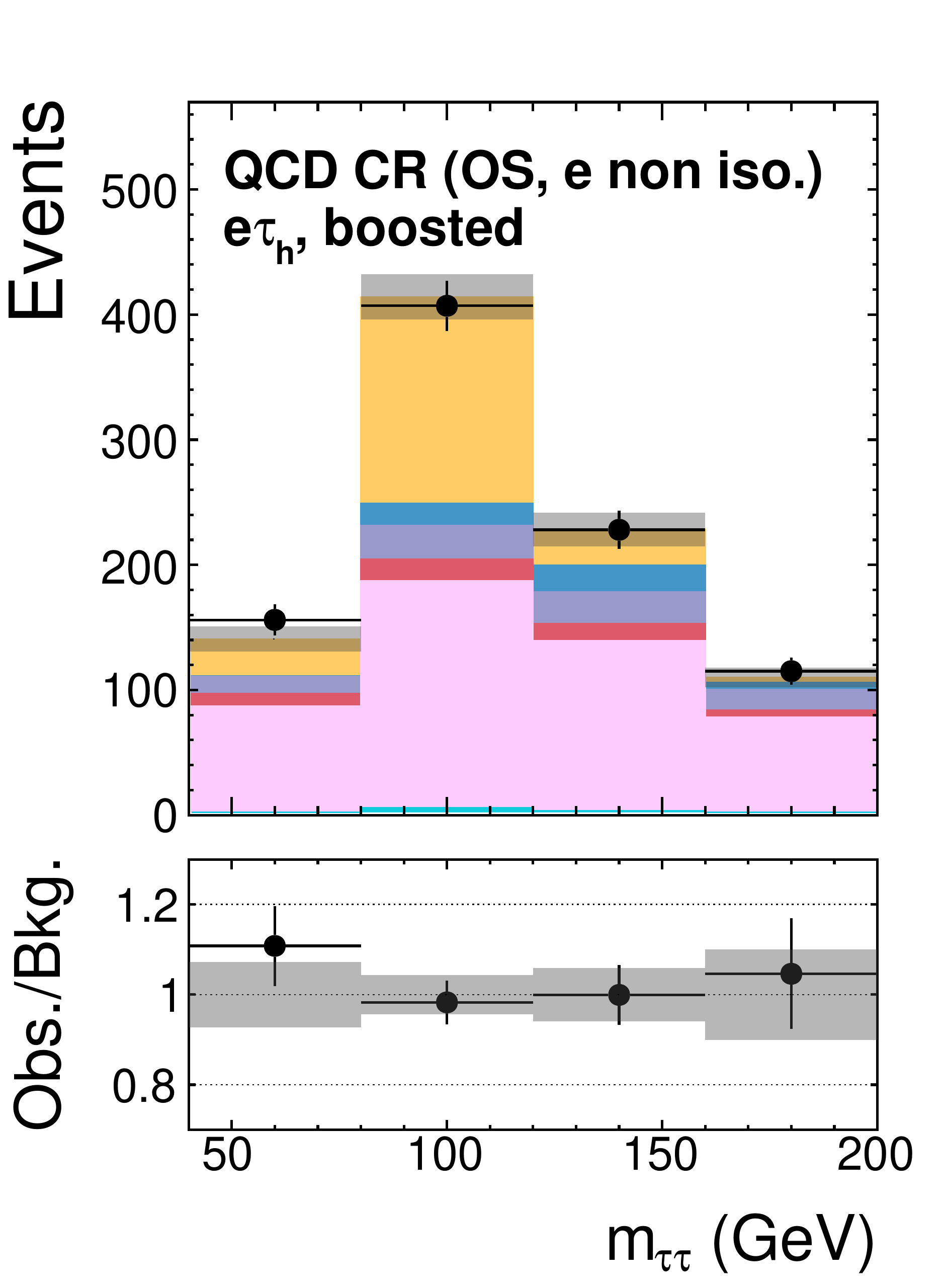}
     \caption{Control regions enriched in the QCD multijet background used in the maximum likelihood fit, together with the signal regions, to extract the results. The normalization of the predicted background distributions corresponds to the result of the global fit. These regions, defined by selecting events with opposite-sign $\ell$ and $\tauh$ candidates with $\ell$ passing inverted isolation conditions,  control the
yields of the QCD multijet background in the $\Pgm\tauh$ and $\Pe\tauh$ channels.  The constraints obtained in the boosted categories are propagated to the VBF categories of the corresponding channels.}
     \label{fig:CR3}
\end{figure*}

The QCD multijet events constitute another important source of reducible background in the $\ell \tauh$ channels, and it is entirely estimated from data. Various control samples are constituted to estimate the shape and the yield of the QCD multijet background in these channels, as explained below:
\begin{enumerate}
\item The raw yield is extracted using a sample where the
$\ell$ and the $\tauh$ candidates have the same sign. Using this sample, the QCD multijet process is estimated from data by subtracting the contribution of the Drell--Yan, \ttbar, diboson,
and $\PW+\text{jets}$ processes.
\item The yield obtained above is corrected to account for differences between the background composition in the same-sign and opposite-sign regions. The extrapolation factor between the same-sign and opposite-sign regions is determined by comparing the yield of the QCD multijet background for events with $\ell$ candidates passing inverted isolation criteria, in the same-sign and opposite-sign regions. It is constrained and measured by adding to the global fit the opposite-sign region where the $\ell$ candidates pass inverted isolation criteria, using the QCD multijet background estimate from the same-sign region with $\ell$ candidates passing inverted isolation criteria. For the same reasons as in the case of the W+jets background, the constraints are also extrapolated to the VBF signal region. Figure~\ref{fig:CR3} shows these control regions for the 0-jet and boosted categories of the $\Pgm\tauh$ and $\Pe\tauh$ channels; the observable is $\mvis$ or $\mtt$ to provide discrimination between the QCD multijet and the $\PZ\to\Pgt\Pgt$ processes.
\item The 2D distributions of the QCD multijet background are estimated from a region with same-sign leptons, as for the yield estimate, but the isolation of the $\ell$ and $\tauh$ candidates is additionally relaxed to reduce the statistical fluctuations in the distributions. Again the contribution of the Drell--Yan, \ttbar, diboson,
and $\PW+ \text{jets}$ processes are subtracted from data to extract the QCD multijet contribution in this region.
\end{enumerate}
The same technique is used in the $\Pe\Pgm$ decay channel, but no control region is included in the fit because QCD multijet events contribute little to the total background in this decay channel.

\begin{figure*}[!htbp]
\centering
     \includegraphics[width=0.24\textwidth]{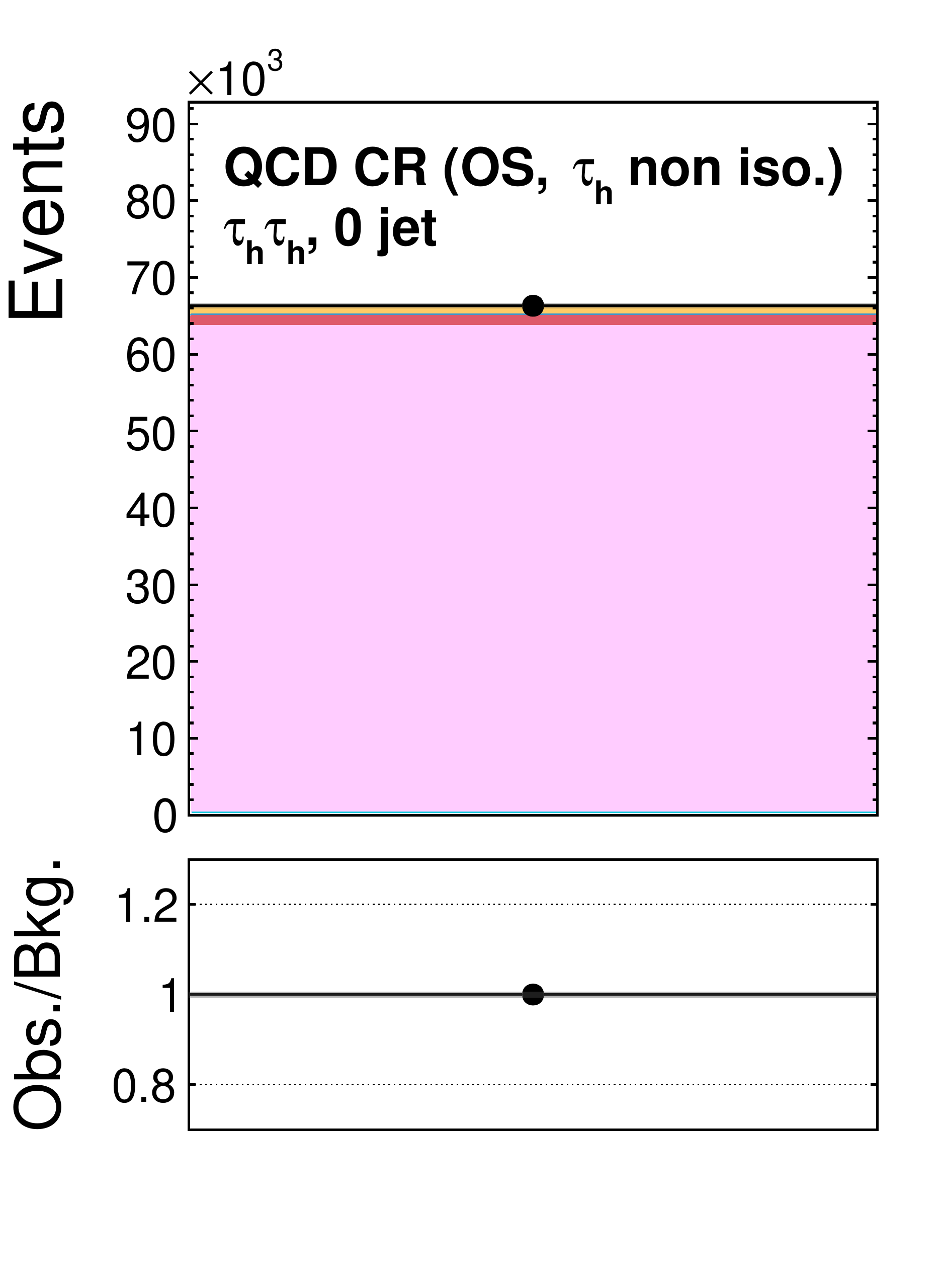}
     \includegraphics[width=0.24\textwidth]{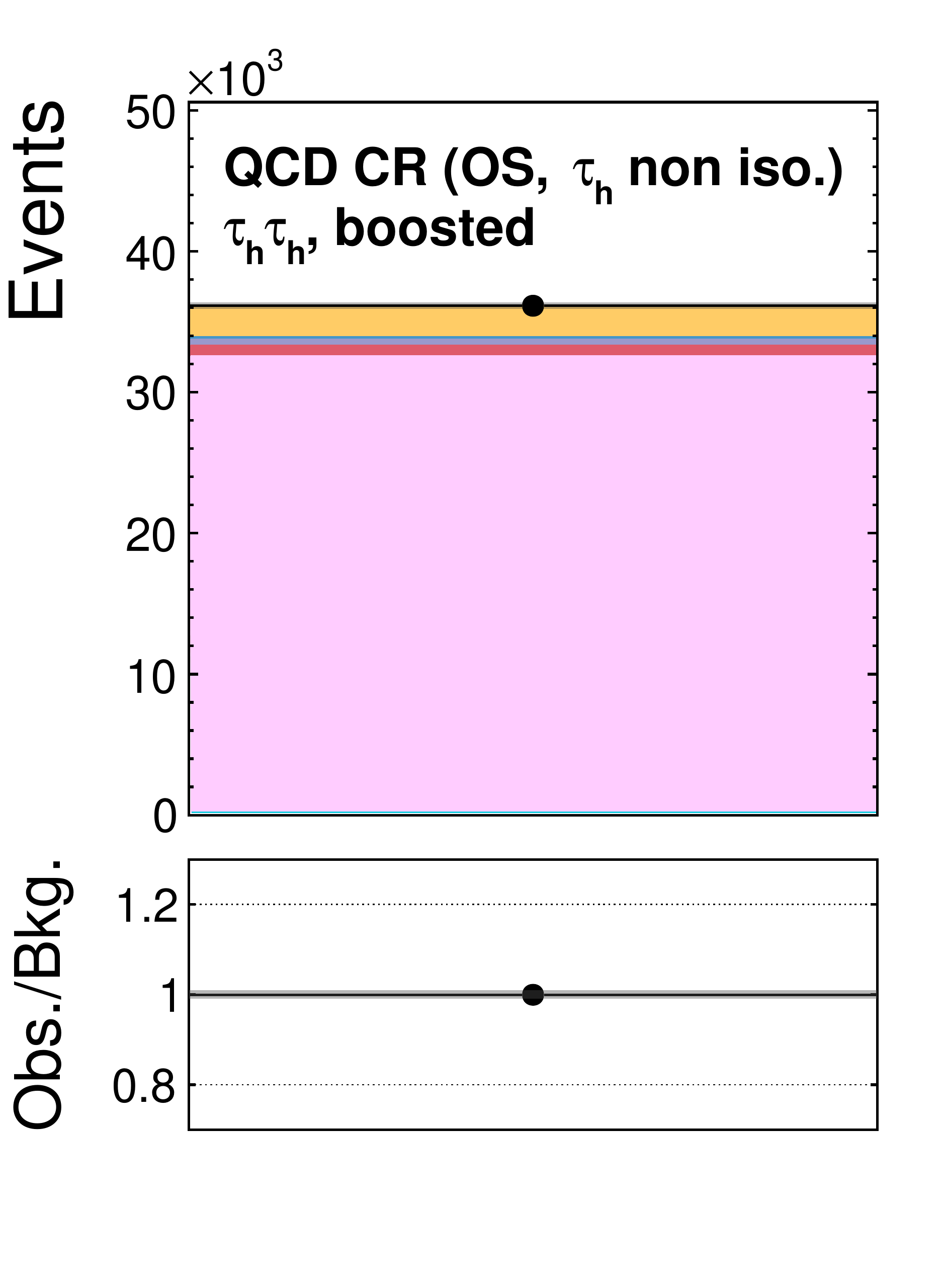}
     \includegraphics[width=0.24\textwidth]{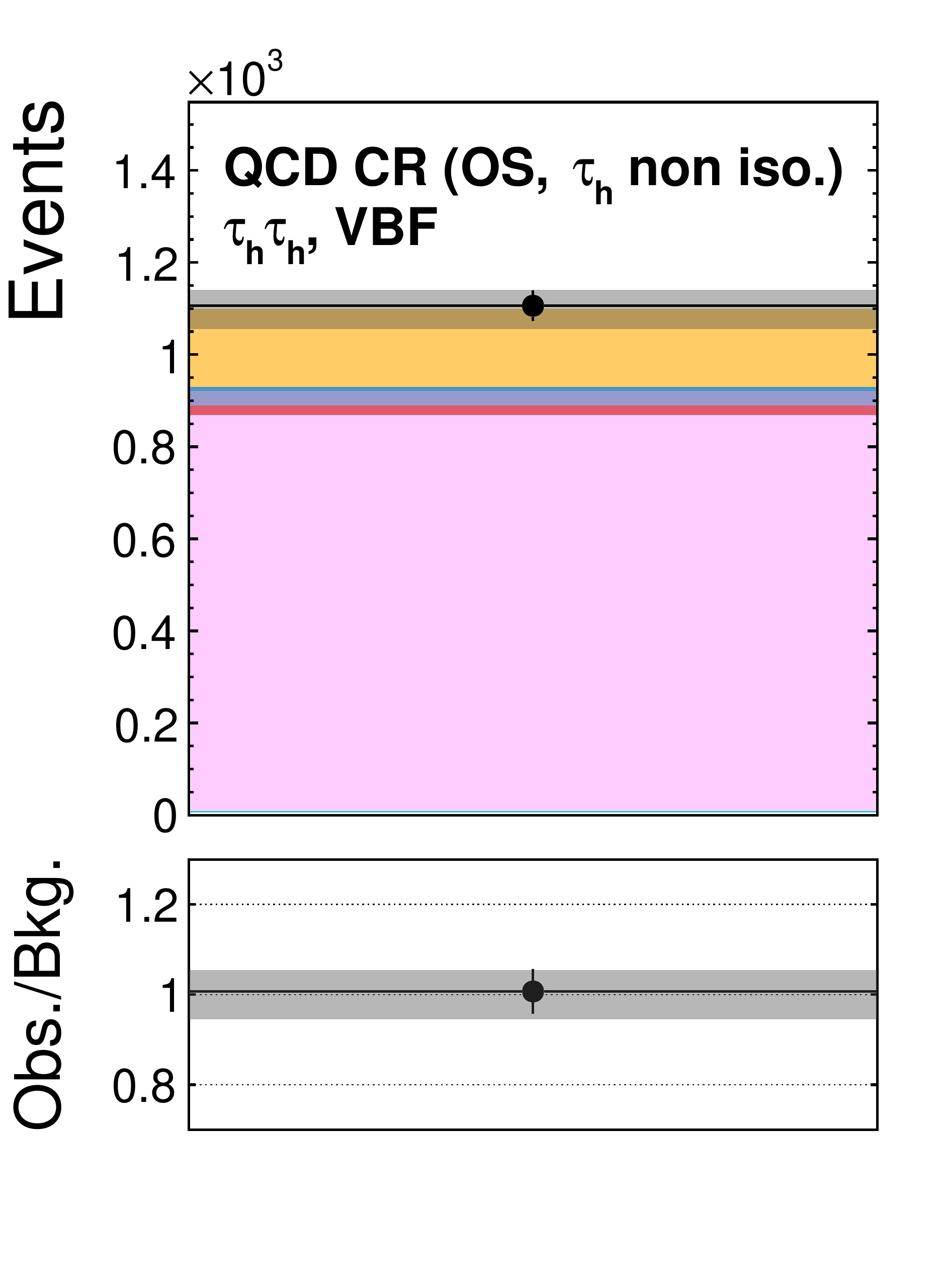}
     \includegraphics[width=0.24\textwidth]{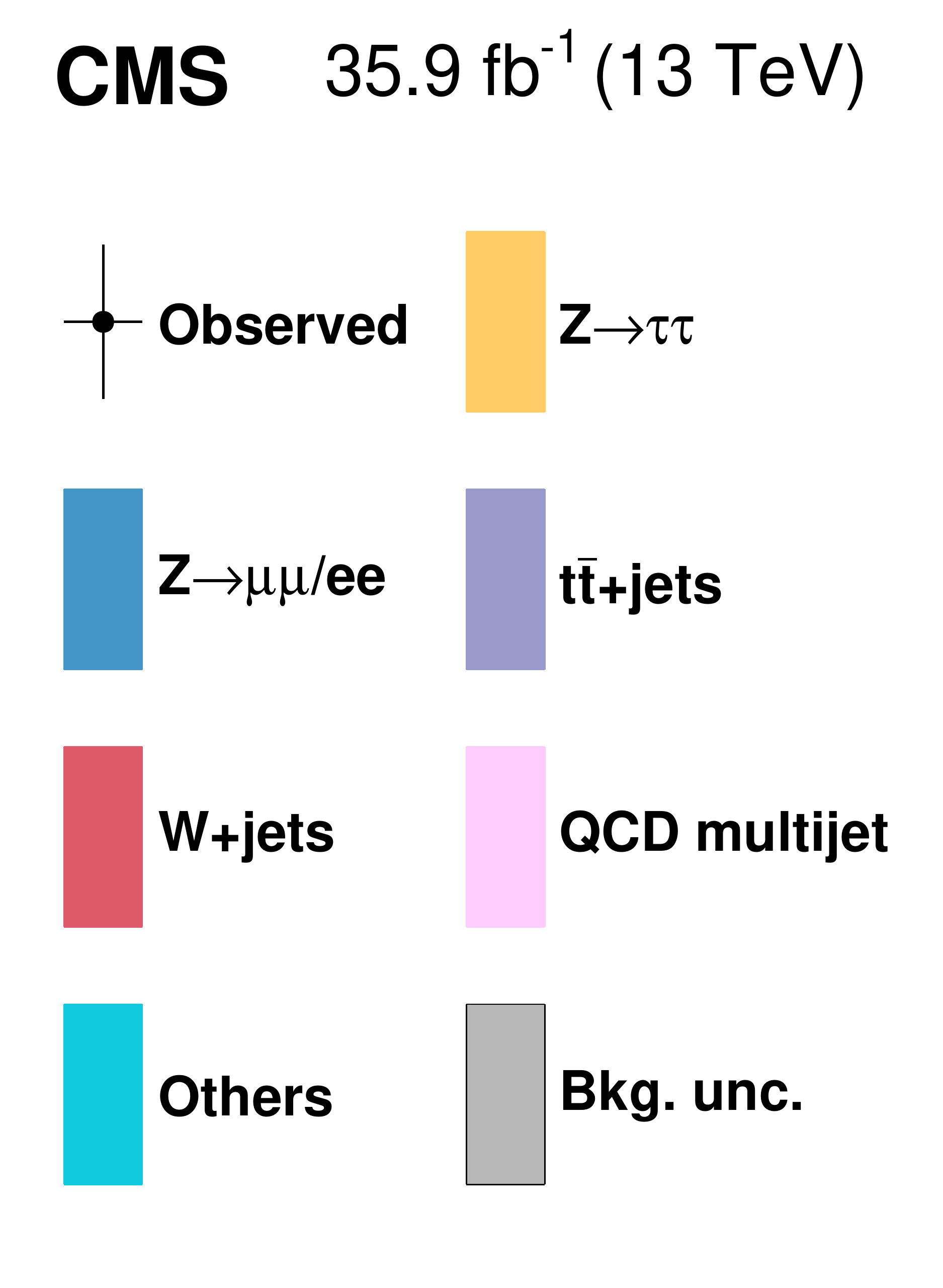}
     \caption{Control regions enriched in the QCD multijet background used in the maximum likelihood fit, together with the signal regions, to extract the results. The normalization of the predicted background distributions corresponds to the result of the global fit. These regions, formed by selecting events with opposite-sign $\tauh$ candidates passing relaxed isolation requirements, control the yields of the QCD multijet background in the $\tauh\tauh$ channel.}
     \label{fig:CR4}
\end{figure*}

In the $\tauh\tauh$ channel, the large QCD multijet background is estimated with a slightly different method, from
a sample composed of events with opposite-sign $\tauh$ satisfying a relaxed isolation requirement, disjoint from the signal region.
In this region, the QCD multijet background shape and yield are obtained
by subtracting the contribution of the
Drell--Yan, \ttbar, and $\PW+\text{jets}$ processes, estimated as explained above, from the data.
The QCD multijet background yield in the signal region is obtained by multiplying
the yield previously obtained in the control region by an extrapolation factor.
The extrapolation factor is measured in events passing identical selection criteria as those in the signal region, and in the relaxed
isolation region, except that the $\tauh$ candidates are required to have the same sign.
The events selected with opposite-sign $\tauh$ candidates passing relaxed isolation requirements form control regions, shown in Fig.~\ref{fig:CR4}, and are used in the fit to extract the results.

\begin{figure}[htb]
\centering
     \includegraphics[width=0.23\textwidth]{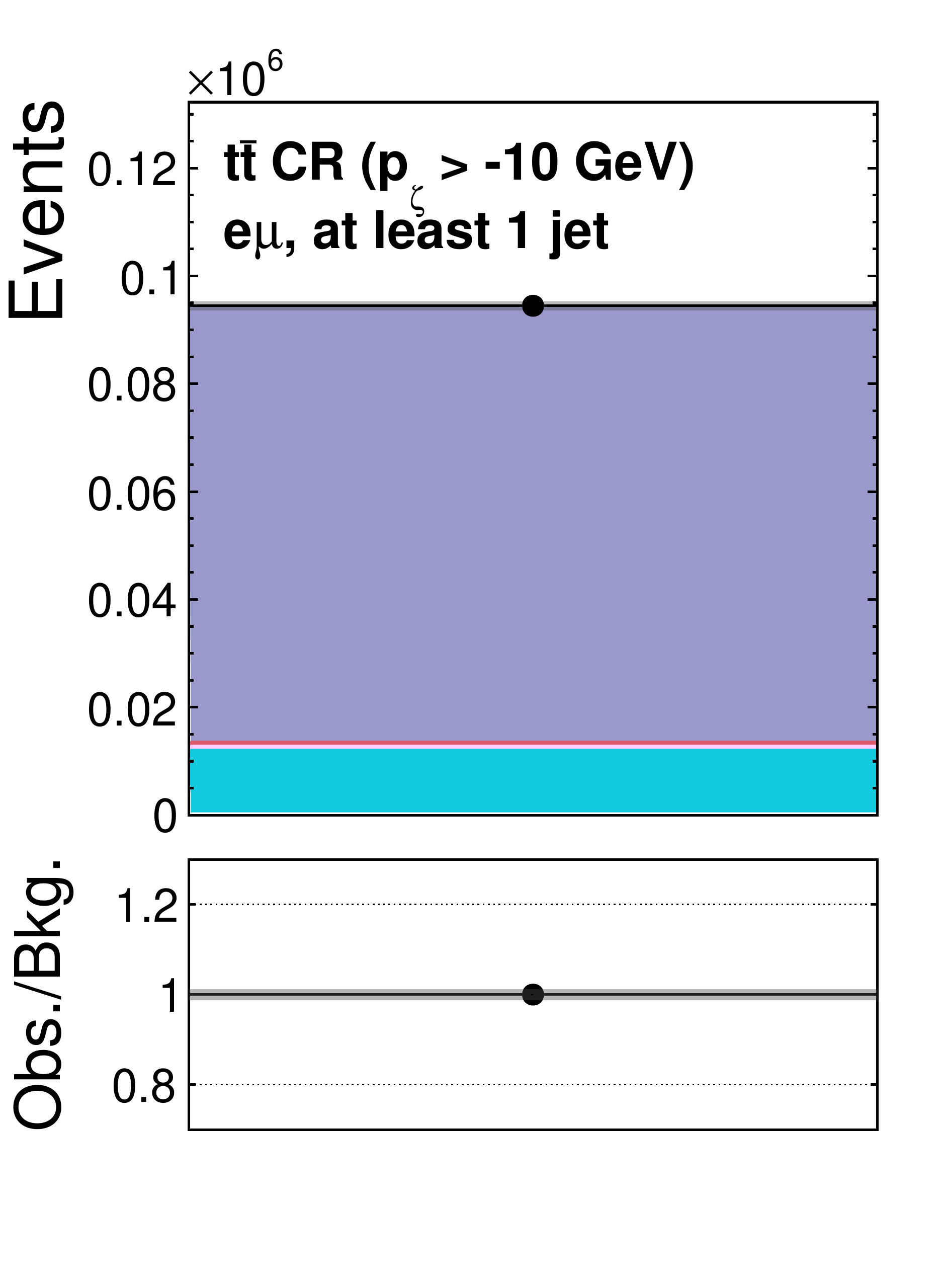}
     \includegraphics[width=0.23\textwidth]{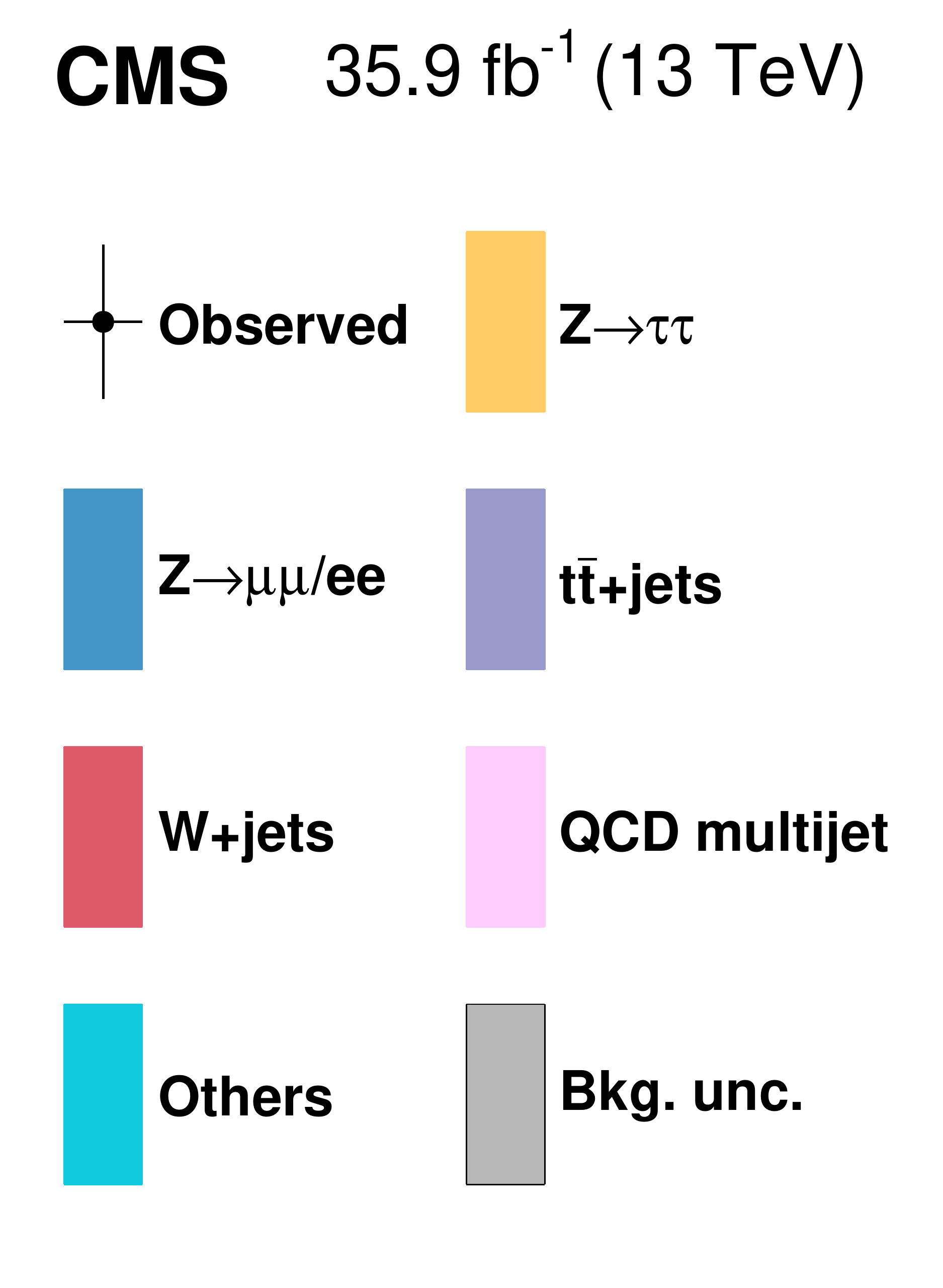}
     \caption{Control region enriched in the $\ttbar$ background, used in the maximum likelihood fit, together with the signal regions, to extract the results. The normalization of the predicted background distributions corresponds to the result of the global fit. This region, defined by inverting the $p_\zeta$ requirement and rejecting events with no jet in the $\Pe\Pgm$ final state,  is used to estimate the
yields of the $\ttbar$ background in all channels.}
     \label{fig:CR2}
\end{figure}

The \ttbar production process is one of the main backgrounds in the $\Pe\Pgm$ channel.
The 2D distributions in all decay channels are predicted by simulation. The normalization is
adjusted to the one observed in a \ttbar-enriched sample orthogonal to the signal region. This control
region, shown in Fig.~\ref{fig:CR2},
is added to the global fit to extract the results, and is defined similarly as the $\Pe\Pgm$ signal region, except that the $p_\zeta$ requirement
is inverted and the events should contain at least one jet.

The contributions from diboson and single top quark production are estimated from simulation, as is the $\hww$ background.

 \section{Systematic uncertainties}
\label{sec:systematics}

\subsection{Uncertainties related to object reconstruction and identification}

The overall uncertainty in the $\tauh$ identification efficiency for genuine $\tauh$ leptons is 5\%, which has been measured with a tag-and-probe method in $\PZ\to\Pgt\Pgt$ events.
This number is not fully correlated among the di-$\Pgt$ channels because the $\tauh$ candidates are required to pass
different working points of the discriminators that reduce the misidentification rate of electrons and muons as $\tauh$ candidates.
The trigger efficiency uncertainty per $\tauh$ candidate amounts to an additional 5\%, which leads to a total trigger uncertainty of 10\% for processes estimated from simulation in the $\tauh\tauh$ decay channel. This uncertainty has also been measured with a tag-and-probe method in $\PZ\to\Pgt\Pgt$ events.

An uncertainty of 1.2\% in the visible energy scale of genuine $\tauh$ leptons affects both the distributions and the
signal and background yields. It is uncorrelated among the 1-prong, 1-prong + $\PGpz$, and
3-prong decay modes.
The magnitude of the uncertainty was determined in $\PZ\to\Pgt\Pgt$ events with one $\Pgt$ lepton decaying hadronically and the other one to a muon, by performing maximum likelihood fits for different values of the visible energy scale of genuine $\tauh$ leptons. Among these events, less than half overlap with the events selected in the $\Pgm\tauh$ channel of this analysis. The fit constrains the visible $\tauh$ energy scale uncertainty to about
0.3\% for all decay modes. The constraint mostly comes from highly populated regions with a high $\tauh$ purity, namely the 0-jet and boosted categories of the $\Pgm\tauh$ and $\tauh\tauh$ channels. The decrease in the size of the uncertainty is explained by the addition of two other decay channels
with $\tauh$ candidates ($\tauh\tauh$ and $\Pe\tauh$), by the higher number of events in the MC simulations, and by the finer categorization that leads to regions with a high $\PZ\to\Pgt\Pgt$ event purity.
Even in the most boosted categories, reconstructed $\tauh$ candidates typically have moderate $\pt$ ($\pt$ less than 100\GeV) and are
found in the barrel region of the detector. As tracks are well measured in the CMS detector for this range of $\pt$,
the visible energy scale of genuine $\tauh$ leptons is fully correlated for all $\tauh$ leptons reconstructed in the same decay mode, irrespective of their $\pt$ and $\eta$. The uncertainties in the visible energy scale for genuine $\tauh$ leptons together contribute an uncertainty of 5\% to the measurement of the signal strength.

In the 0-jet category of the $\Pgm\tauh$ and $\Pe\tauh$ channels, the relative contribution of $\tauh$ in a given
reconstructed decay mode is allowed to fluctuate by 3\% to account for the possibility that the reconstruction and
identification efficiencies are different for each decay mode. This uncertainty has been measured in a region enriched in $\PZ\to\Pgt\Pgt$ events with one $\Pgt$ lepton decaying hadronically and the other one decaying to a muon, by comparing the level of agreement in exclusive bins of the reconstructed $\tauh$ decay mode, after adjusting the inclusive normalization of the $\PZ\to\Pgt\Pgt$ simulation to its best-fit value. The effect of migration between the reconstructed $\tauh$ decay modes is negligible in other categories, where
all decay modes are treated together.

For events where muons or electrons are misidentified as $\tauh$ candidates, essentially $\PZ\to \Pgm\Pgm$ events in the $\Pgm\tauh$ decay channel and $Z\to \Pe\Pe$ events in the $\Pe\tauh$ decay channel, the $\tauh$ identification leads to rate uncertainties of 25 and 12\%, respectively, per reconstructed $\tauh$ decay mode. Using $\mvis$ and the reconstructed $\tauh$ decay mode as the observables in the 0-jet category of the $\Pgm\tauh$ and $\Pe\tauh$ channels helps reduce the uncertainty after the signal extraction fit: the uncertainty in the rate of muons or electrons misidentified as $\tauh$ becomes of the order of 5\%. The energy scale uncertainty for muons or electrons
 misidentified as $\tauh$ candidates is 1.5 or 3\%, respectively, and is uncorrelated between reconstructed $\tauh$ decay
modes. The fit constrains these uncertainties to about one third of their initial values. For events where quark- or gluon-initiated jets are misidentified as $\tauh$ candidates, a linear uncertainty that increases by 20\% per 100\GeV in $\tauh$ $\pt$ accounts for a potential mismodeling of the jet$\to\tauh$ misidentification rate as a
function of the $\tauh$ $\pt$ in simulations. The uncertainty has been determined from a region enriched in $\PW+\text{jets}$ events, using events with a muon and a $\tauh$ candidate in the final state, characterized by a large transverse mass between the $\ptmiss$ and the muon~\cite{Khachatryan:2015dfa,CMS-PAS-TAU-16-002}.

In the decay channels with muons or electrons, the uncertainties in the muon and electron identification, isolation, and trigger efficiencies lead to the rate uncertainty of 2\% for both muons and electrons.
The uncertainty in the electron energy scale, which amounts to 2.5\% in the endcaps and 1\% in the barrel of the detector, is relevant only in the $\Pe\Pgm$ decay channel, where it affects the final distributions.
In all channels, the effect of the uncertainty in
the muon energy scale is negligible.

The uncertainties in the jet energy scale depend on the \pt and $\eta$ of the jet~\cite{CMS-JME-10-011}.
They are propagated to the computation of the number of jets, which affects the repartition of events between the 0-jet, VBF, and boosted categories, and to the computation of $\mjj$, which is one of the observables in the VBF category.

The rate uncertainty related to discarding events with a b-tagged jet in the $\Pe\Pgm$ decay channel is up to
5\% for the $\ttbar$ background. The uncertainty in the mistagging rate of gluon and light-flavor jets is negligible.

The \ptvecmiss scale uncertainties~\cite{CMS-JME-12-002}, which are computed event-by-event, affect the normalization of various processes through the event selection, as well as their distributions through the propagation of these uncertainties to the di-$\Pgt$ mass $\mtt$. The \ptvecmiss scale uncertainties arising from unclustered energy deposits in the detector come from four independent sources related to the tracker, ECAL, HCAL, and forward calorimeters subdetectors. Additionally, \ptvecmiss scale uncertainties related to the uncertainties in the jet energy scale measurement, which lead to uncertainties in the \ptvecmiss calculation, are taken into account. The combination of both sources of uncertainties in the \ptvecmiss scale leads to an uncertainty of about 10\% in the measured signal strength.

\subsection{Background estimation uncertainties}

The $\PZ\to\Pgt\Pgt$ background yield and distribution are corrected based on the agreement between data and the background prediction in a control region enriched in the $\PZ\to\Pgm\Pgm$ events, as explained in Section~\ref{sec:background_estimation}.
The extrapolation uncertainty related to kinematic differences in the selections in the signal and control regions ranges between 3 and 10\%, depending
on the category. In addition, shape uncertainties related to the uncertainties in the applied corrections are considered; they reach 20\% for some ranges of $\mjj$ in the VBF category. These uncertainties arise from the different level of agreement between data and simulation in the $\PZ\to\Pgm\Pgm$ control region obtained when varying the threshold on the muon $\pt$.

The uncertainties in the $\PW+ \text{jets}$ event yield determined from the control regions in the $\Pgm\tauh$ and $\Pe\tauh$ channels account for the statistical uncertainty of the observed data, the statistical uncertainty of the $\PW+ \text{jets}$ simulated sample, and the systematic uncertainties associated with background processes in these control regions. Additionally, an uncertainty in the extrapolation  of the constraints from the high-\MT ($\MT>80\GeV$) control regions to the low-\MT ($\MT<50\GeV$) signal regions is additionally taken into account. The latter ranges from 5 to 10\%, and is obtained by comparing the \MT distributions of simulated and observed $\PZ\to\Pgm\Pgm$ events where one of the muons is removed and the \ptvecmiss adjusted accordingly, to mimic $\PW+\text{jets}$ events. The reconstructed invariant mass of the parent boson in the rest frame is multiplied by the ratio of the $\PW$ and $\PZ$ boson masses before removing the muon.
In the $\tauh\tauh$ and $\Pe\Pgm$ channels, where the $\PW+\text{jets}$ background is estimated from simulation, the uncertainty in the yield of this small background is equal to 4 and 20\%, respectively. The larger value for the $\Pe\Pgm$ channel includes uncertainties in the misidentification rates of jets as electrons and muons, whereas the uncertainty in the misidentification rate of jets as $\tauh$ candidates in the $\tauh\tauh$ channel is accounted for by the linear uncertainty as a function of the $\tauh$ $\pt$ described earlier.

The uncertainty in the QCD multijet background yield in the $\Pe\Pgm$ decay channel ranges from 10 to 20\%, depending on the category. It corresponds to the
uncertainty in the extrapolation factor from the same-sign to opposite-sign region, measured in events with anti-isolated leptons.
In the $\Pgm\tauh$ and $\Pe\tauh$ decay channels, uncertainties from the fit of the control regions with leptons passing relaxed isolation conditions are
considered, together with an additional 20\% uncertainty that accounts for the extrapolation from the relaxed-isolation control region to the isolated signal region.
In the $\tauh\tauh$ decay channel, the uncertainty in the QCD mutlijet background yield is a combination of the uncertainties obtained from fitting the dedicated control regions with $\tauh$ candidates passing relaxed isolation criteria, and of extrapolation uncertainties to the signal region ranging from 3 to 15\% and accounting for limited disagreement between prediction and data in signal-free regions with various loose isolation criteria.

The yield of events in a $\ttbar$-enriched region is added to the maximum likelihood fit to control the normalization of this process
in the signal region, as explained in Section~\ref{sec:background_estimation}. The uncertainty from the fit in the control region is automatically propagated to the signal regions, resulting in an uncertainty of about 5\% on the $\ttbar$ cross section. Per-channel uncertainties related to the object reconstruction and identification are considered when extrapolating from the $\Pe\Pgm$ final state to the others. The $\ttbar$ simulation is corrected for differences in the top quark $\pt$ distributions observed between data and simulation, and an uncertainty in the correction is taken into account.

The combined systematic uncertainty in the background yield arising from diboson and single top quark production processes is estimated to be 5\%
on the basis of recent CMS measurements~\cite{Khachatryan:2016tgp,Sirunyan:2016cdg}.

\subsection{Signal prediction uncertainties}

The rate and acceptance uncertainties for the signal processes related to the theoretical calculations are due to uncertainties in the PDFs, variations of the QCD renormalization and factorization scales,
and uncertainties in the modelling of parton showers.
The magnitude of the rate uncertainty depends on the production process and on the event category.

The inclusive uncertainty related to the PDFs amounts to 3.2, 2.1, 1.9, and 1.6\%, respectively, for the $ \cPg\cPg\PH $, VBF, $\PW\PH$, and $\PZ\PH$ production modes~\cite{deFlorian:2016spz}. The
corresponding uncertainty for the variation of the renormalization and factorization scales is 3.9, 0.4, 0.7, and 3.8\%, respectively~\cite{deFlorian:2016spz}.
The acceptance uncertainties related to the particular selection criteria used in this analysis are less than 1\% for the $ \cPg\cPg\PH $ and VBF productions for the PDF
uncertainties. The acceptance uncertainties for the VBF production in the renormalization and factorization scale uncertainties are also less than 1\%, while the corresponding uncertainties for the $ \cPg\cPg\PH $ process are treated as shape uncertainties as the uncertainty increases
linearly with $\pth$ and $\mjj$.

The \pt distribution of the Higgs boson in the {\POWHEG2.0} simulations is tuned to match more closely
the next-to-NLO (NNLO) plus
next-to-next-to-leading-logarithmic (NNLL) prediction in the
\textsc{HRes2.1} generator~\cite{deFlorian:2012mx,Grazzini:2013mca}.
The acceptance changes with the variation of the parton shower tune in \HERWIG++ 2.6 samples~\cite{Bellm:2013hwb} are considered as additional uncertainties, and amount to up to 7\% in the boosted category. The theoretical uncertainty in the branching fraction of the Higgs boson to $\Pgt$ leptons is equal to 2.1\%~\cite{deFlorian:2016spz}.

The theoretical uncertainties in the signal production depend on the jet multiplicity; this effect is included by following the prescriptions in Ref.~\cite{Stewart:2011cf}. This effect needs to be taken into account because the definitions of the three categories used in the analysis are based partially on the number of reconstructed jets. Additional uncertainties for boosted Higgs bosons, related to the treatment of the top quark mass in the calculations, are considered for signal events with $\pth>150\GeV$.

Theory uncertainties in the signal prediction contribute an uncertainty of 10\% to the measurement of the signal strength.

\subsection{Other uncertainties}

The uncertainty in the integrated luminosity amounts to 2.5\%~\cite{CMS-PAS-LUM-17-001}.

Uncertainties related to the finite number of simulated events, or to the limited number of events in data control regions, are taken into account. They are considered for all bins of the distributions used to extract the results if the uncertainty is larger than 5\%. They are uncorrelated across different samples, and across bins of a single distribution. Taken together, they contribute an uncertainty of about 12\% to the signal strength measurement, coming essentially from the VBF category, where the background templates are less
populated than in the other categories.

The systematic uncertainties considered in the analysis are summarized in Table~\ref{tab:uncertainties}.

\begin{table*}[!ht]
\centering
\topcaption{Sources of systematic uncertainty. If the global fit to the signal and control regions, described in the next section, significantly constrains these uncertainties, the values of the uncertainties after the global fit are indicated in the third column. The acronyms CR and ID stand for control region and identification, respectively.}
\newcolumntype{x}{D{,}{\text{--}}{2.2}}
\begin{tabular}{llx}
Source of uncertainty & Prefit & \multicolumn{1}{c}{Postfit (\%) }\\
\hline
 $\tauh$ energy scale                & 1.2\% in energy scale & 0.2,0.3 \\
 $\Pe$ energy scale               & 1--2.5\%  in energy scale & 0.2,0.5\\
 $\Pe$ misidentified as $\tauh$ energy scale & 3\% in energy scale & 0.6,0.8 \\
 $\Pgm$ misidentified as $\tauh$ energy scale & 1.5\% in energy scale &  0.3,1.0\\
 Jet energy scale               & Dependent upon $\pt$ and $\eta$ & \multicolumn{1}{c}{\NA} \\
 \ptvecmiss energy scale              & Dependent upon $\pt$ and $\eta$ & \multicolumn{1}{c}{\NA}  \\[\cmsTabSkip]
 $\tauh$ ID \& isolation & 5\% per $\tauh$ & \multicolumn{1}{c}{3.5} \\
 $\tauh$ trigger & 5\% per $\tauh$ & \multicolumn{1}{c}{3} \\
 $\tauh$ reconstruction per decay mode & 3\% migration between decay modes & \multicolumn{1}{c}{2} \\
 $\Pe$ ID \& isolation \& trigger  &   2\% & \multicolumn{1}{c}{\NA} \\
 $\Pgm$ ID \& isolation \& trigger & 2\% & \multicolumn{1}{c}{\NA} \\
 $\Pe$ misidentified as $\tauh$ rate   & 12\%  & \multicolumn{1}{c}{5} \\
 $\Pgm$ misidentified as $\tauh$ rate  & 25\%  & 3,8 \\
 Jet misidentified as $\tauh$ rate     & 20\% per 100\GeV $\tauh$ $\pt$ & \multicolumn{1}{c}{15}  \\[\cmsTabSkip]
 $\PZ\to\Pgt\Pgt/\ell\ell$ estimation & Normalization: 7--15\% & 3,15 \\
                             & Uncertainty in $m_{\ell\ell/\Pgt\Pgt}$, $\pt(\ell\ell/\Pgt\Pgt)$,  & \multicolumn{1}{c}{\NA} \\
                             & and $\mjj$ corrections & \\[\cmsTabSkip]
 $\PW+ \text{jets}$ estimation & Normalization ($\Pe\Pgm$, $\tauh\tauh$): 4--20\% &  \multicolumn{1}{c}{\NA} \\
                               & Unc. from CR ($\Pe\tauh$, $\Pgm\tauh$): $\simeq$5--15& \multicolumn{1}{c}{\NA} \\
                               & Extrap. from high-$m_T$ CR ($\Pe\tauh$, $\Pgm\tauh$): 5--10\% & \multicolumn{1}{c}{\NA}  \\[\cmsTabSkip]
QCD multijet estimation        & Normalization ($\Pe\Pgm$): 10--20\% & 5,20\% \\
                               & Unc. from CR ($\Pe\tauh$, $\tauh\tauh$, $\Pgm\tauh$): $\simeq$5--15\% & \multicolumn{1}{c}{\NA} \\
                               & Extrap. from anti-iso. CR ($\Pe\tauh$, $\Pgm\tauh$): 20\% & 7,10 \\
                               & Extrap. from anti-iso. CR ($\tauh\tauh$): 3--15\% & 3,10 \\[\cmsTabSkip]
 Diboson normalization & 5\% & \multicolumn{1}{c}{\NA}  \\[\cmsTabSkip]
 Single top quark normalization  & 5\% & \multicolumn{1}{c}{\NA} \\[\cmsTabSkip]
 $\ttbar$ estimation & Normalization from CR: $\simeq$5\% & \multicolumn{1}{c}{\NA} \\
                     & Uncertainty on top quark $\pt$ reweighting & \multicolumn{1}{c}{\NA}  \\[\cmsTabSkip]
 Integrated luminosity     & 2.5\% & \multicolumn{1}{c}{\NA} \\
 b-tagged jet rejection ($\Pe\Pgm$) & 3.5--5.0\% & \multicolumn{1}{c}{\NA} \\
 Limited number of events                & Statistical uncertainty in individual bins & \multicolumn{1}{c}{\NA}  \\[\cmsTabSkip]
 Signal theoretical uncertainty  & Up to 20\% & \multicolumn{1}{c}{\NA} \\
\hline
\end{tabular}
\label{tab:uncertainties}
\end{table*}

\section{Results}
\label{sec:results}

\begin{figure*}[htbp]
\centering
     \includegraphics[width=1.0\textwidth]{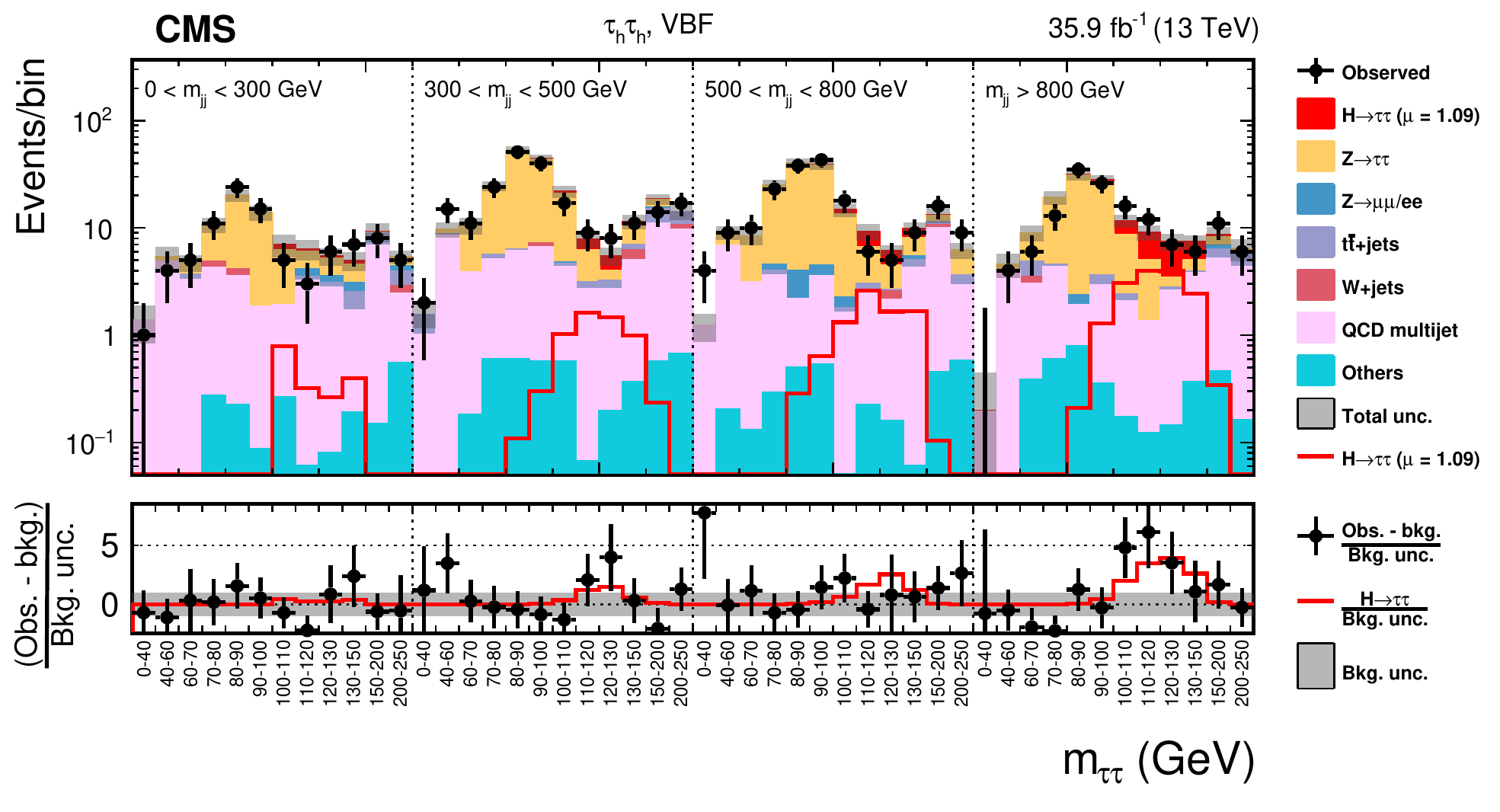}
     \caption{Observed and predicted 2D distributions in the VBF category of the $\tauh\tauh$ decay channel.
The normalization of the predicted background distributions corresponds to the result of the global fit.
The signal distribution is normalized to its best fit signal strength. The background histograms are stacked. The ``Others" background contribution includes events from diboson and single top quark production, as well as Higgs boson decays to a pair of $\PW$ bosons. The background uncertainty band accounts for all sources of background uncertainty, systematic as well as statistical, after the global fit. The signal is shown both as a stacked filled histogram and an open overlaid histogram.}
     \label{fig:mass_tt_0jet}
\end{figure*}

\begin{figure*}[htbp]
\centering
     \includegraphics[width=1.0\textwidth]{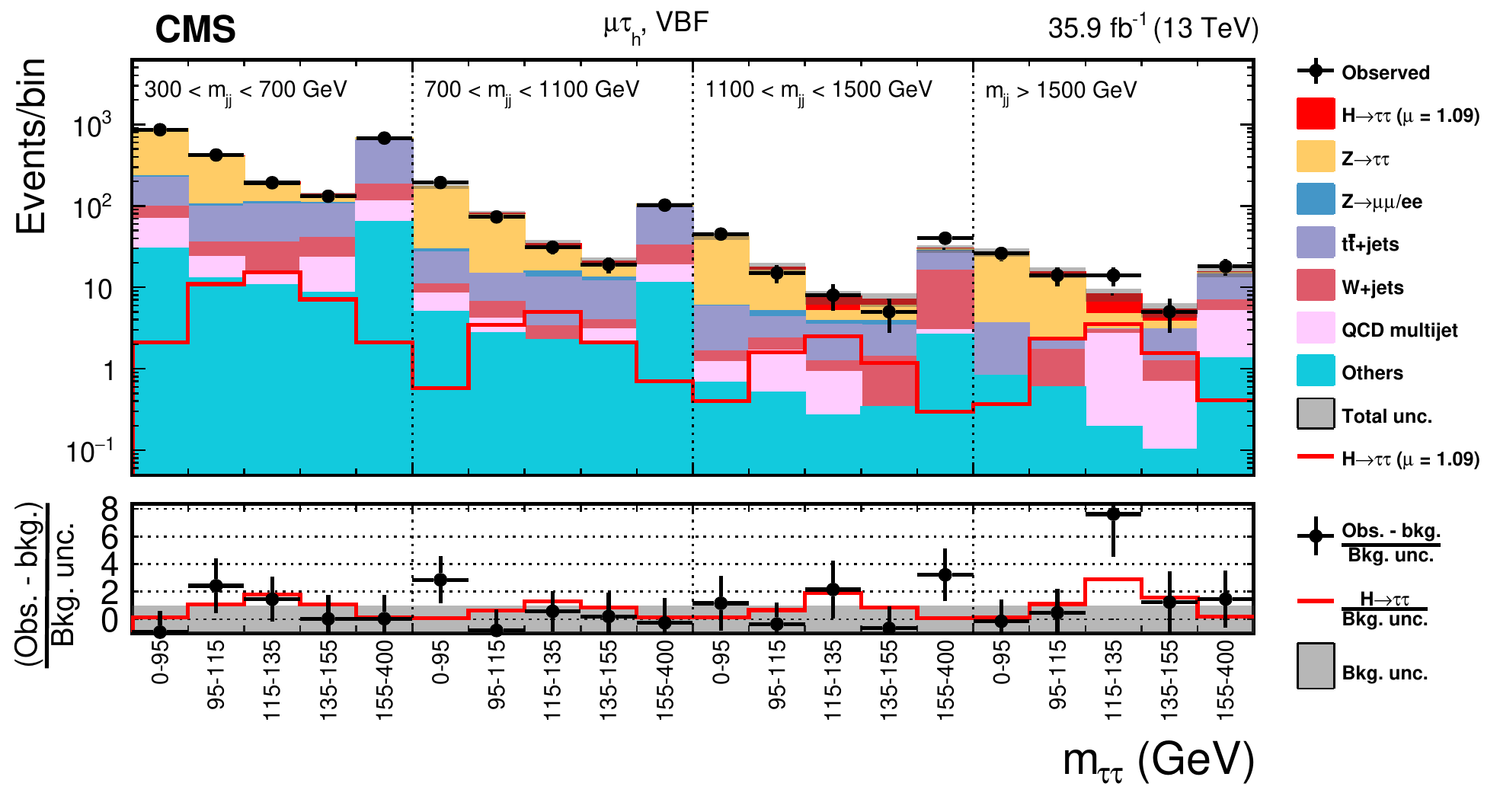}
     \caption{Observed and predicted 2D distributions in the VBF category of the $\Pgm\tauh$ decay channel. The description of the histograms is the same as in Fig.~\ref{fig:mass_tt_0jet}.}
     \label{fig:mass_tt_vbf}
\end{figure*}

\begin{figure*}[htbp]
\centering
     \includegraphics[width=1.0\textwidth]{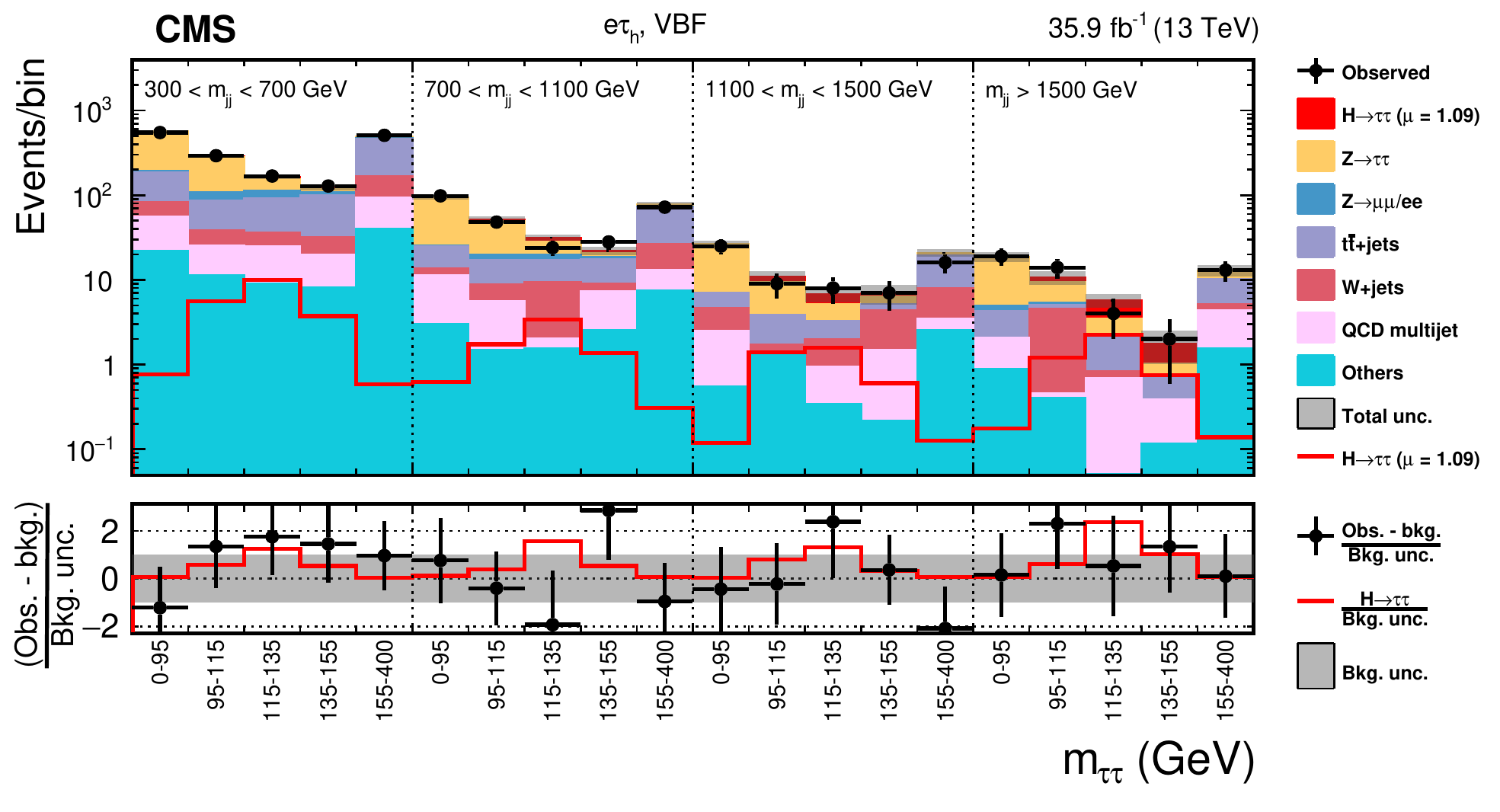}
     \caption{Observed and predicted 2D distributions in the VBF category of the $\Pe\tauh$ decay channel. The description of the histograms is the same as in Fig.~\ref{fig:mass_tt_0jet}.}
     \label{fig:mass_tt_boosted}
\end{figure*}

\begin{figure*}[htbp]
\centering
     \includegraphics[width=1.0\textwidth]{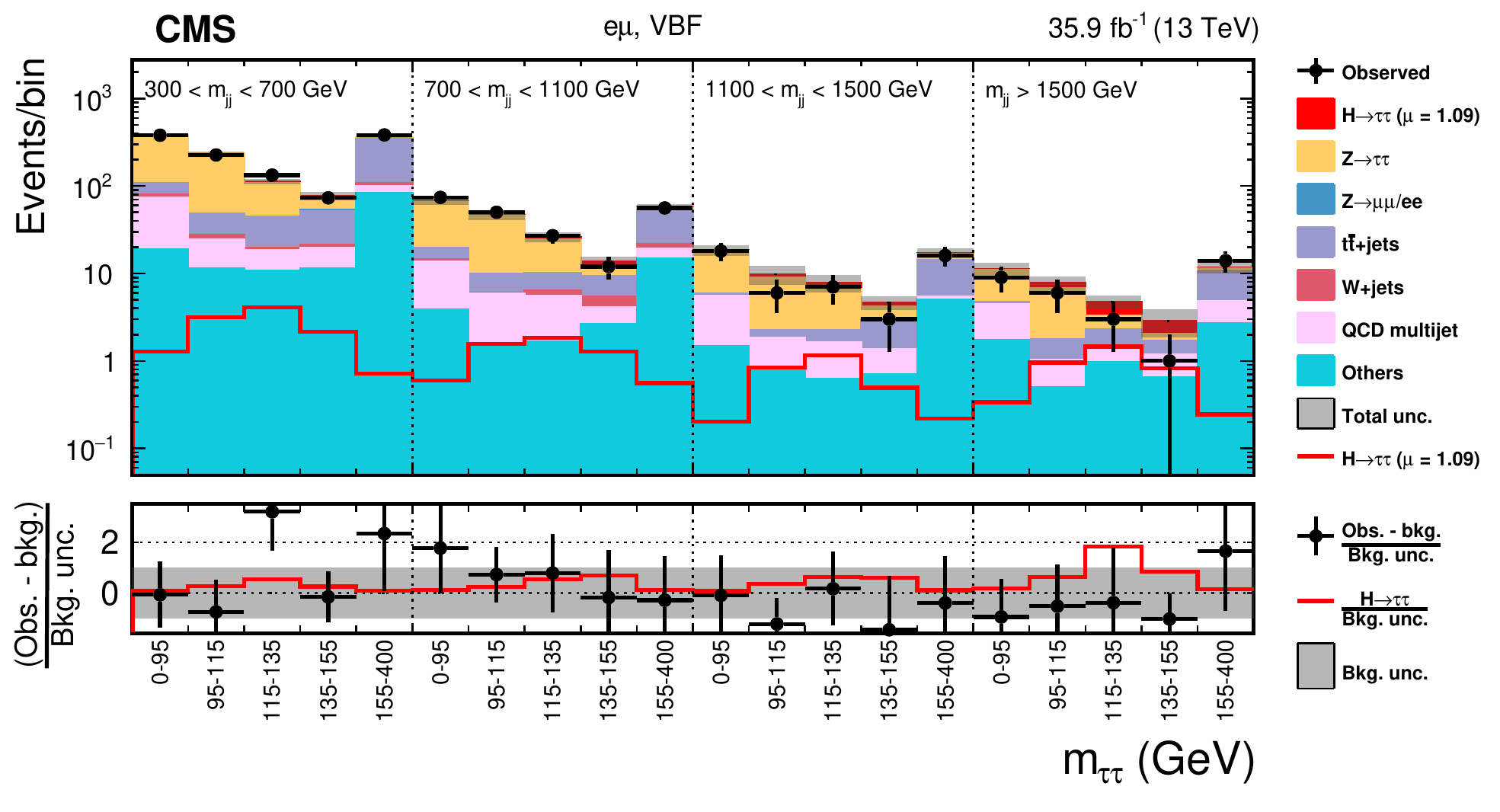}
     \caption{Observed and predicted 2D distributions in the VBF category of the $\Pe\Pgm$ decay channel. The description of the histograms is the same as in Fig.~\ref{fig:mass_tt_0jet}.}
     \label{fig:mass_mt_0jet}
\end{figure*}

\begin{figure*}[htbp]
\centering
     \includegraphics[width=1.0\textwidth]{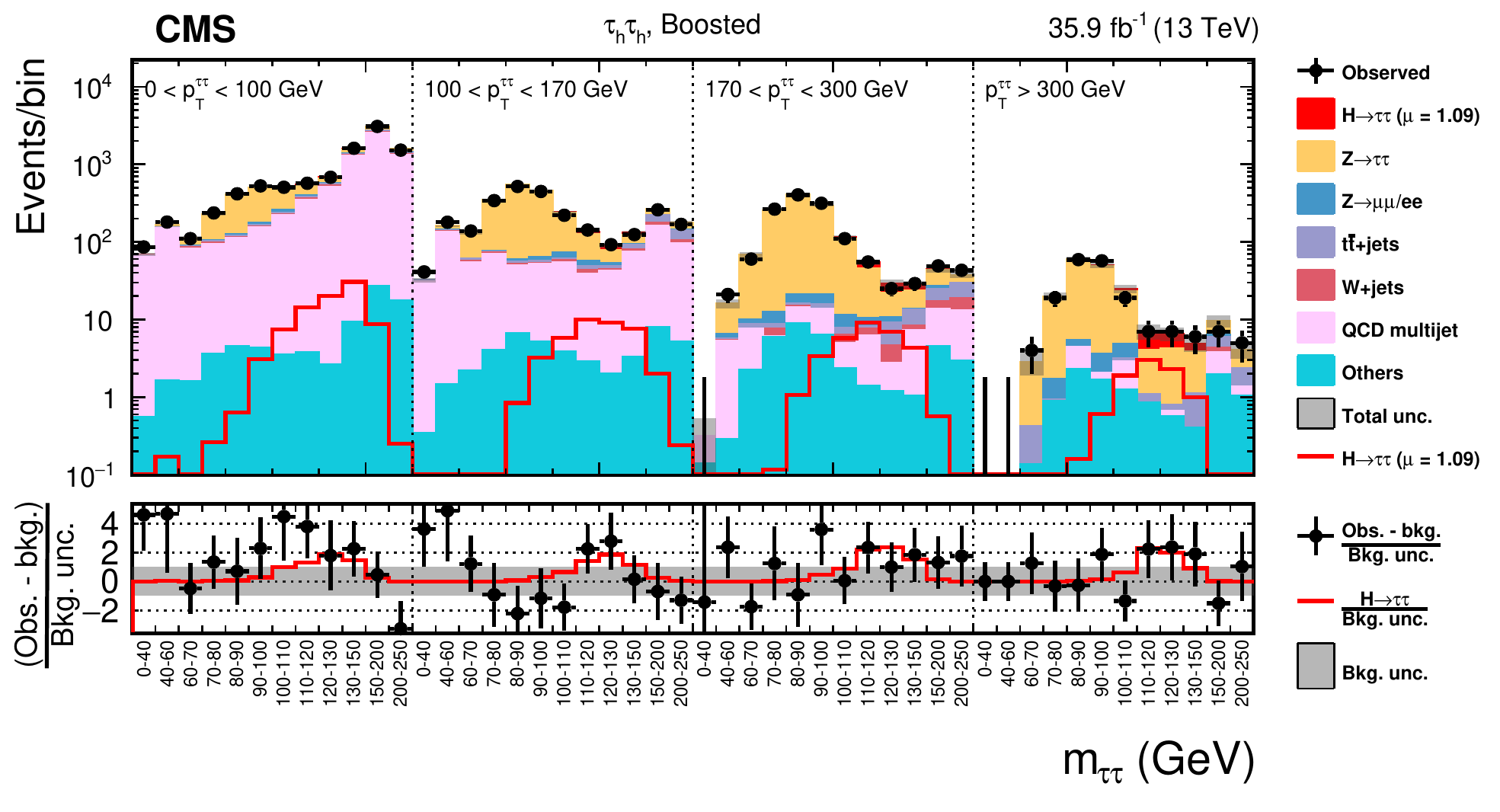}
     \caption{Observed and predicted 2D distributions in the boosted category of the $\tauh\tauh$ decay channel. The description of the histograms is the same as in Fig.~\ref{fig:mass_tt_0jet}.}
     \label{fig:mass_mt_vbf}
\end{figure*}

\begin{figure*}[htbp]
\centering
     \includegraphics[width=1.0\textwidth]{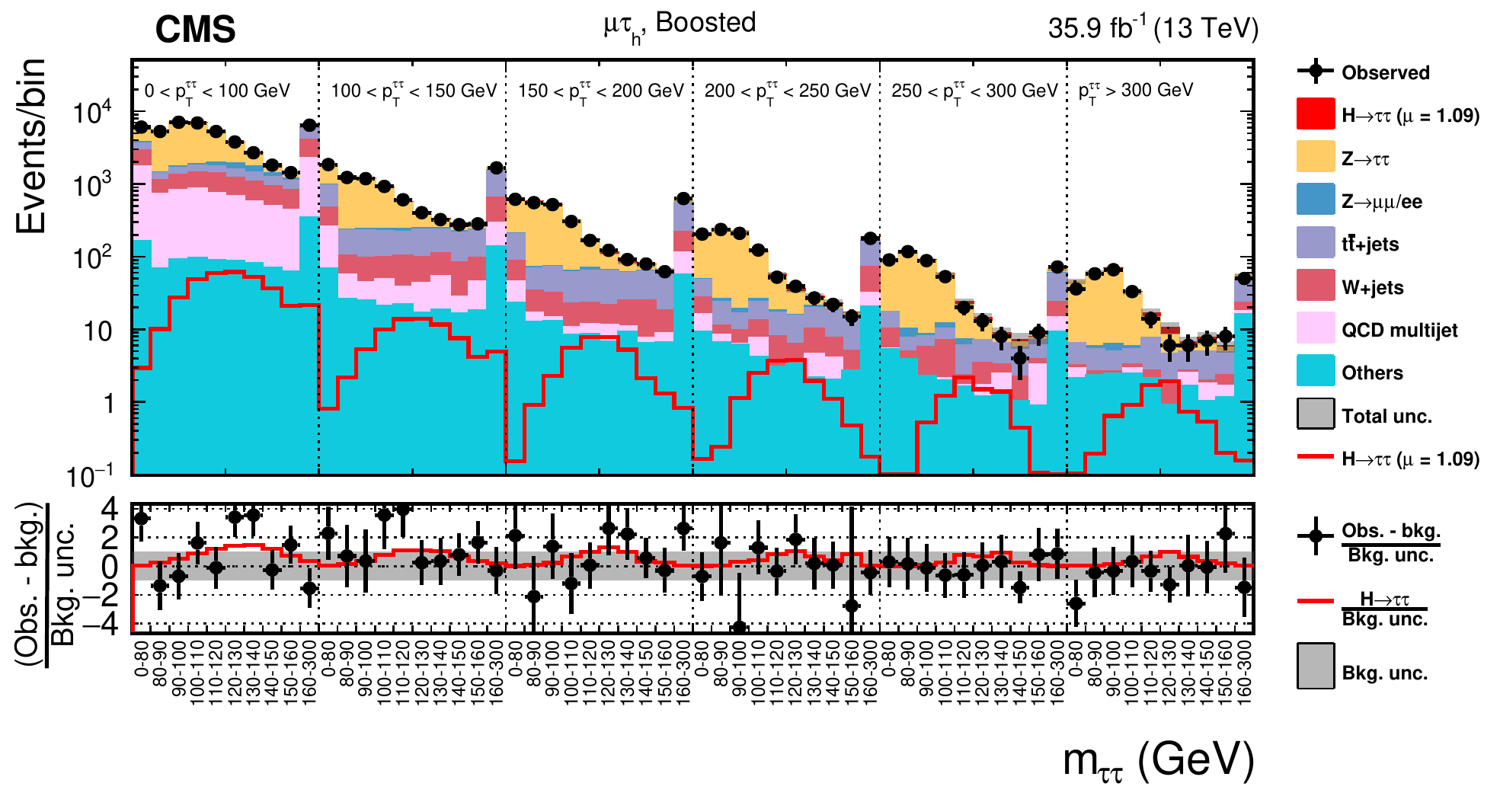}
     \caption{Observed and predicted 2D distributions in the boosted category of the $\Pgm\tauh$ decay channel. The description of the histograms is the same as in Fig.~\ref{fig:mass_tt_0jet}.}
     \label{fig:mass_mt_boosted}
\end{figure*}

\begin{figure*}[htbp]
\centering
     \includegraphics[width=1.0\textwidth]{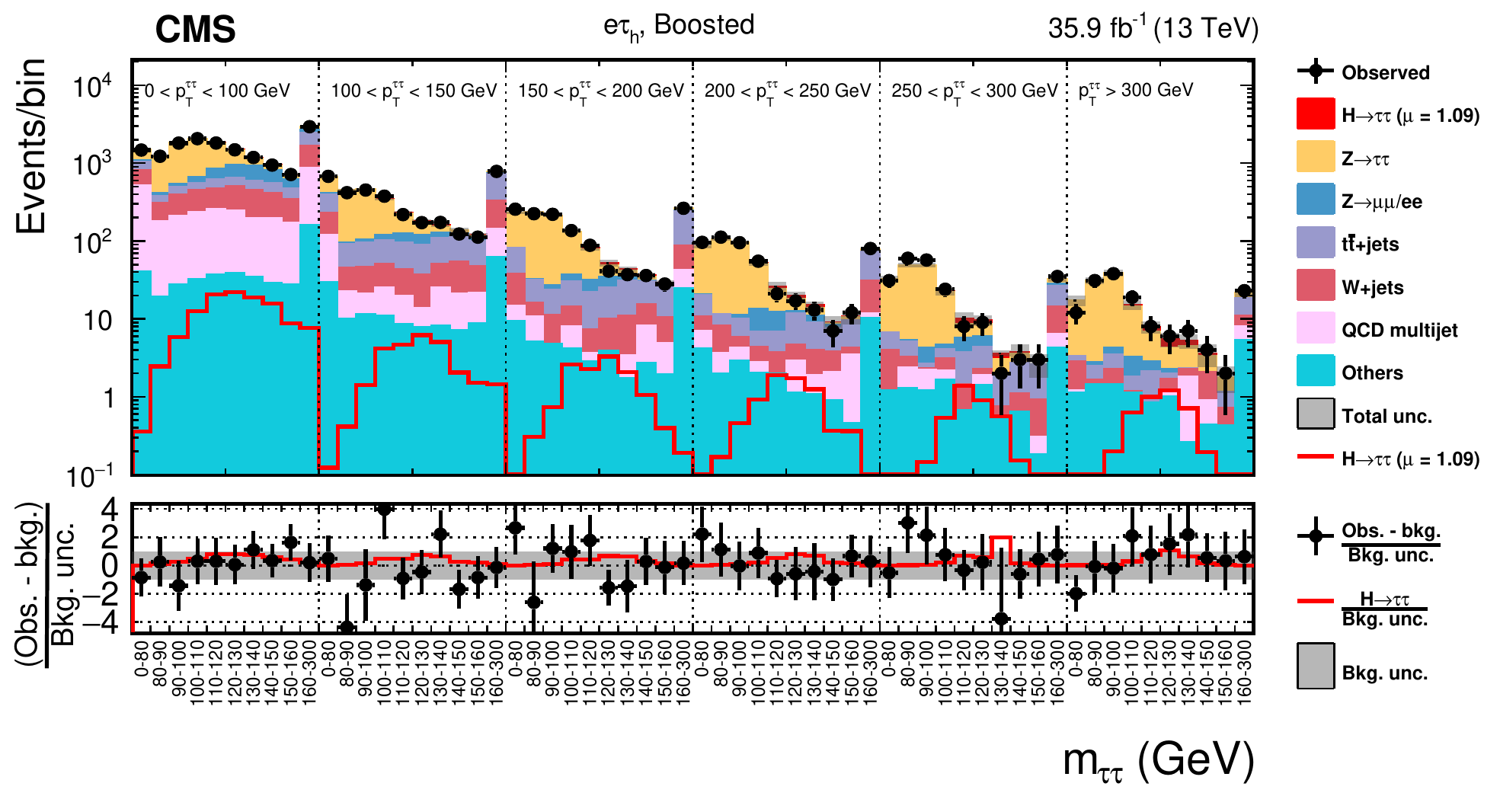}
     \caption{Observed and predicted 2D distributions in the boosted category of the $\Pe\tauh$ decay channel. The description of the histograms is the same as in Fig.~\ref{fig:mass_tt_0jet}.}
     \label{fig:mass_et_0jet}
\end{figure*}

\begin{figure*}[htbp]
\centering
     \includegraphics[width=1.0\textwidth]{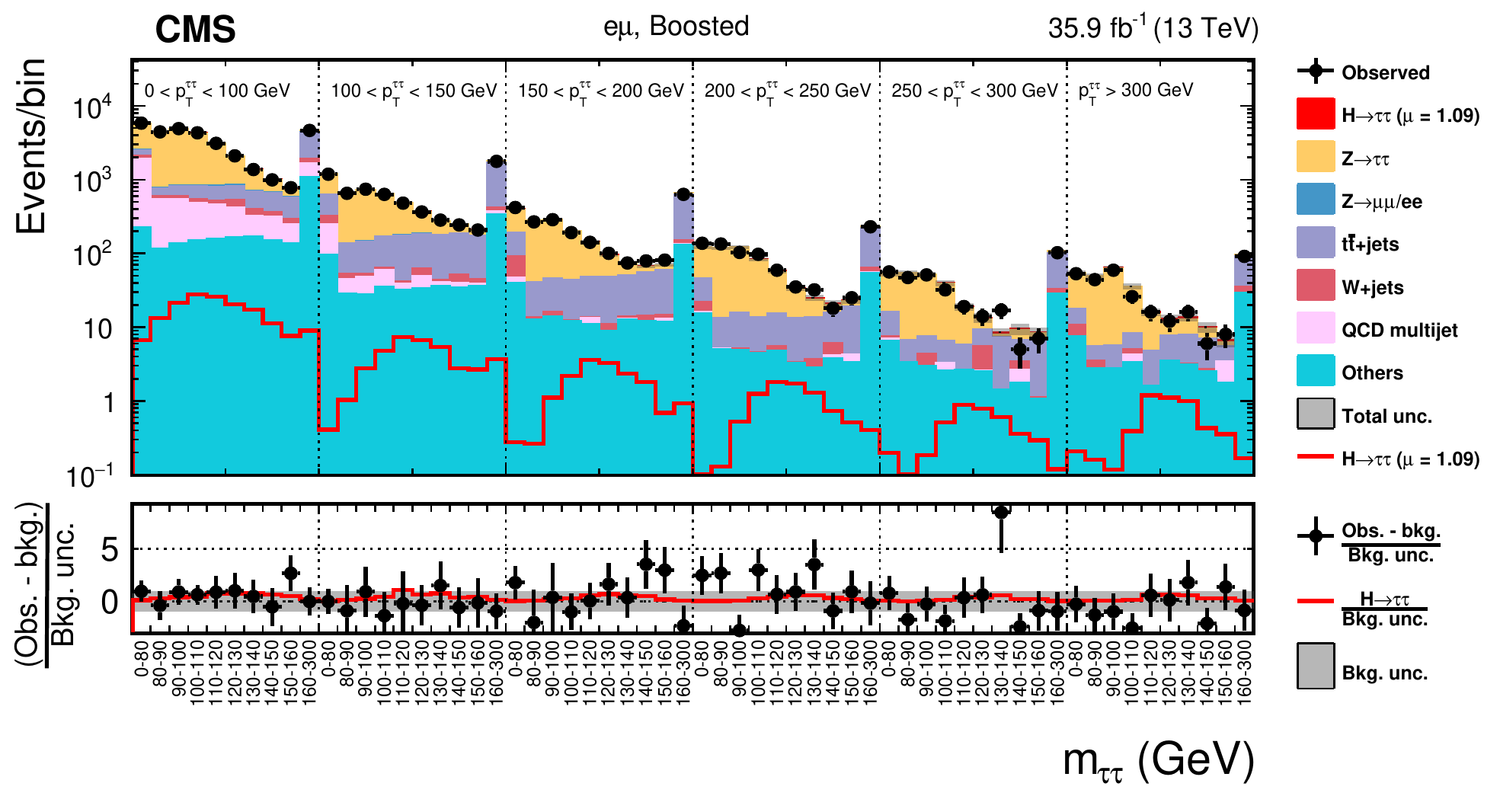}
     \caption{Observed and predicted 2D distributions in the boosted category of the $\Pe\Pgm$ decay channel. The description of the histograms is the same as in Fig.~\ref{fig:mass_tt_0jet}.}
     \label{fig:mass_et_vbf}
\end{figure*}

\begin{figure*}[htbp]
\centering
     \includegraphics[width=1.0\textwidth]{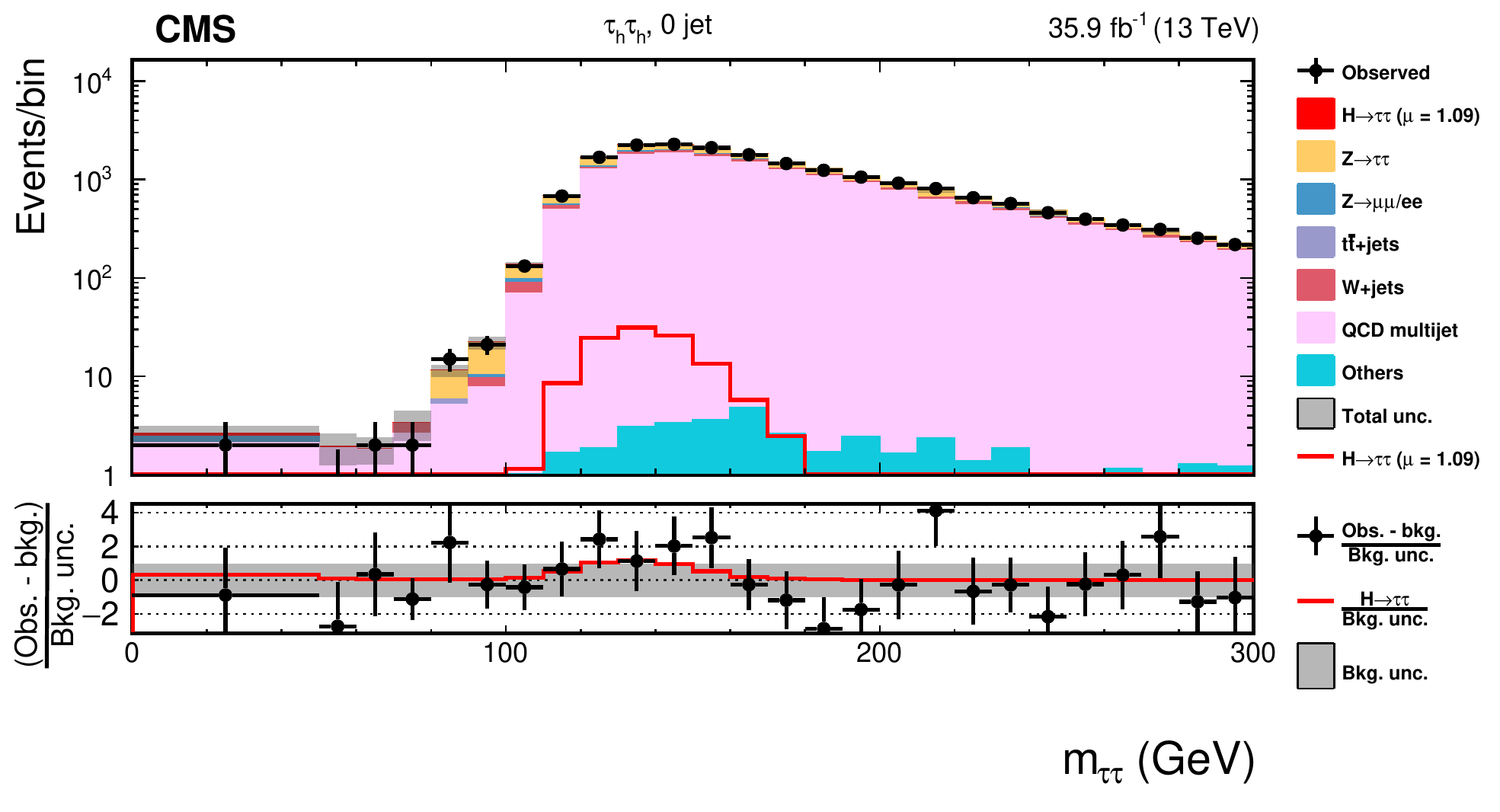}
     \caption{Observed and predicted distributions in the 0-jet category of the $\tauh\tauh$ decay channel. The description of the histograms is the same as in Fig.~\ref{fig:mass_tt_0jet}.}
     \label{fig:mass_et_boosted}
\end{figure*}

\begin{figure*}[htbp]
\centering
     \includegraphics[width=1.0\textwidth]{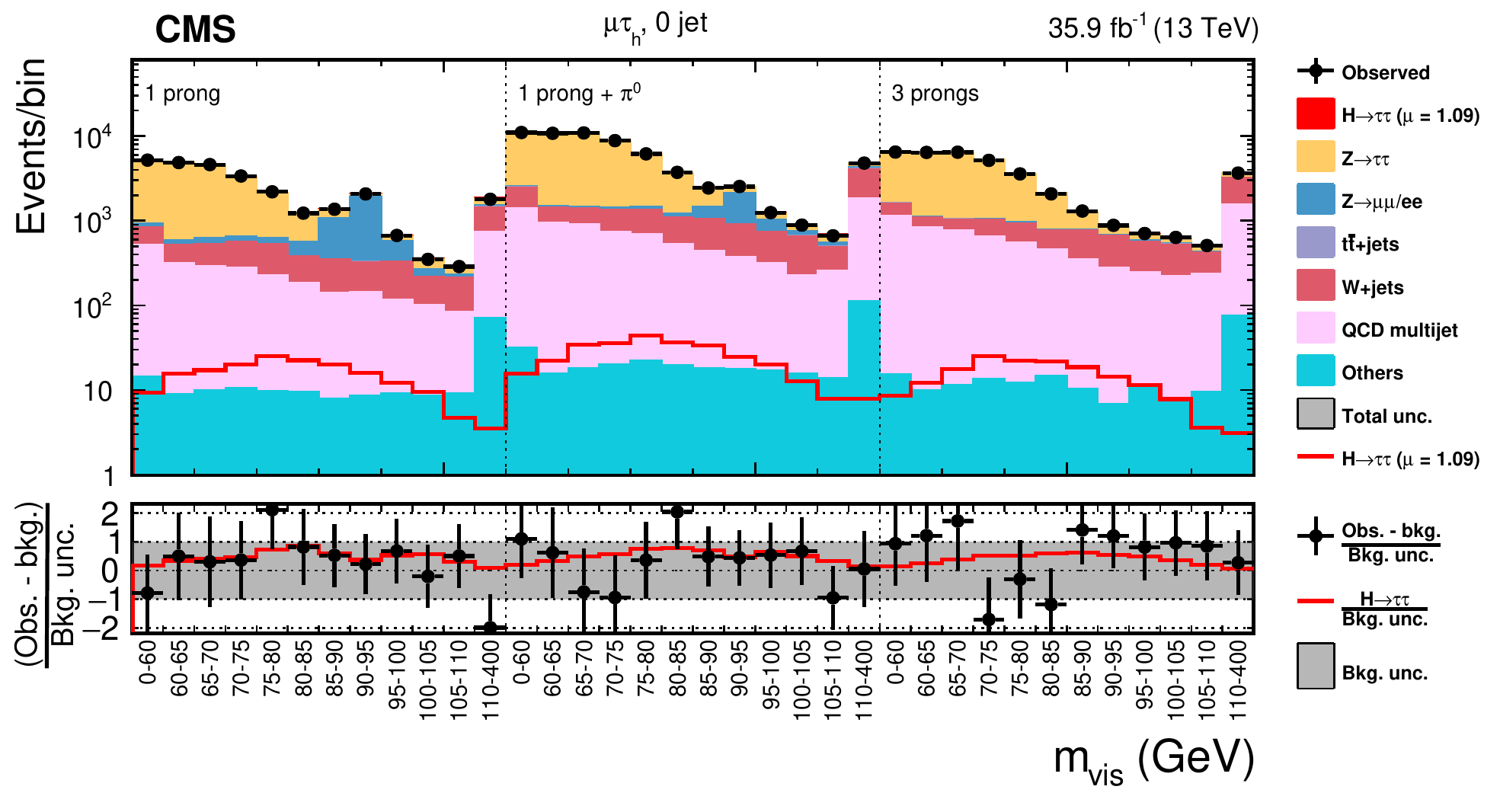}
     \caption{Observed and predicted 2D distributions in the 0-jet category of the $\Pgm\tauh$ decay channel. The description of the histograms is the same as in Fig.~\ref{fig:mass_tt_0jet}.}
     \label{fig:mass_em_0jet}
\end{figure*}

\begin{figure*}[htbp]
\centering
     \includegraphics[width=1.0\textwidth]{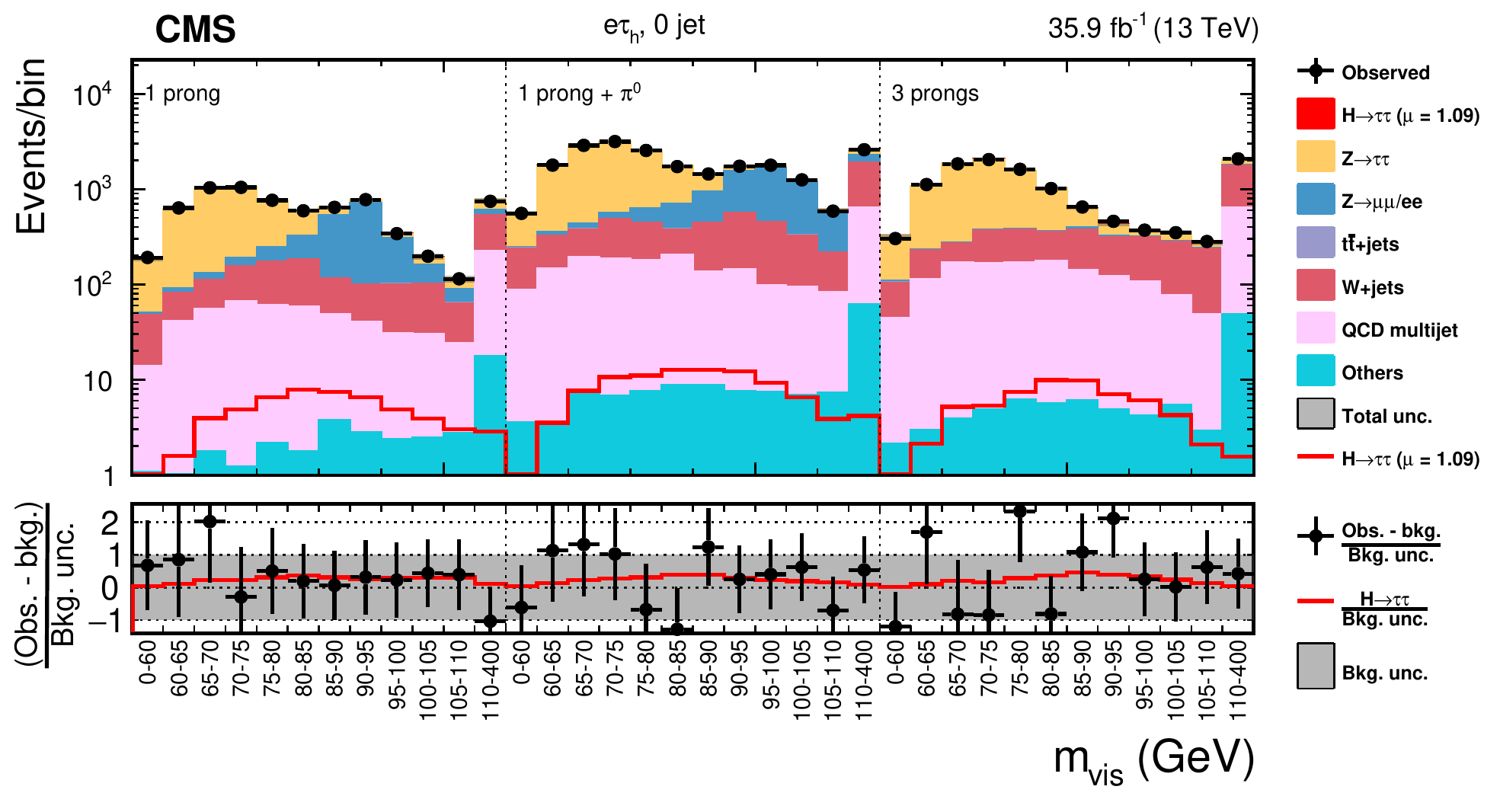}
     \caption{Observed and predicted 2D distributions in the 0-jet category of the $\Pe\tauh$ decay channel. The description of the histograms is the same as in Fig.~\ref{fig:mass_tt_0jet}.}
     \label{fig:mass_em_vbf}
\end{figure*}

\begin{figure*}[htbp]
\centering
     \includegraphics[width=1.0\textwidth]{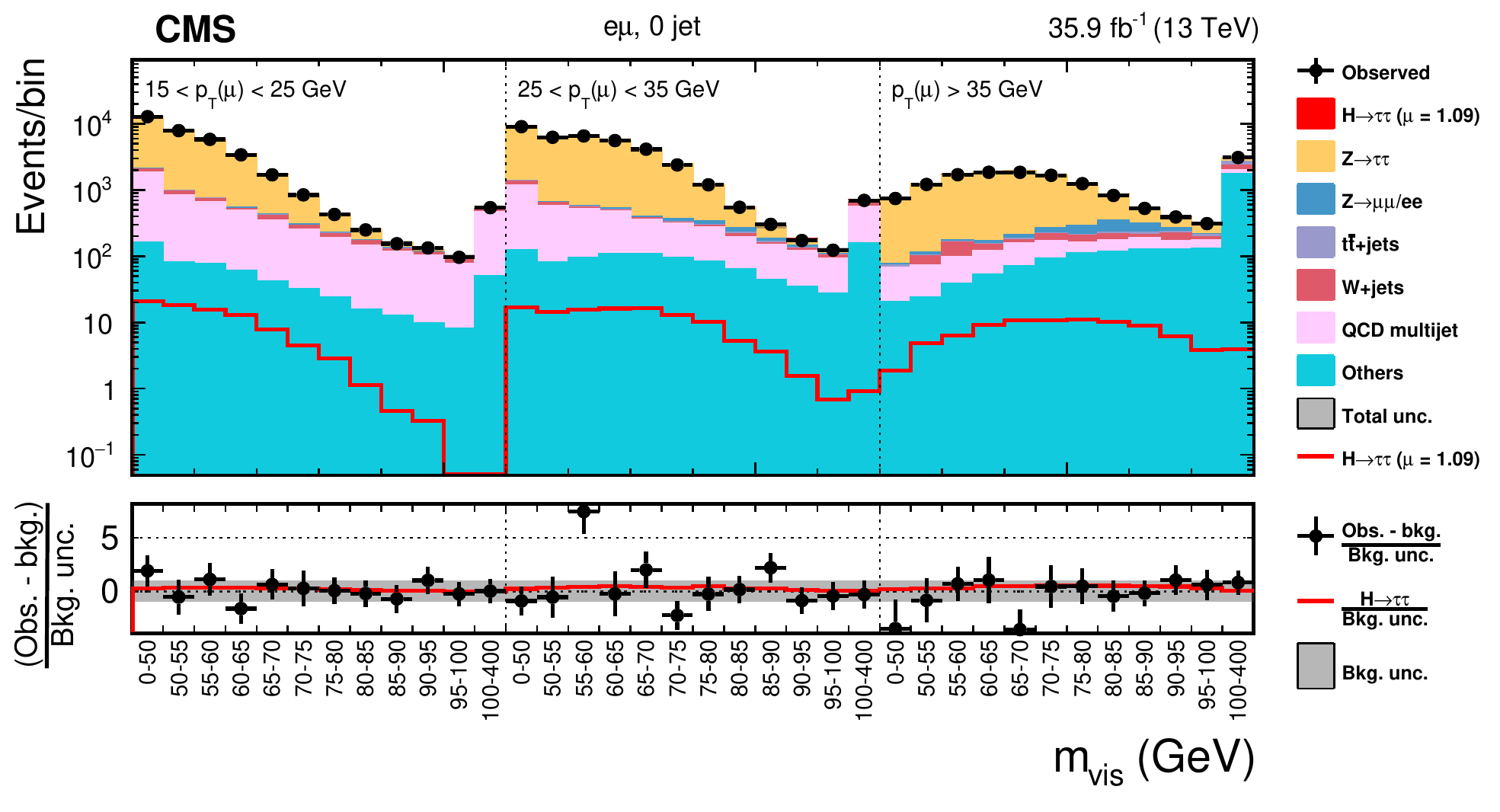}
     \caption{Observed and predicted 2D distributions in the 0-jet category of the $\Pe\Pgm$ decay channel. The description of the histograms is the same as in Fig.~\ref{fig:mass_tt_0jet}.}
     \label{fig:mass_em_boosted}
\end{figure*}

The extraction of the results involves a global maximum likelihood fit based on 2D distributions in all channels, shown in Figs.~\ref{fig:mass_tt_0jet}--\ref{fig:mass_em_boosted}, together with the control regions for the
$\ttbar$, QCD multijet, and $\PW +\text{jets}$ backgrounds. The choice of the binning is driven by the statistical precision of the background and data templates, leading to wider bins in the poorly-populated VBF category. The most sensitive category, VBF, is shown first and is followed by the boosted and 0-jet categories.
The signal prediction for a Higgs boson with $\mH = 125.09\GeV$ is normalized to its best fit cross section times branching fraction.
The background distributions are adjusted to the results of the global maximum likelihood fit.

The 2D distributions of the final discriminating variables obtained for each category and each channel in the signal regions, along with the control regions, are combined in a binned likelihood involving the expected and observed numbers of events in each bin.
The expected number of signal events is the one predicted for the production of
a SM Higgs boson of mass $\mH=125.09\GeV$ decaying into a pair of $\Pgt$ leptons,
multiplied by a signal strength modifier $\mu$ treated as a free parameter in the fit.

The systematic uncertainties are represented by nuisance parameters that are varied in the fit according to their probability density functions.
A log-normal probability density function is assumed for the nuisance parameters affecting the event yields of the various background contributions, whereas systematic uncertainties that affect the shape of the distributions are represented by nuisance parameters whose variation results in a continuous perturbation of the spectrum~\cite{Conway-PhyStat} and which are assumed to have a Gaussian probability density function.
Overall, the statistical uncertainty in the observed event yields is the dominant source of uncertainty for all combined results.

\begin{figure}[htb]
  \centering
    \includegraphics[width=\cmsFigWidth]{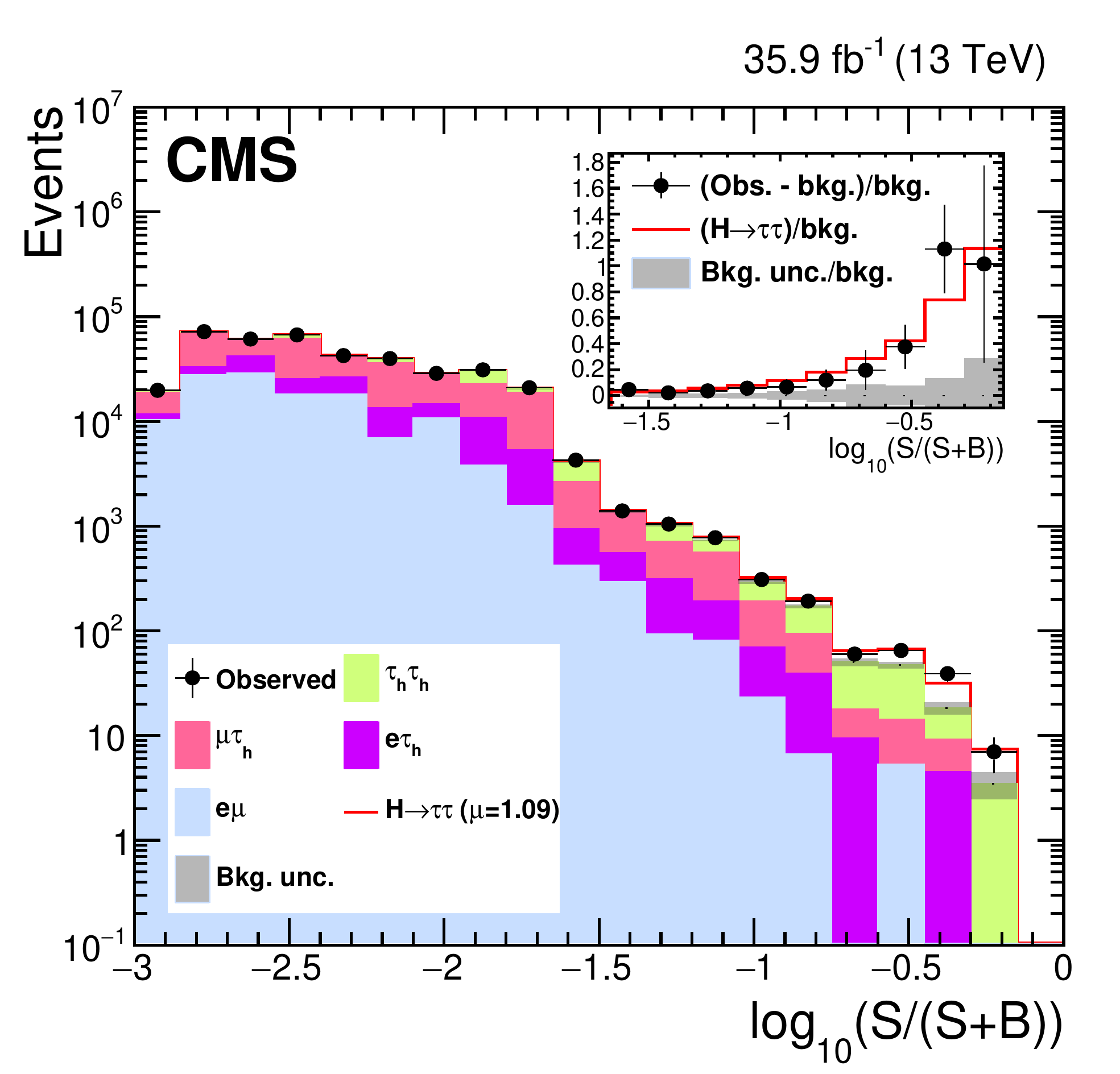}
   \caption{Distribution of the decimal logarithm of the ratio between the expected signal and the sum of expected signal and expected background in each bin of the mass distributions used to extract the results, in all signal regions. The background contributions are separated by decay channel. The inset shows the corresponding difference between the observed data and expected background distributions divided by the background expectation, as well as the signal expectation divided by the background expectation.
   }
    \label{fig:sb}

\end{figure}

Grouping events
in the signal region by their decimal logarithm of the ratio of the signal ($S$) to signal-plus-background ($S+B$) in each bin (Fig.~\ref{fig:sb}), an excess of observed events with respect to the SM background expectation is clearly visible in the most sensitive bins of the analysis. The expected background and signal contributions, as well as the observed number of events, are indicated per process and category in Table~\ref{tab:sb} for the bins with $\log_{10}(S/(S+B))>-0.9$. The channel that contributes the most to these bins is $\tauh\tauh$.

\begin{table*}
\centering
\topcaption{Background and signal expectations, together with the number of observed events, for bins in the signal region for which $\log_{10}(S/(S+B))>-0.9$, where
$S$ and $B$ are, respectively, the number of expected signal events for a Higgs boson with a mass $\mH = 125.09$\GeV and of expected background events, in those bins. The background uncertainty accounts for all sources of background uncertainty, systematic as well as statistical, after the global fit. The contribution from ``other backgrounds" includes events from diboson and single top quark production. The contribution from Higgs boson decays to a pair of $\PW$ bosons is zero in these bins.}
\label{tab:sb}
\newcolumntype{x}{D{,}{\,\pm\,}{4.2}}
\begin{tabular}{lxxxx}
Process & \multicolumn{1}{c}{$\Pe\Pgm$} &  \multicolumn{1}{c}{$\Pe\tauh$} &  \multicolumn{1}{c}{$\Pgm\tauh$} &  \multicolumn{1}{c}{$\tauh\tauh$} \\
\hline
$\PZ \to\Pgt\Pgt$ & 5.8, 2.2 & 21.2, 3.3 &34.6, 4.9 & 89.1, 6.9 \\
$\PZ \to \Pe\Pe/\Pgm\Pgm$ & 0.0, 0.0 & 2.9, 0.2 &3.7, 0.2 & 5.0, 0.2 \\
$\ttbar$+jets & 1.9, 0.1 & 10.4, 0.3 &22.2, 1.8 & 13.9, 0.5 \\
$\PW+\text{jets}$ & 0.8, 0.02 & 4.0, 0.3 &6.6, 1.3 & 7.6, 0.8 \\
QCD multijet & 2.1, 0.3 & 3.3, 2.5 &5.0, 1.3 & 35.5, 2.1 \\
Other backgrounds & 1.4, 0.1 & 5.2, 0.2 &6.1, 0.2 & 7.3, 0.2 \\[\cmsTabSkip]
$\Pg\Pg\PH, \PH \to\Pgt\Pgt$ & 0.6, 0.1 & 5.0, 0.6 &6.0, 0.6 & 27.4, 2.1  \\
VBF $\PH \to\Pgt\Pgt$ & 2.8, 0.3 & 5.1, 0.5 &12.55, 1.0 & 17.5, 1.0 \\
V$\PH, \PH \to\Pgt\Pgt$ & 0.0, 0.0 & 0.3, 0.0 &0.2, 0.0 & 1.3, 0.1  \\[\cmsTabSkip]
Total backgrounds & 12.1, 2.2 & 46.5, 4.1 &77.7, 5.5 & 156.2, 7.3 \\
Total signal & 3.4, 0.4 & 10.9, 0.8 &19.2, 1.4 & 48.3, 2.6  \\
Observed &  \multicolumn{1}{c}{11} &  \multicolumn{1}{c}{54} &  \multicolumn{1}{c}{91} &  \multicolumn{1}{c}{207}  \\
\hline
\end{tabular}
\end{table*}

\begin{figure}[htb]
  \centering
    \includegraphics[width=0.48\textwidth]{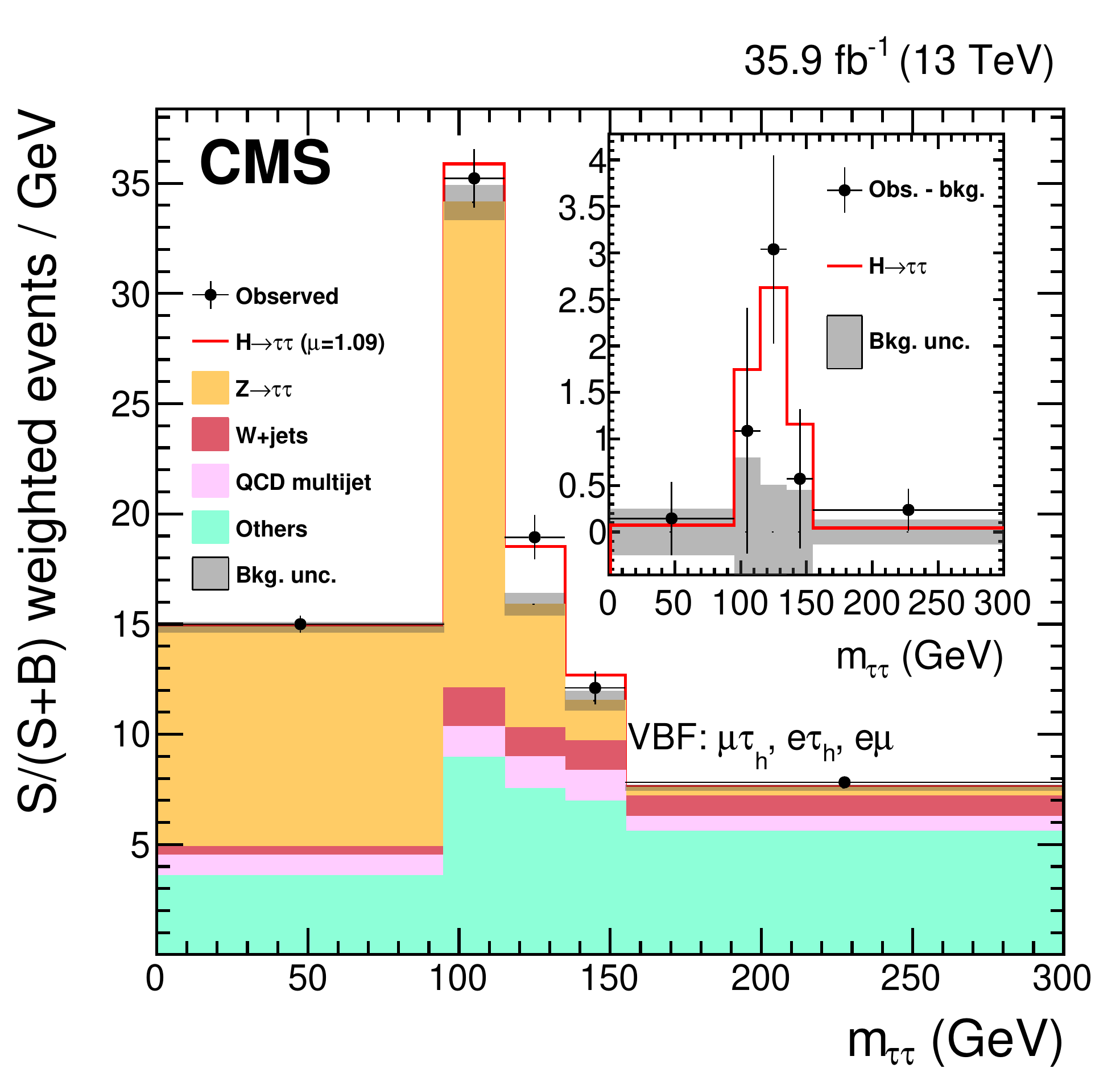}
    \includegraphics[width=0.48\textwidth]{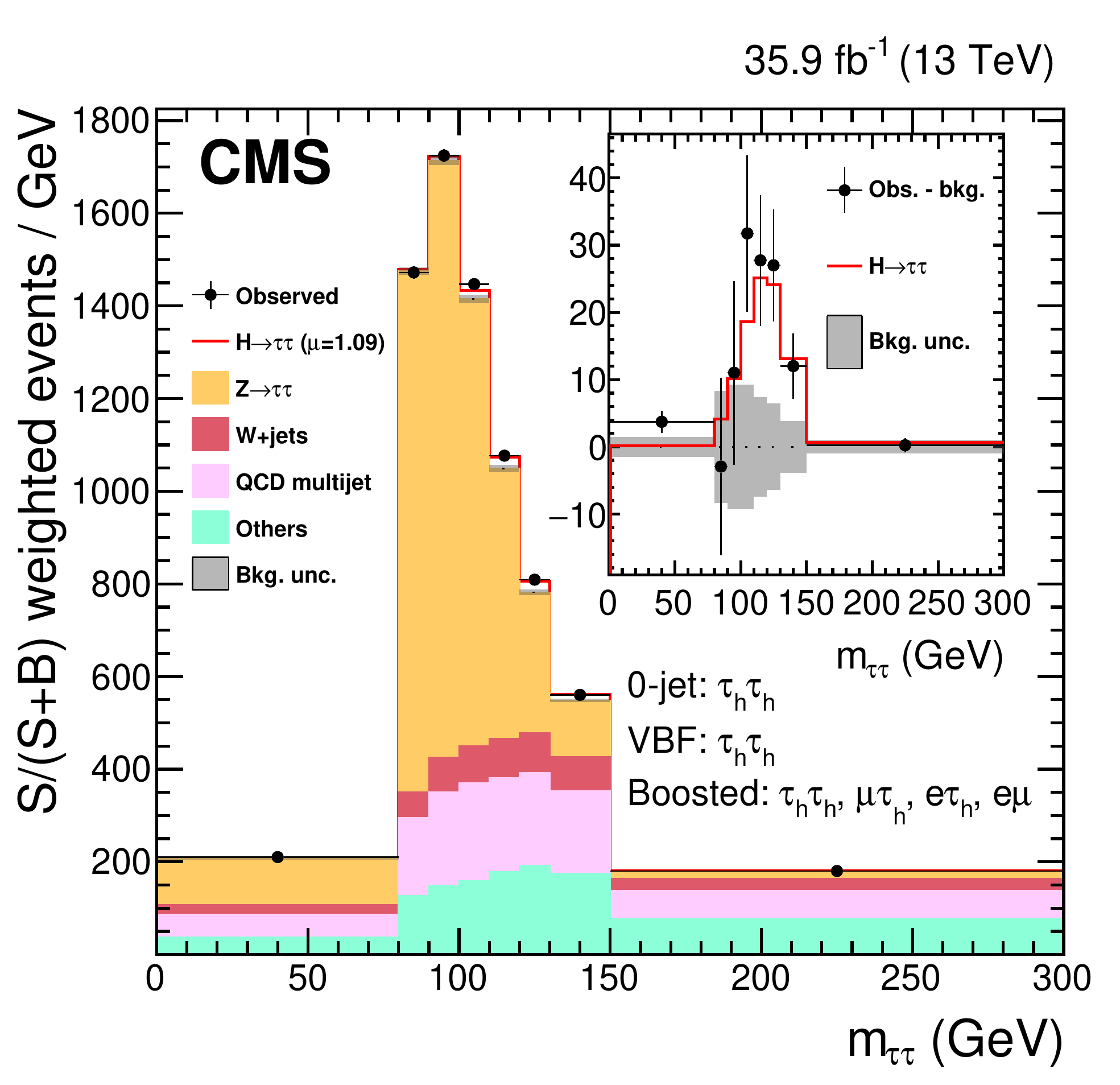}
   \caption{Combined observed and predicted $\mtt$ distributions. The \cmsLeft~pane includes the VBF category of the $\Pgm\tauh$, $\Pe\tauh$ and $\Pe\Pgm$ channels, and the \cmsRight~pane includes all other channels that make use of $\mtt$ instead of $\mvis$ for the signal strength fit. The binning reflects the one used in the 2D distributions, and does not allow merging of the two figures. The normalization of the predicted background distributions corresponds to the result of the global fit, while the signal is normalized to its best fit signal strength. The mass distributions for a constant range of the second dimension of the signal distributions are weighted according to $S/(S+B)$, where $S$ and $B$ are computed, respectively, as the signal or background contribution in the mass distribution excluding the first and last bins. The ``Others" background contribution includes events from diboson, $\ttbar$, and single top quark production, as well as Higgs boson decay to a pair of $\PW$ bosons and $\PZ$ bosons decaying to a pair of light leptons. The background uncertainty band accounts for all sources of background uncertainty, systematic as well as statistical, after the global fit. The inset shows the corresponding difference between the observed data and expected background distributions, together with the signal expectation. The signal yield is not affected by the reweighting.
   }
    \label{fig:massweighted}

\end{figure}

An excess of observed events relative to the background expectation is also visible in Fig.~\ref{fig:massweighted}, where every mass distribution for a constant range of the second dimension of the signal distributions has been summed with a weight of $S/(S+B)$  to increase the contribution of the most sensitive distributions. In this case, $S$ and $B$ are computed, respectively, as the signal or background contribution in the mass distribution excluding the first and last bins, in which the amount of signal is negligible. The signal regions that use $\mvis$ instead of $\mtt$, namely the 0-jet category of the $\Pgm\tauh$, $\Pe\tauh$ and $\Pe\Pgm$ channels, are not included. The two panes of Fig.~\ref{fig:massweighted} group the compatible bins of Figs.~\ref{fig:mass_tt_0jet}--\ref{fig:mass_em_boosted}.

\begin{figure}[!ht]
  \centering
    \includegraphics[width=0.5\textwidth]{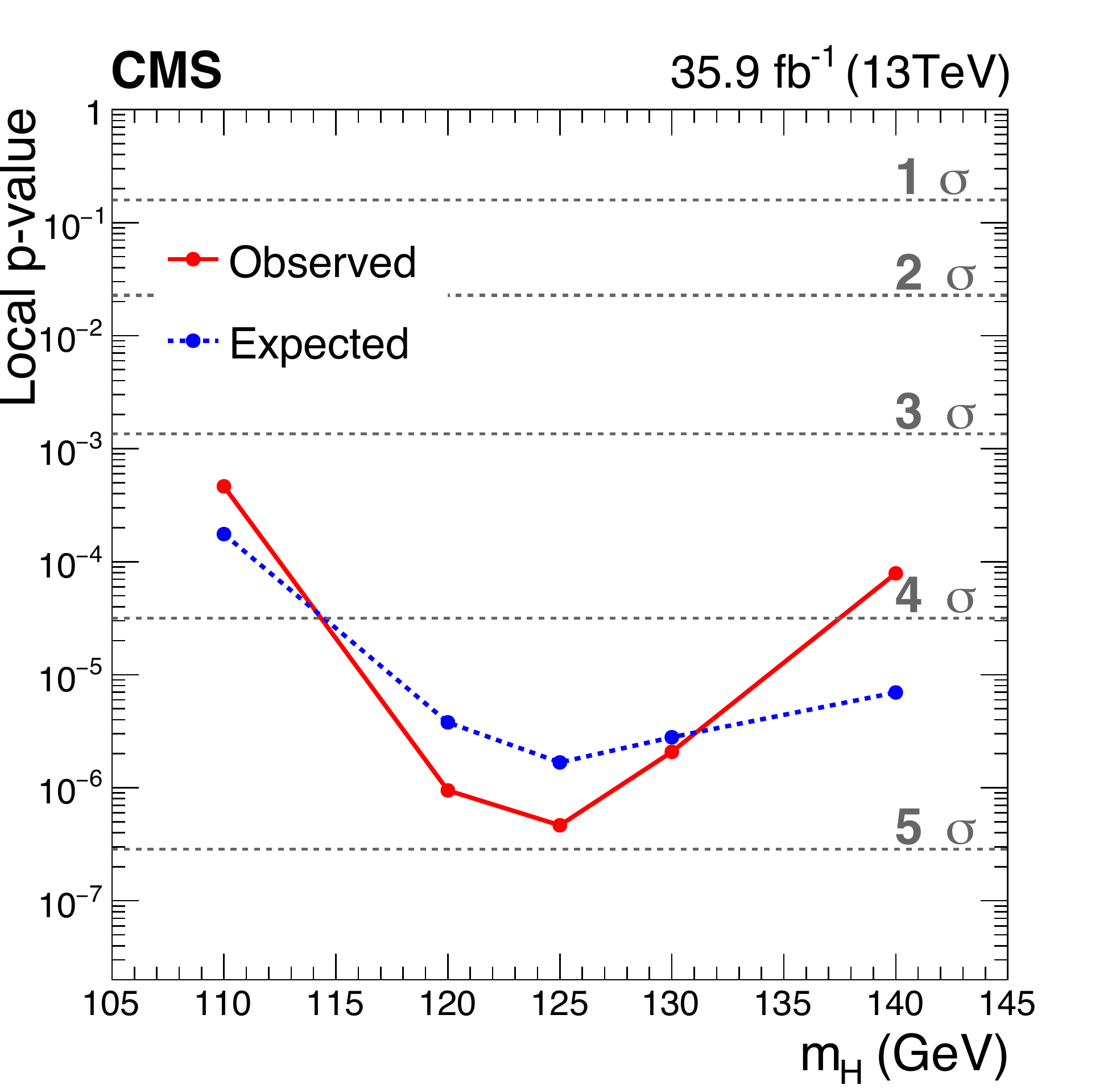}
   \caption{Local ${p}$-value and significance as a function of the SM Higgs boson mass hypothesis. The observation (red, solid) is compared to the expectation (blue, dashed) for a Higgs boson with a mass $\mH = 125.09$\GeV. The background includes Higgs boson decays to pairs of $\PW$ bosons, with $\mH = 125.09$\GeV.}
    \label{fig:pvalue}

\end{figure}

The excess in data is quantified by calculating the corresponding local $p$-value using a profile likelihood ratio test statistic~\cite{LHC-HCG-Report,Chatrchyan:2012tx,Junk,Read:2002hq}.
As shown in Fig.~\ref{fig:pvalue}, the observed significance for a SM Higgs boson with $\mH = 125.09$\GeV is 4.9 standard deviations, for an expected
significance of 4.7 standard deviations.

\begin{figure}[!ht]
  \centering
    \includegraphics[width=0.49\textwidth]{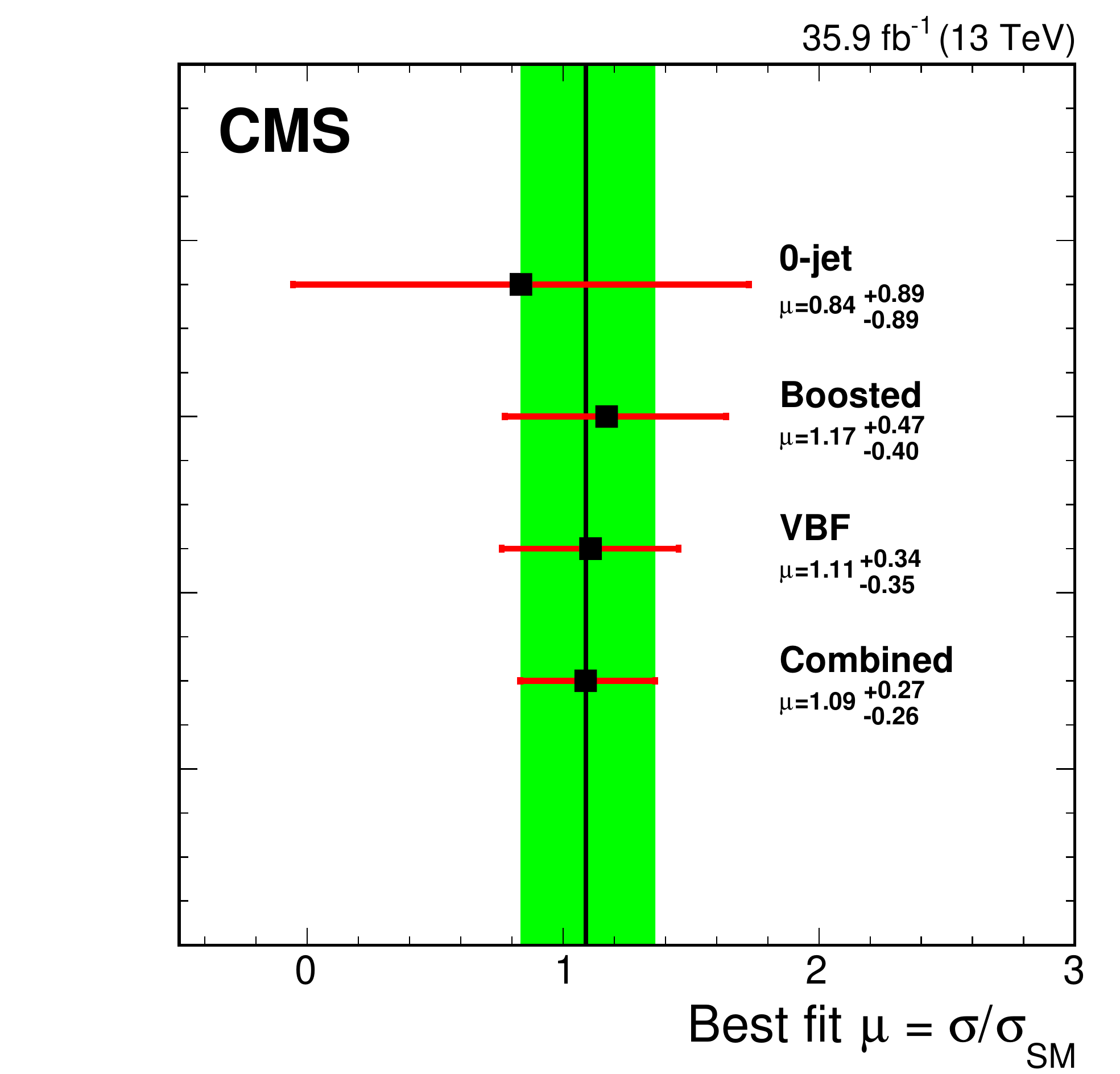}
    \includegraphics[width=0.49\textwidth]{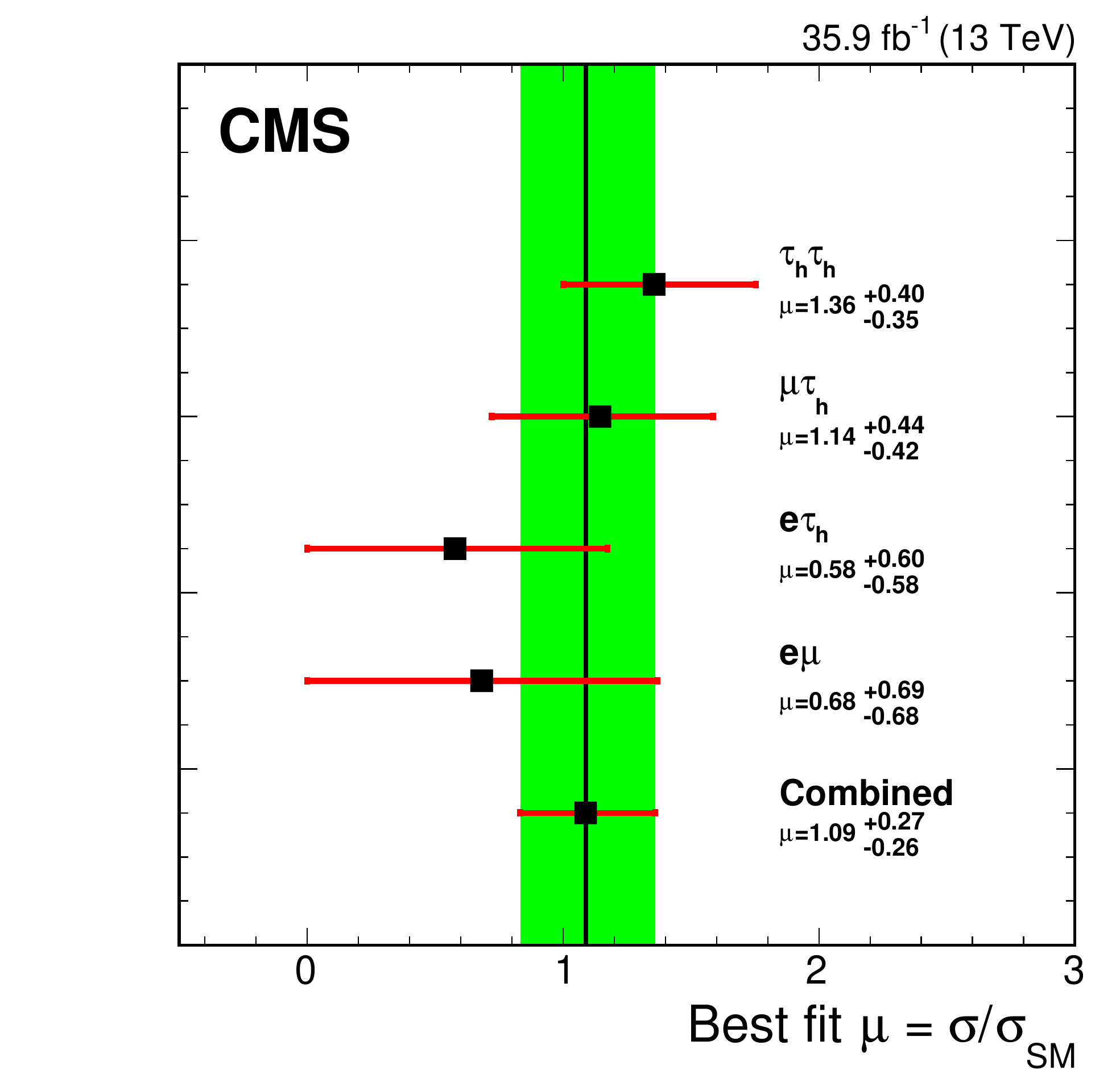}
   \caption{Best fit signal strength per category (\cmsLeft) and channel (\cmsRight), for $\mH = 125.09$\GeV. The constraints from the global fit are used to extract each of the individual best fit signal strengths. The combined best fit signal strength is $\mu = 1.09 ^{+0.27} _{-0.26}$.}
    \label{fig:muvalue}

\end{figure}

The corresponding best fit value for the signal strength $\mu$ is $1.09 ^{+0.27} _{-0.26}$ at $\mH = 125.09\GeV$. The uncertainty in the best fit signal strength can be decomposed into four components: theoretical uncertainties, bin-by-bin statistical uncertainties on the backgrounds, other systematic uncertainties, and the statistical uncertainty. In this format, the best fit signal strength is $\mu = 1.09^{+0.15}_{-0.15}\stat{}^{+0.16}_{-0.15}\syst{}^{+0.10}_{-0.08}\thy{}^{+0.13}_{-0.12}$ (bin-by-bin).
The individual best fit signal strengths per channel and per category, using the constraints obtained on the systematic uncertainties through the global fit, are given in Fig.~\ref{fig:muvalue}; they demonstrate the channel- and category-wise consistency of the observation with the SM Higgs boson hypothesis.

\begin{figure}[!ht]
  \centering
    \includegraphics[width=\cmsFigWidth]{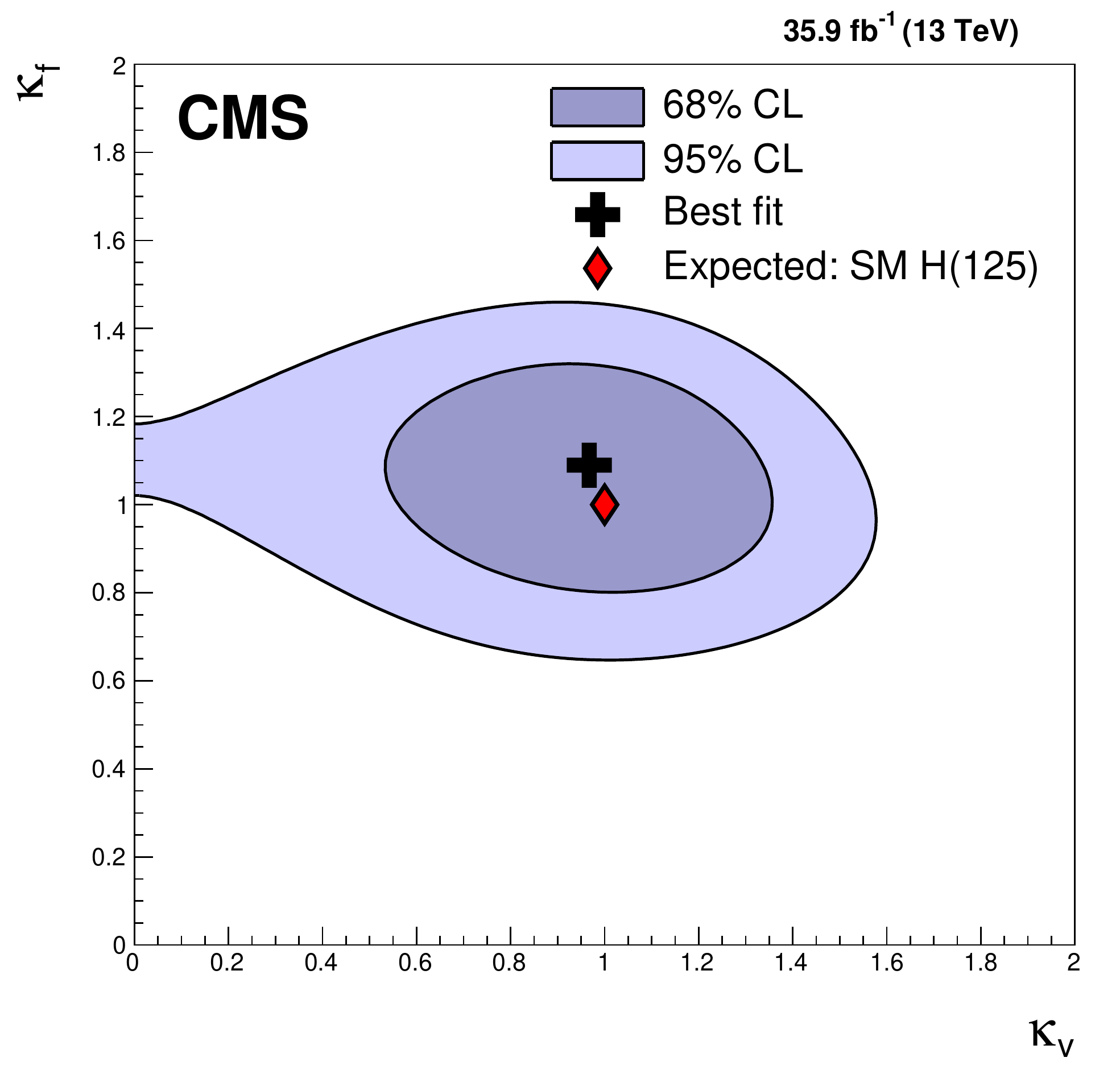}
   \caption{Scan of the negative log-likelihood difference as a function of $\kappa_V$ and $\kappa_f$, for $\mH = 125.09$\GeV.  All nuisance parameters are profiled for each point. For this scan, the $\Pp\Pp\to \PH\to\PW\PW$ contribution is treated as a signal process.}
    \label{fig:kVkf}

\end{figure}

A likelihood scan is performed for $\mH=125.09$\GeV in the ($\kappa_\mathrm{V}$,$\kappa_\mathrm{f}$) parameter space, where $\kappa_\mathrm{V}$ and $\kappa_\mathrm{f}$ quantify, respectively, the ratio between the measured and the SM value for the couplings of the Higgs boson to vector bosons and fermions, with the methods described in Ref.~\cite{Chatrchyan:2014nva}. For this scan only, Higgs boson decays to pairs of $\PW$ bosons are considered as part of the signal. All nuisance parameters are profiled for each point of the scan. As shown in Fig.~\ref{fig:kVkf}, the observed likelihood contour is consistent with the SM expectation of $\kappa_\mathrm{V}$ and $\kappa_\mathrm{f}$ equal to unity.

The results are combined with the results of the search for $\PH\to\Pgt\Pgt$ performed with the data collected with the CMS detector at center-of-mass energies of 7 and 8\TeV~\cite{Khachatryan:2014jba}, using a common signal strength for all data taking periods. All uncertainties are considered as fully uncorrelated between the different center-of-mass energies. The combination leads to an observed and an expected significance of 5.9 standard deviations. The corresponding best fit value for the signal strength $\mu$ is $0.98\pm 0.18$ at $\mH = 125.09\GeV$. This constitutes the most significant direct measurement of the coupling of the Higgs boson to fermions by a single experiment.

\section{Summary}

A measurement of the $\PH\to\Pgt\Pgt$ signal strength, using events recorded in proton-proton collisions by the CMS experiment at the LHC in 2016 at a center-of-mass energy of 13\TeV, has been presented. Event categories are designed to target Higgs boson signal events produced by gluon or vector boson fusion.
The results are extracted via maximum likelihood fits in two-dimensional planes, and give an observed significance for Higgs boson decays to $\Pgt$ lepton pairs of 4.9 standard deviations, to be compared with an expected significance of 4.7 standard deviations. The combination with the corresponding measurement performed at center-of-mass energies of 7 and 8\TeV with the CMS detector leads to the first observation by a single experiment of decays of the Higgs boson to pairs of $\Pgt$ leptons, with a significance of 5.9 standard deviations.

\begin{acknowledgments}
We congratulate our colleagues in the CERN accelerator departments for the excellent performance of the LHC and thank the technical and administrative staffs at CERN and at other CMS institutes for their contributions to the success of the CMS effort. In addition, we gratefully acknowledge the computing centers and personnel of the Worldwide LHC Computing Grid for delivering so effectively the computing infrastructure essential to our analyses. Finally, we acknowledge the enduring support for the construction and operation of the LHC and the CMS detector provided by the following funding agencies: BMWFW and FWF (Austria); FNRS and FWO (Belgium); CNPq, CAPES, FAPERJ, and FAPESP (Brazil); MES (Bulgaria); CERN; CAS, MoST, and NSFC (China); COLCIENCIAS (Colombia); MSES and CSF (Croatia); RPF (Cyprus); SENESCYT (Ecuador); MoER, ERC IUT, and ERDF (Estonia); Academy of Finland, MEC, and HIP (Finland); CEA and CNRS/IN2P3 (France); BMBF, DFG, and HGF (Germany); GSRT (Greece); OTKA and NIH (Hungary); DAE and DST (India); IPM (Iran); SFI (Ireland); INFN (Italy); MSIP and NRF (Republic of Korea); LAS (Lithuania); MOE and UM (Malaysia); BUAP, CINVESTAV, CONACYT, LNS, SEP, and UASLP-FAI (Mexico); MBIE (New Zealand); PAEC (Pakistan); MSHE and NSC (Poland); FCT (Portugal); JINR (Dubna); MON, RosAtom, RAS, RFBR and RAEP (Russia); MESTD (Serbia); SEIDI, CPAN, PCTI and FEDER (Spain); Swiss Funding Agencies (Switzerland); MST (Taipei); ThEPCenter, IPST, STAR, and NSTDA (Thailand); TUBITAK and TAEK (Turkey); NASU and SFFR (Ukraine); STFC (United Kingdom); DOE and NSF (USA).

\hyphenation{Rachada-pisek} Individuals have received support from the Marie-Curie program and the European Research Council and Horizon 2020 Grant, contract No. 675440 (European Union); the Leventis Foundation; the A. P. Sloan Foundation; the Alexander von Humboldt Foundation; the Belgian Federal Science Policy Office; the Fonds pour la Formation \`a la Recherche dans l'Industrie et dans l'Agriculture (FRIA-Belgium); the Agentschap voor Innovatie door Wetenschap en Technologie (IWT-Belgium); the Ministry of Education, Youth and Sports (MEYS) of the Czech Republic; the Council of Science and Industrial Research, India; the HOMING PLUS program of the Foundation for Polish Science, cofinanced from European Union, Regional Development Fund, the Mobility Plus program of the Ministry of Science and Higher Education, the National Science Center (Poland), contracts Harmonia 2014/14/M/ST2/00428, Opus 2014/13/B/ST2/02543, 2014/15/B/ST2/03998, and 2015/19/B/ST2/02861, Sonata-bis 2012/07/E/ST2/01406; the National Priorities Research Program by Qatar National Research Fund; the Programa Clar\'in-COFUND del Principado de Asturias; the Thalis and Aristeia programs cofinanced by EU-ESF and the Greek NSRF; the Rachadapisek Sompot Fund for Postdoctoral Fellowship, Chulalongkorn University and the Chulalongkorn Academic into Its 2nd Century Project Advancement Project (Thailand); and the Welch Foundation, contract C-1845. \end{acknowledgments}

\clearpage

\bibliography{auto_generated}

\cleardoublepage \appendix\section{The CMS Collaboration \label{app:collab}}\begin{sloppypar}\hyphenpenalty=5000\widowpenalty=500\clubpenalty=5000\textbf{Yerevan Physics Institute,  Yerevan,  Armenia}\\*[0pt]
A.M.~Sirunyan, A.~Tumasyan
\vskip\cmsinstskip
\textbf{Institut f\"{u}r Hochenergiephysik,  Wien,  Austria}\\*[0pt]
W.~Adam, F.~Ambrogi, E.~Asilar, T.~Bergauer, J.~Brandstetter, E.~Brondolin, M.~Dragicevic, J.~Er\"{o}, M.~Flechl, M.~Friedl, R.~Fr\"{u}hwirth\cmsAuthorMark{1}, V.M.~Ghete, J.~Grossmann, J.~Hrubec, M.~Jeitler\cmsAuthorMark{1}, A.~K\"{o}nig, N.~Krammer, I.~Kr\"{a}tschmer, D.~Liko, T.~Madlener, I.~Mikulec, E.~Pree, D.~Rabady, N.~Rad, H.~Rohringer, J.~Schieck\cmsAuthorMark{1}, R.~Sch\"{o}fbeck, M.~Spanring, D.~Spitzbart, W.~Waltenberger, J.~Wittmann, C.-E.~Wulz\cmsAuthorMark{1}, M.~Zarucki
% ...
\vskip\cmsinstskip
% ...
1:~~Also at Vienna University of Technology, Vienna, Austria\\
% ...

\end{sloppypar}
\end{document}